\documentclass[11pt]{article}

\usepackage{makecell}

\usepackage[utf8]{inputenc}
\usepackage{amssymb,amsmath,amsfonts,eurosym,geometry,ulem,graphicx,caption,color,setspace, comment,footmisc,caption,pdflscape,array, booktabs, dcolumn, threeparttable, rotating, afterpage, pdfpages, geometry, titlesec, csquotes, rotating, longtable, verbatimbox, multirow, afterpage, enumitem, subcaption}
\usepackage{psfrag}
\usepackage{dsfont}
\usepackage{afterpage}

\graphicspath{{./}{./Figures/}{./Tables/}{./Tables/appendix_tables/}}

\usepackage{siunitx}
\newcolumntype{d}{S[input-symbols = ()]}
\usepackage{tabularx}

\usepackage[hyphens]{url}
\usepackage[dvipsnames]{xcolor}
\usepackage{todonotes}
\usepackage{csquotes}

\usepackage{epigraph}

\usepackage{ntheorem}

\makeatletter
\newcounter{subhyp} 
\let\savedc@hyp\c@hyp

\newcommand{\normhyp}{%
  \let\c@hyp\savedc@hyp % revert to the old one
  \renewcommand\thehyp{\arabic{hyp}}%
} 
\makeatother
 
\usepackage[hashEnumerators,smartEllipses]{markdown}

\usepackage{lscape}
\usepackage[colorlinks,%
citecolor=blue, 
filecolor=black,%
linkcolor=black,%
urlcolor=blue]{hyperref} % for linking between references, figures, TOC, etc in the pdf document
\usepackage[toc, page, title]{appendix}
\usepackage{titletoc}

\newcommand{\dobib}{
    \bibliography{reference}
}

\usepackage[english]{babel}
\usepackage[style=apa, sorting=nyt,natbib=true,backend=biber,uniquename=false]{biblatex}
\AtEveryBibitem{%
  \clearfield{annotation}%
  \clearfield{note}%
}

\newcolumntype{H}{>{\setbox0=\hbox\bgroup}c<{\egroup}@{}}

\newcolumntype{Y}{>{\centering\arraybackslash}X}

\newcommand{\sym}[1]{{#1}}

\begin{document}
\renewcommand{\dobib}{}

\title{Bread Upon the Waters: Corporate Science and the Benefits from Follow-On Public Research}

\author{Dror Shvadron\thanks{University of Toronto, Rotman School of Management. E-mail: \href{mailto:dror.shvadron@rotman.utoronto.ca}{dror.shvadron@rotman.utoronto.ca}. \\ I am particularly indebted to Ashish Arora, Wesley M. Cohen, Sharon Belenzon, and Bronwyn H. Hall for extensive advice and guidance. For thoughtful suggestions, I would like to thank Matt Marx, Ines Black, Sharique Hasan, Brian Silverman, Brent Goldfarb, Micha\"{e}l Bikard, Daniel Souza (discussant), Arvids Zeidonis (discussant), Markus Perkmann (discussant), Keyvan Vakili (discussant), Annamaria Conti (discussant), Scott Stern (discussant), Felix Poege (discussant) and numerous participants at the Workshop on The Organization, Economics and  Policy of Scientific Research 2023, the Open Innovation in Science Research Conference 2023, Strategy Science Conference 2023, the CCC Doctoral Conference 2023, the Roundtable for Engineering Entrepreneurship Research 2023, and the Academy of Management Annual Meeting 2024. Lastly, I would like to thank my fellow colleagues at the Fuqua Strategy Area.
}}
\date{\today}

\begin{titlepage}
\maketitle

\begin{abstract}
\noindent

\noindent Why do firms produce scientific research and make it available to the public, including their rivals? Prior literature has emphasized the tension between imitation risks from disclosure and scientists' preferences for publication. This study examines an additional managerial consideration: the value of follow-on research conducted by external scientists building upon firms' publications. Using data on U.S. public firms' scientific publications from 1990 to 2012, and a novel instrumental variable based on quasi-random journal issue assignment, I find that accumulation of follow-on research is associated with increased subsequent scientific investments, improved patenting outcomes, and greater employee retention by the originating firms. Benefits are more pronounced for firms with complementary assets and those operating in emerging research fields. Beyond serving as direct input into innovation, follow-on research provides external validation of internal research programs, helping managers allocate resources under conditions of scientific uncertainty. These findings demonstrate that firms benefit when their scientific disclosures inspire follow-on research by the broader scientific community.

\end{abstract}

\setcounter{page}{0}
\thispagestyle{empty}
\end{titlepage}

\pagebreak \newpage

\doublespacing

% Content
\epigraph{\textit{Cast your bread upon the waters, for you will find it after many days. - Ecclesiastes 11}}{}
\section{Introduction}

Why do firms produce scientific research and make it available to the public, including their rivals? I find that firms' scientific publications are infrequently cited by their own patents. These publications do, however, influence external scientific research, which these firms later utilize. While prior literature has highlighted the benefits of corporate science  \citep{FlemingSorenson2004ScienceMapTechnological, AroraBelenzonSheer2021KnowledgeSpilloversCorporate,CohenLevinthal1990AbsorptiveCapacityNew, Rosenberg1990WhyFirmsBasic}, the publication of findings has been primarily associated with the risk of imitation \citep{Stern2004ScientistsPayBe, SauermannRoach2014NotAllScientists, GansMurrayStern2017ContractingDisclosureScientific}. This study shows that alongside these risks, publications can generate benefits through follow-on research by academics building upon firms' published findings. 
Such follow-on research increases firms' employee retention, drives subsequent investments in science, and improves patenting outcomes both by providing inputs for innovation and by validating research quality.
Firms that possess complementary assets are more likely to benefit. 
These findings suggest that when firms can shape the knowledge environment in which they operate, participation in public research can be a strategic choice that drives subsequent performance \citep{HelfatKaulKetchenJr.EtAl2023RenewingResourcebasedViewa, GavettiHelfatMarengo2017SearchingShapingQuest}.
I contribute to understanding the determinants of firms' participation in public research and how the scientific community drives corporate innovation \citep{Jaffe1989RealEffectsAcademic, Mansfield1991AcademicResearchIndustrial, CohenNelsonWalsh2002LinksImpactsInfluence}.

Publishing scientific findings entails significant risks for firms, as disclosure enables competitors to learn about research capabilities, technological approaches, and facilitates imitative entry \citep{Rosenberg1990WhyFirmsBasic, AroraBelenzonSheer2021KnowledgeSpilloversCorporate}. Despite these risks, firms often permit (and promote) publication \citep{HendersonCockburn1994MeasuringCompetenceExploring}. 
Researcher employees highly value open science participation and some will accept lower wages in exchange for publishing opportunities \citep{Stern2004ScientistsPayBe, SauermannRoach2014NotAllScientists}. Moreover, allowing them to publish can drive their innovative performance \citep{SauermannCohen2010WhatMakesThem}.
The literature thus established that firms' disclosure strategies emerge from bargaining between managers seeking to minimize competitive risks and scientists preferring greater openness \citep{GansMurrayStern2017ContractingDisclosureScientific}. This framework implies that publications, by enhancing scientists' academic recognition, may increase their labor market mobility and ease job switching. Overlooked in this analysis, however, are the effects publications might have on the broader scientific ecosystem.

Firms have various channels by which they can influence public research. Prior literature has focused primarily on direct ties \citep{PerkmannSalandraTartariEtAl2021AcademicEngagementReview}, such as geographic proximity \citep{Sohn2021HowLocalIndustry}, funding \citep{BabinaHeHowellEtAl2023CuttingInnovationEngine}, and collaborations \citep{CockburnHenderson1998AbsorptiveCapacityCoauthoring, BikardVakiliTeodoridis2018WhenCollaborationBridges, BaruffaldiPoege2022StarsHowFirms}. %Firms can establish these ties by, for example, attending and sponsoring academic conferences \citep{BaruffaldiPoege2022StarsHowFirms}. 
Beyond these direct relationships, the disclosure of findings through scientific publications represents another channel of influence. By publishing scientific papers, firms can influence the direction of scientific research even without establishing direct ties with specific academics. The literature has also established that corporate scientific publications correlate with higher market values \citep{SimethCincera2016CorporateScienceInnovation}, and that knowledge spillovers from corporate R\&D can positively affect firm performance \citep{BloomSchankermanReenen2013IdentifyingTechnologySpillovers, Belenzon2012CumulativeInnovationMarket, AlnuaimiGeorge2016AppropriabilityRetrievalKnowledge}. 
 
Less clear is whether publications mobilize external research in ways that are valuable for the originating firm \citep{AlexyGeorgeSalter2013CuiBonoSelectivea}, and if so, through what mechanisms and with what  implications for subsequent scientist retention and  R\&D performance.

In this paper, I ask how firms benefit from external research that builds upon their own investments in science. I define participation in public research as the decision to invest in science and publicly disclose scientific findings. The prospect of valuable follow-on research from external sources can incentivize firms to participate in public research in the first place.  I use the DISCERN database on scientific publications and patents of publicly listed U.S.-based firms between 1990 and 2012 \citep*{AroraBelenzonSheer2020DISCERNDukeInnovationa}, matched to data from Microsoft Academic Graph (MAG), Dimensions.ai, the American Men and Women of Science (AMWS) directory, and several complementary datasets.\footnote{I thank Hansen Zhang for sharing a match between AMWS and DISCERN data. I thank Bernardo Dionisi for sharing a match of DISCERN to Microsoft Academic Graph.} I measure external follow-on research using scientific citations through multiple generations. Using these data, I test whether external follow-on research drives subsequent scientist retention, scientific publishing,  hiring of academics, and patenting by the originating firms. I then explore the conditions that moderate these effects and the mechanisms that enable them.

An example illustrates the potential value of external follow-on findings. In 1986, two IBM researchers, M\"uller and Bednorz, made a breakthrough discovery. They were the first to find a material that behaves as a superconductor in high temperatures. According to IBM's website:

\begin{quote}
    \textit{\small The scientific community shook. Scientists from across the world reproduced, modified and improved M\"uller and Bednorz’s process at a breakneck pace—reigniting global interest in superconductors and accelerating superconductor development. Based on M\"uller and Bednorz’s discovery, scientists soon developed materials that\ldots opened the door for a multitude of practical applications}.\footnote{IBM’s 100 Icons of Progress. ``High-Temperature Superconductors.'' Retrieved from \url{https://www.ibm.com/ibm/history/ibm100/us/en/icons/hightempsuperconductors/}.}
\end{quote}

Indeed, in sixteen of its patents, IBM cited the original paper by \citet{BednorzMuller1986PossibleHighTcSuperconductivity}. The paper was cited by 6,166 publications by authors unaffiliated with IBM, which generated roughly 64,000 second-generation and 381,000 third-generation citations. Beyond IBM's patents that directly cited the original paper, 563 additional IBM patents cited these follow-on publications. Based on patent value estimates from  \citet{KoganPapanikolaouSeruEtAl2017TechnologicalInnovationResource}, IBM's private value associated with the original sixteen patents was \$146 million. An additional \$1.9 billion is associated with IBM's patents related to the follow-on research. IBM's ability to capture value from external follow-on research contributes to this latter figure.

External follow-on research can serve as an input into firms' subsequent innovation, expanding capabilities beyond internal expertise. Academic contributions can develop downstream applications of firms' upstream science or broaden basic science in response to downstream challenges. Greater opportunities for knowledge recombination \citep{FlemingSorenson2004ScienceMapTechnological} make the firm's scientists more productive, increasing the value of retaining them. Obtaining such inputs through publication can be viewed as a method for broad knowledge search \citep{LeiponenHelfat2010InnovationObjectivesKnowledge}. Direct channels of accessing external expertise, such as redirecting existing collaborators toward new topics by funding, face significant constraints \citep{Myers2020ElasticityScience}. By making their work widely available, firms instead allow unaffiliated academics to self-select into producing follow-on research when it aligns with their existing interests. This reveals researchers working on related topics and facilitates new direct ties through hiring and future collaborations  \citep{PerkmannSalandraTartariEtAl2021AcademicEngagementReview}.

In addition, external follow-on research can be valuable even when not used as an input. 
Under uncertainty in scientific research \citep{AzoulayStuartWang2014MatthewEffectFable, JinMaUzzi2021ScientificPrizesExtraordinary}, corporate R\&D managers observe follow-on research to infer the quality of individuals' work and promising research trajectories.
These signals can therefore redirect managerial attention and influence subsequent R\&D investments by the firm. Moreover, external validation increases the firm's incentive to retain validated scientists, as their work has demonstrated value. This argument is consistent with the view that external validation of research quality can reduce uncertainty around nascent research initiatives \citep{CockburnHenderson1998AbsorptiveCapacityCoauthoring}.\footnote{In a recent blog post, a past employee at Google Brain claimed that ``Google's researcher promotion criteria were for some time linked to external recognition of research significance.'' (Lee, B. K. Why Did Google Brain Exist? April 2023. Retrieved from \url{www.moderndescartes.com/essays/why_brain/}.) The author further argued that, despite the inclusion of senior researchers on promotion committees, these panels have a limited capacity to accurately assess the scientific quality of research.}

Descriptively, I show that the relation between corporate science and patenting is more substantial than that which is revealed by only considering patents that directly cite the firms' science. In my data, only 7\% of corporate publications were cited internally by patents of the same firm. However, by observing external follow-on research, I find that an additional 33\% of firms' publications were eventually indirectly cited by the same firms' patents. %It is important to note that these citations require a long time horizon. On average, a third-generation follow-on paper is cited by a patent 13 years after the publication of the original paper. 
Academics' citations to corporate publications exhibit greater textual similarity to those focal publications than the citing authors' prior works do, consistent with corporate publications influencing the direction of external academics' research. 
Furthermore, publications that lead to extensive follow-on research tend to find more varied applications in the originating firm's subsequent inventions, extending beyond the immediate context of the original paper. These findings suggest a close relation between firms' investments in science and their inventive activities and highlight the potential role of external research in supporting this relation. %In addition, the findings suggest that external research might require years to evolve to the point where it is internally useful for the originating firm. 

A key challenge in identifying the effects of follow-on research is to control for unobserved variables. For example, highly promising findings likely attract both more scientific interest and greater firm investments. Conversely, academics might sometimes avoid building on corporate research, potentially biasing estimates downward. To account for this source of endogeneity, I implement a new instrumental variable that exploits exogenous variation in attention paid to a given publication. Papers are typically grouped into journal issues quasi-randomly. Within the same journal and year, some issues include publications by more prominent authors than others. At least until readership moved online, these issues were likely to draw more attention from fellow academics. This increased attention drove citations to other papers in the same journal issue. To measure prominence, I calculate the H-index of authors in journal issues, identifying the top two prominent authors per issue (excluding focal authors). I use the sum of their H-indexes as an instrument for follow-on research when estimating effects on firms' scientific and innovation outcomes. 

In the main analysis, I apply a two-stage least square estimation to identify the effect of follow-on research on firms' investments in science and patenting outcomes. First, I find that external follow-on research drives subsequent scientific investments and patenting by the originating firms. It also increases retention of the firms' researcher employees, consistent with follow-on research both enhancing scientists' productivity and signaling their quality. Second, it increases firms' engagement with the scientific community, through collaborations and hiring of scientists. Taken together, I interpret these results as evidence for the positive value of follow-on research for firms' scientific and innovative efforts. Next, I examine the moderators and mechanisms that drive the positive effects of follow-on research. I find suggestive evidence that follow-on research is useful both as an input and as providing quality validation for internal research. Lastly, I explore the conditions that enable firms to benefit from follow-on research. I find more substantial effects in areas where firms own intellectual property (IP) rights and have internal research capabilities. Follow-on research is also more valuable in nascent scientific domains and areas where government funding is readily available.

I contribute to literature on firms' disclosure strategies. Despite imitation risks \citep{Rosenberg1990WhyFirmsBasic, AroraBelenzonSheer2021KnowledgeSpilloversCorporate}, firms permit publication because scientists value open science and accept lower wages in exchange \citep{Stern2004ScientistsPayBe, SauermannRoach2014NotAllScientists}.
From this perspective, disclosure strategies emerge from bargaining between managers seeking to minimize competitive risks and scientists preferring greater openness \citep{GansMurrayStern2017ContractingDisclosureScientific}. Rather than viewing these interests as purely competing, I show they can be complementary: publications generate value through their influence on the broader scientific ecosystem. I show that follow-on research increases scientist retention, countering the concern that publications primarily facilitate employee departure. These findings suggest the benefits side of the bargaining equation includes not only scientist satisfaction but also tangible returns through external research mobilization.

More broadly, I contribute to understanding how firms benefit from investments in science. Prior work has shown that such investments enhance combinative capabilities by extending knowledge bases and providing direct inputs into invention \citep{KogutZander1992KnowledgeFirmCombinative, FlemingSorenson2004ScienceMapTechnological, AroraBelenzonSheer2021KnowledgeSpilloversCorporate}, while also improving absorptive capacity \citep{CohenLevinthal1990AbsorptiveCapacityNew, Rosenberg1990WhyFirmsBasic}. These views consider how firms are affected by \textit{already-existing} external knowledge. However, given the magnitude of resources and outputs produced by the scientific community, it is evident that, on top of the benefits of better \textit{access} to public research, firms can benefit from the ability to \textit{influence} academics' research agenda. By participating in public research, firms can potentially influence the future content produced within the scientific information networks in ways that are privately beneficial. This implies that a well-developed and responsive scientific community constitutes a valuable resource that incentivizes firms to participate in public research.

My third contribution is to the literatures on open innovation and knowledge spillovers. \citet{Chesbrough2003OpenInnovationNew} argued that opening internal resources to external use and adopting external technologies can improve the performance of innovative firms. Later works, focusing on open-source software, showed that selective revealing could drive valuable external contributions (e.g., \citet{Henkel2006SelectiveRevealingOpen, DahlanderWallin2006ManUnlockingCommunities, AlexyGeorgeSalter2013CuiBonoSelectivea}). Studies on knowledge spillovers from patent disclosures have shown the potential value of reabsorbing external developments \citep{YangPhelpsSteensma2010LearningWhatOthers, Belenzon2012CumulativeInnovationMarket, AlnuaimiGeorge2016AppropriabilityRetrievalKnowledge, YangSteensma2014WhenFirmsRely}. However, while the disclosure of patented inventions is mandated by law, the decision to engage with the scientific community is a strategic choice \citep{AlexyWestKlapperEtAl2018SurrenderingControlGain}. This paper supports the view that opening up internal knowledge through participation in public research could be a strategic choice that drives down R\&D costs, enhances the value of complementary assets, and allows firms to mitigate uncertainty associated with their investments in research.

\section{Theoretical Development and Related Literature}

\subsection{Firms' Participation in Public Research}

Firms participate in public research by conducting scientific research and publicly disclosing their findings. Investments in science create knowledge that firms can use as an input into invention  \citep{AroraBelenzonSheer2021KnowledgeSpilloversCorporate}. Even when not used directly, scientific research enables firms to overcome the limitations of incremental search   \citep{FlemingSorenson2004ScienceMapTechnological} and improves their combinative capabilities \citep{KogutZander1992KnowledgeFirmCombinative, Arthur2011NatureTechnologyWhat}. In addition, investments in science enhance firms' absorptive capacity, or their ability to identify, assimilate, and exploit \textit{already-existing} external knowledge \citep{CohenLevinthal1990AbsorptiveCapacityNew, Rosenberg1990WhyFirmsBasic}. Studies have found a positive relationship between firms' investments in scientific research and various measures of firm performance (e.g., \citet{DengLevNarin1999ScienceTechnologyPredictors, SimethCincera2016CorporateScienceInnovation}). 

The publication of findings is typically integral to scientific endeavors. In some settings, firms encourage the disclosure of research findings \citep{Hicks1995PublishedPapersTacit, CockburnHenderson1998AbsorptiveCapacityCoauthoring}, despite potential risks of unintended knowledge spillovers and reduced R\&D returns \citep{Arrow1962EconomicWelfareAllocation, DasguptaDavid1994NewEconomicsScience, Nelson1959SimpleEconomicsBasic}. Various incentives for firms to publish have been identified \citep{RotoloCameraniGrassanoEtAl2022WhyFirmsPublish}, including  IP strategy  \citep{BakerMezzetti2005DisclosureStrategyPatent}, reputation \citep{PolidoroTheeke2012GettingCompetitionScience, HarhoffHenkelvonHippel2003ProfitingVoluntaryInformation, BaruffaldiSimethWehrheim2023AsymmetricInformationDisclosure}, and marketing strategies \citep{SimethRaffo2013WhatMakesCompanies}. 

A key perspective emphasizes that publication decisions emerge from negotiations between firms and their research employees, driven by the conflicting preferences of each party. Scientists have strong preferences for publishing their research findings, reflecting both career incentives and intrinsic motivation for contributing to scientific knowledge. This ``taste for science'' is sufficiently strong that scientists accept lower salaries in exchange for greater publication rights \citep{Stern2004ScientistsPayBe, SauermannCohen2010WhatMakesThem}. Firms, in contrast, prefer to limit disclosure to protect their ability to appropriate returns from their research investments. The resulting publication practices reflect the resolution of this tension through bargaining \citep{GansMurrayStern2017ContractingDisclosureScientific}. This perspective implies that external citations to employees' science are potentially harmful to the firm, both through knowledge spillovers to competitors and by enabling employees to signal their quality to outside labor markets, increasing the likelihood of turnover.

Alongside the risks of spillovers and employee turnover, firms' engagement with the scientific community through publication may generate countervailing benefits. I suggest that participation in public research, through both investments and publication, serves as a channel for firms to access the resources of the scientific community in ways that can offset some of the costs described above.

\subsection{Engaging with the Scientific Community}

Firms' participation in public research can influence the pace and trajectory of scientific advances beyond their organizational boundaries. There are various channels  through which firms can engage with academics \citep{CohenNelsonWalsh2002LinksImpactsInfluence}. Many of these channels, and the focus of prior literature, require direct interpersonal ties. For example, university-industry research collaborations (UIC) enable firms to direct academic inquiry in their favor and benefit from academics' expertise and resources \citep{BikardVakiliTeodoridis2018WhenCollaborationBridges}.\footnote{See \citet{PerkmannSalandraTartariEtAl2021AcademicEngagementReview} for recent literature reviews regarding UICs.} Other direct channels, such as geographic proximity, funding, conference participation, and corporate hiring, are additional channels by which firms can influence academia \citep{Sohn2021HowLocalIndustry, BabinaHeHowellEtAl2023CuttingInnovationEngine, BaruffaldiPoege2022StarsHowFirms}. These engagements can steer external scientific inquiry toward areas relevant to firms' needs, shaping both immediate technological applications and the long-term scientific agenda.

While direct ties with academics allow firms to influence their work, corporate publications serve as an additional channel of engagement. Publication can lead to follow-on research, defined as science produced outside the firm that is informed by the findings firms disclose \citep{Hicks1995PublishedPapersTacit}.\footnote{\citet{Hicks1995PublishedPapersTacit} referred to this as  ``collateral research.''} In some cases, follow-on research creates downstream ``scientific steps,'' cumulative findings that open up potential applications of the underlying science \citep{DasguptaDavid1994NewEconomicsScience, David1998CommonAgencyContracting, AhmadpoorJones2017DualFrontierPatented}. In other cases, corporate science informs academics about downstream demand and technical problems that require upstream exploration. Follow-on research could emerge from distant geographic locations and a wide range of knowledge domains. In this sense, corporate publications as a channel of engagement differ from direct channels because they offer a broad, undirected search \citep{LeiponenHelfat2010InnovationObjectivesKnowledge}.

This broad search mechanism is particularly valuable given the high switching costs scientists face when changing research directions \citep{Myers2020ElasticityScience} and the increasing specialization of scientific work \citep{Jones2009BurdenKnowledgeDeath}. Building on corporate publications does not require academics to make dramatic shifts in their research agendas. Rather, these publications serve as a discovery mechanism that brings relevant corporate research into view for academics already working on related problems. By broadcasting findings through publication, firms enable academics to discover relevant work within their existing research programs that they might otherwise overlook. This work can inform and advance their ongoing inquiries without requiring costly reorientation.

\subsection{Firms' Benefits from External Follow-On Research}

Aside from the risks of knowledge spillovers to competitors, there are cases where knowledge spillovers from firms' R\&D investments can also have positive effects on the originating firms. For example, knowledge can spill over to technologically-related firms not competing in the same product market \citep{BloomSchankermanReenen2013IdentifyingTechnologySpillovers}. Several studies  explored firms' ability to benefit from spilled knowledge through reabsorption. 
\citet{YangPhelpsSteensma2010LearningWhatOthers} used a sample of patents originating from telecommunications firms and showed a positive relationship between the spillover knowledge pool and firms' innovative activities. 
\citet{Belenzon2012CumulativeInnovationMarket} studied how firms' ability to reabsorb knowledge spillovers from patents affects firm-level performance outcomes. Subsequently, \citet{AlnuaimiGeorge2016AppropriabilityRetrievalKnowledge} and \citet{YangSteensma2014WhenFirmsRely} studied how technology, firm, and industry characteristics interact with the ability to reabsorb knowledge. However, unlike scientific publication, these positive effects might be partly unintentional, as patent law mandates the disclosure of information once an inventor seeks patent protection of intellectual property rights.

Participation in public research, as a case of selective knowledge revealing \citep{AlexyGeorgeSalter2013CuiBonoSelectivea}, can be seen as a deliberate choice to engage in strategic openness \citep{AlexyWestKlapperEtAl2018SurrenderingControlGain, Henkel2006SelectiveRevealingOpen}. Openness, through the influence on external actors, can reduce R\&D costs and increase value appropriation from complementary assets owned by the firm. Two recent works explore these arguments. Studying corporate patents in artificial intelligence (AI), \citet{Shen2022SpreadSeedWin} compares the cumulative innovation that follows patent-paper pairs to that of patents that lack a corresponding scientific publication. The results suggest that scientific publications broadcast the firms' inventions to a broader audience of external inventors. Also focusing on the AI industry, 
\citet{JeeSohn2023FirmCreationProprietary} find correlational evidence that corporate publications influence the knowledge spillover pool and subsequently benefit firms' patenting outcomes. 

Several mechanisms make external follow-on research beneficial for the originating firm. First, it can be useful as input into subsequent R\&D activities. In many cases, academics have the expertise and resources that the originating firms lack. By inducing follow-on research, firms positioned to appropriate value from publicly-available knowledge can benefit from tapping into these resources. As a result of follow-on research, these firms will likely continue to invest in related research and experience better patenting outcomes. This increased research productivity and access to new ideas makes continued employment more attractive to scientists who value intellectual challenge \citep{SauermannCohen2010WhatMakesThem}, strengthening the firm's ability to retain employees working in areas that generate substantial external follow-on research. 
Second, Follow-on research can facilitate valuable direct collaborations by identifying potential partners \citep{AlexyGeorgeSalter2013CuiBonoSelectivea}. Scientists producing relevant follow-on work are prime candidates for collaborations and recruitment, potentially leading to research partnerships, joint inventions, and academic hiring by the originating firm.\footnote{See Appendix \ref{sec:studycases} for further details and examples from the data.}

Even when follow-on research is not used as input by the firm or does not lead to other forms of engagement, it can signal the technical promise of research trajectories and guide R\&D managers' resource allocation decisions.\footnote{This aligns with findings on the impact of scientists' status \citep{AzoulayStuartWang2014MatthewEffectFable} and scientific prizes \citep{JinMaUzzi2021ScientificPrizesExtraordinary} on attention and research growth. Within firms, high-status individuals can similarly affect resource distribution \citep{PratoFerraro2018StarstruckHowHiring}.} Using external validation through citations is particularly efficient because monitoring research quality internally is notoriously costly \citep{CockburnHenderson1998AbsorptiveCapacityCoauthoring}. Such signals can inform decisions about future R\&D investments and employee retention, as firms seek to retain their most productive scientists. The value of follow-on research as a signal is likely greatest when there is significant uncertainty about the work being done by researchers within the firm.

Several factors influence firms' ability to benefit from follow-on research. When appropriability regimes are weak, control of specialized complementary assets becomes critical for capturing value from innovation \citep{Teece1986ProfitingTechnologicalInnovation}. Ownership of complementary assets, particularly patents protecting related intellectual property, enhances value appropriation. Therefore, publications that are part of patent-paper pairs \citep{MarxFuegi2022RelianceSciencePatenting} are likely to yield more significant effects on firm outcomes. Moreover, persistent internal research capabilities enhance a firm's ability to absorb follow-on research, whereas outsourcing research to universities may hinder this ability in subsequent years \citep{BikardVakiliTeodoridis2018WhenCollaborationBridges}.

The characteristics of the scientific domain also affect the value of follow-on research. In mature, crowded fields, an additional publication by the focal firm may have minimal impact, whereas in nascent, under-explored areas, follow-on research can be substantially more valuable. Lastly, the availability of government funding to academics influences the public good nature of science \citep{BabinaHeHowellEtAl2023CuttingInnovationEngine}, potentially reducing barriers for originating firms in assimilating follow-on research into subsequent innovation.

\section{Data}

\subsection{\label{sec:sample}Sample Construction}

I combine data from several sources: (i) corporate scientific publications, patents, and accounting information from the Duke Innovation \& SCientific Enterprises Research Network (DISCERN, \citet{AroraBelenzonSheer2020DISCERNDukeInnovationa}); (ii) scientific publications and citations data from Microsoft Academic Graph \citep{SinhaShenSongEtAl2015OverviewMicrosoftAcademic}; additional publications and disambiguated author data from Dimensions.ai; 
(iii) patent citations to scientific publications from the Reliance on Science in Patenting project \citep{MarxFuegi2022RelianceSciencePatenting}; and (iv) data on renowned scientists from the American Men and Women of Science (AMWS) directory.

The DISCERN dataset comprises 582,107 journal articles and conference proceedings from Web of Science (WoS) associated with U.S. publicly-traded firms (1980-2015). I created a crosswalk between WoS, Dimensions, and MAG, yielding 471,153 records of firms' publications.\footnote{463,027 unique publications, as some are coauthored by researchers from multiple firms.} The sample is limited to 1990-2012 due to data constraints. For the main analysis, I exclude conference proceedings and special issue publications to align with the instrumental variable's logic.\footnote{See Section \ref{sec:iv} for details. Analysis of conference proceedings is in Appendix Section \ref{sec:proceedings}.} After removing observations with zero citations,\footnote{Because the main specification uses $\ln(\text{Follow-On Research})$, observations with zero external citations (4\% of the sample) are excluded, restricting the  estimates to the intensive margin among cited publications. Table \ref{tbl:poisson_iv} presents PPML estimates that retain these observations; results are qualitatively similar.} missing data and singletons resulting from fixed effects inclusion, the final publication-level sample consists of 164,516 observations (156,486 unique publications) matched to 1,522 firms.

\subsection{Variables and Measures}

See Appendix \ref{sec:data_appendix} for additional details regarding sources and data construction.

\subsubsection{Explanatory Variable}

I measure follow-on research by observing citations from outside the firm to the focal publication. I count up to three generations of citations (the first generation being direct citations). In cases where several routes connect the focal publication to the citing paper, I count the shortest route.\footnote{See Appendix Figure \ref{fig:followon} for an illustration.} Most citations to firms' publications originate from academics employed by universities and other public research institutions. However, I also include citations from researchers at other firms.\footnote{See Appendix Section \ref{sec:fo_univ_corp} for a split analysis. }

\subsubsection{Dependent Variables}
To identify the effect of follow-on research on originating firms' subsequent scientific and innovative performance, I employ several measures across three key areas: scientific production, hiring of scientists, and patenting activity. For scientific production, I examine three related measures: indications for corporate authors' scientific publications published after the focal paper, corporate authors' publications involving collaborations with university academics (University-Industry Collaborations, UIC), and their conference proceedings. To proxy for employee retention, I observe whether researchers who authored the focal publication continue to publish at least ten years after the focal publication.\footnote{This measure serves as a proxy for retention that depends on continued active publication. Researchers who remain employed at the firm but do not publish would not be captured by this measure.}

To assess the impact on the hiring of scientists, I identify AMWS scientists, whose work is relevant to the focal publication, hired by the firm after the focal publication year. Relevance is determined by concept overlap between the focal publication and the scientist's publications.\footnote{Concepts are extracted by Dimensions from titles and abstracts using a pointwise mutual information algorithm. See Appendix \ref{sec:data_appendix} for details.} This approach yields accurate employment years but likely underestimates total hiring activities due to data limitations. Out of 20,552 individuals identified as employed by DISCERN firms, 6,673 were matched to publications with extracted concepts.

To explore the effect of follow-on research on patenting activity, I employ two approaches. First, I identify patents by the focal paper's authors assigned to the originating firm, considering both patents filed at least three years and five years after the focal publication year. Second, identify the firm's patents that cite the focal publication in non-patent literature (NPL) citations. Additionally, I identify the firm's patents that cite either the focal publication or the follow-on research in NPL citations.\footnote{Patent citations to scientific articles typically reflect knowledge flows and the use of science in the invention process \citep{RoachCohen2012LensPrismPatent}. However, the count of patents citing follow-on research may be mechanically related to the availability of follow-on publications to cite, potentially overestimating the relationship between follow-on research and internal use in subsequent inventive activity.} These measures collectively provide a comprehensive view of how follow-on research impacts the originating firm's subsequent scientific and innovative activities.

\subsubsection{\label{sec:hindex}Control Variables}

I conceptualize the number of citations a publication receives as a function of four variables: the scientific content of the publication, the journal and publication year, the authors' identity (as scientific prominence drives attention), and peer effects across authors whose publications appear in the same journal issue (see Section \ref{sec:iv} for details). To proxy for authors' prominence, I employ the H-index, a widely popular metric of academic impact and productivity, calculated at the time of publication.\footnote{The H-index is calculated by identifying all papers published by the author prior to year $t$ and counting citations received up to that year. After sorting papers by descending citation counts, the H-index is defined as $h=\max {i \in \mathbb{N}: f(i) \geq i}$, where $f(i)$ is the citation count for the publication in position $i$. For instance, an author with five publications having citation counts of 33, 20, 8, 4, and 1 would have an H-index of 4.} For multi-authored publications, I use the highest H-index among the authors as the control variable.

I include the focal paper author's H-index to control for their prominence. As discussed in Section \ref{sec:iv_results}, I use the sum of the two highest H-indexes among all other authors in the same journal issue as an instrumental variable, which is arguably exogenous to the focal publication's scientific content after controlling for journal and year differences. To control for variations between journals at different points in time, I incorporate a complete set of journal-year fixed effects. Lastly, I include firm fixed effects to account for time-invariant differences across firms.

\section{\label{sec:res_desc}Descriptive Analysis}

Corporate scientific research and firms' patenting activities are intricately related. Table \ref{tbl:discern_desc_pubs} categorizes corporate publications into three groups: directly cited by internal patents, indirectly linked through citations to external follow-on research, and unlinked to the firm's patents. Only about 7\% of firms' publications are directly cited by patents.\footnote{Analysis uses the complete sample of DISCERN publications with a MAG identifier, focusing on papers published prior to 2000 to account for truncation in later years.} While this might suggest a disconnect between scientific investments and patenting, tracking patent citations to external follow-on research reveals a more comprehensive view. Approximately 40\% of publications are tied to patents, either through direct citation or indirectly via patents citing external follow-on research. This indicates a close science-invention connection within firms and underscores the potential benefit of follow-on research in establishing this link.\footnote{Appendix Section \ref{app:desc_pats} shows a similar result from the patent perspective, with over 40\% of corporate science-based patents directly or indirectly related to firms' public research contributions.}

\begin{table}[htbp]

\scriptsize %footnotesize

\centering

\caption{\label{tbl:discern_desc_pubs}
{Firms' Scientific Publications Cited by Own Patents}}

\begin{threeparttable}

\begin{tabularx}{\textwidth}{l>{\raggedleft\arraybackslash}X>{\raggedleft\arraybackslash}X>{\raggedleft\arraybackslash}X>{\raggedleft\arraybackslash}X}
\toprule
\multicolumn{1}{l}{}  & \multicolumn{2}{c}{All Publications} & \multicolumn{2}{c}{Years 1980-2000} \\ 
\cmidrule(lr){2-3} \cmidrule(lr){4-5}
\multicolumn{1}{l}{}  & Count & Percent & Count & Percent \\ 
\midrule
 Firm's patent directly cites paper & $26,741$ & $5.68\%$ & $15,718$ & $6.99\%$ \\ 

\multicolumn{5}{l}{Firm's patent cites follow-on research}\\
  \hspace{2em} 1st Generation & $24,133$ & $5.12\%$ & $16,318$ & $7.26\%$ \\ 
  \hspace{2em} 2nd Generation & $38,514$ & $8.17\%$ & $27,761$ & $12.34\%$ \\ 
  \hspace{2em} 3rd Generation & $42,723$ & $9.07\%$ & $30,531$ & $13.58\%$ \\ 
  \cmidrule(lr){2-5}
   \hspace{2em} All & $105,370$ & $22.36\%$ & $74,610$ & $33.18\%$ \\ 
  \midrule 
  Cited by a firm's patent (directly or indirectly) & $132,111$ & $28.04\%$ & $90,328$ & $40.17\%$ \\
 Not cited by a firm's patent & $339,042$ & $71.96\%$ & $134,553$ & $59.83\%$ \\ 
Total & $471,153$ & $100.00\%$ & $224,881$ & $100.00\%$ \\ 
\bottomrule
\end{tabularx}
\begin{tablenotes}[flushleft] \scriptsize
\item \textit{Note.} This table summarizes corporate scientific publications cited by firms' own patents (NPL), encompassing 471,153 publications. Publications are categorized by the shortest route to a firm's patent citation. 26,741 (5.7\%) are directly cited by the firm's patents, while an additional 105,370 (22.3\%) are indirectly cited through external publications cited by the firm's patents, up to the 3rd generation. 339,042 (72\%) remain uncited by the firm's patents. To address truncation, columns 3 and 4 focus on publications up to 2000, revealing that approximately 40\% of these are eventually cited by the firms' patents.
\end{tablenotes}
\end{threeparttable}

\end{table}

\begin{figure}[h]
\caption{\label{fig:cumnpl}Direct and Indirect Patent Citations to Firms' Publications}
\captionsetup{width=0.8\textwidth}
\centering
\includegraphics[width=0.6\textwidth]{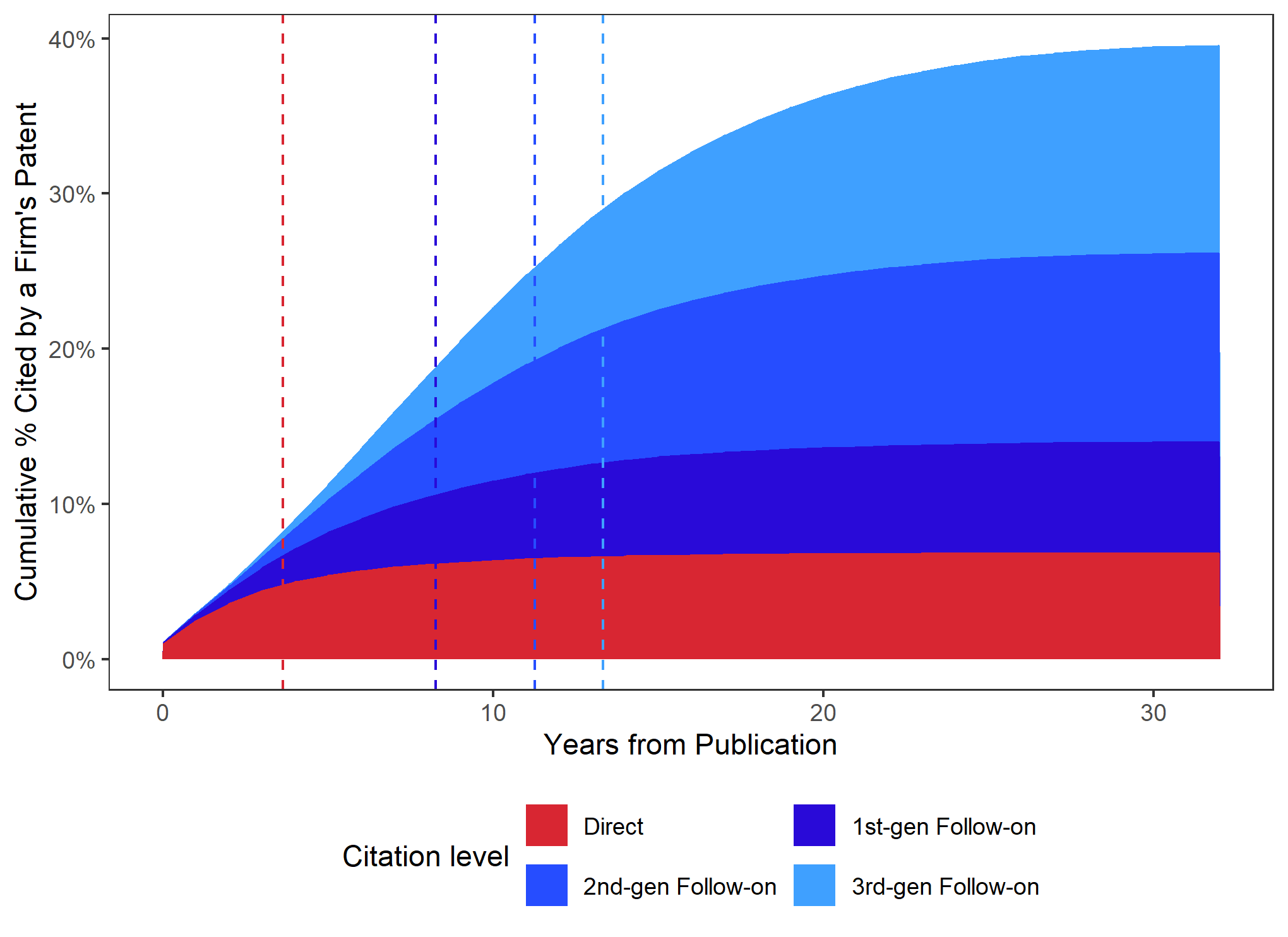}
\caption*{\footnotesize \textit{Note:} This figure shows the cumulative percentage of corporate publications cited by the originating firms' patents. Direct citations are patents by the originating firm that directly cite the publication. Citations to follow-on research are patent citations to external follow-on research that cites the focal publication. A patent is counted once for each focal publication based on the shortest citation route. Citation timing is based on patent filing years. The dotted lines represent the average times to citations.}
\end{figure}

Figure \ref{fig:cumnpl} examines the timing of direct and indirect patent citations. Direct citations predominantly occur within a few years of the focal paper's publication, with an average lag of 3.5 years. Indirect citations, referencing follow-on research, typically appear later as this research accumulates. The average times for indirect citations are 8.2, 11.2, and 13.4 years for first-, second-, and third-generation follow-on research, respectively. This generational progression in citation timing highlights the cumulative nature of scientific knowledge and its enduring impact on firms' innovative activities. Consequently, evaluating the relationship between a firm's scientific research and inventive output based solely on direct, short-term citations may substantially underestimate the long-term value of corporate investments in science. 

To further examine how follow-on research is internalized by the firm, I analyze overlap between patent inventors and the authors of cited scientific publications, as an indication of cases where the firm may have hired or engaged the external academic as a consultant.\footnote{Appendix Section \ref{app:desc_pats} provides details. I thank an anonymous reviewer for suggesting this analysis.} Among patent citations to follow-on research, 2.65\% include at least one same-name inventor–author match, and 11.5\% of patents contain at least one such citation, indicating that while this represents a minority of cases, direct personal or contractual ties to external scientists constitute an important channel through which firms internalize scientific knowledge.

Further evidence suggests that citations from external academics often coincide with adjustments in their research direction, and that the extent of follow-on research correlates with broader utilization of firms' scientific findings in their subsequent inventions.
To quantify the first relationship, I compare the textual similarity between the citing papers and the focal publication with the similarity between the focal publication and citing author's nearest prior work.\footnote{Textual similarity is measured using cosine similarity between embeddings generated by Google's \texttt{gemini-embedding-001} model.} Table \ref{tbl:similarity}, Panel A, presents these results, showing that citing papers are more textually similar to the focal publication than are the citing authors' earlier works. This is consistent with citations being associated with at least some change in research focus.
To quantify the second relationship, I analyzed the textual similarity between cited paper abstracts and their citing patents.\footnote{This analysis uses a subsample of publications with complete metadata (journal issue, publication date, and abstract).} Table \ref{tbl:similarity} Panel B presents an inverse relationship between follow-on research extent and patent-publication textual similarity.\footnote{For consistency, only the first citing patent (based on filing date) is considered in the textual similarity analysis.} This pattern persists even when considering only patents directly citing the focal publication (Columns 3 and 4). The findings suggest that scientific publications generating extensive follow-on research are more likely to find diverse applications in the originating firm's inventions, extending beyond the original paper's specific language and immediate context.\footnote{Establishing causal evidence for these effects would require observing the counterfactual: what would have occurred had the firm not published. I leave this to future work.}

\begin{table}[htbp]
\centering
\begin{threeparttable}
\caption{Textual Similarity Analysis}
\label{tbl:similarity}
\small
\begin{tabularx}{\textwidth}{X*{4}{c}}
\toprule
\multicolumn{5}{l}{\textbf{Panel A: Citing Publications vs. Previous Publications by Same Authors}} \\
                    &\multicolumn{4}{c}{Textual Similarity to Focal Publication}                                                 \\\cmidrule(lr){2-5}
                    &\multicolumn{1}{c}{OLS}&\multicolumn{1}{c}{OLS}&\multicolumn{1}{c}{OLS}&\multicolumn{1}{c}{OLS}\\
&(1)&(2)&(3)&(4)\\ \midrule
Citing Publication  &       0.118\sym{***}&       0.118\sym{***}&       0.118\sym{***}&       0.118\sym{***}\\
                    &     (0.000)         &     (0.000)         &     (0.000)         &     (0.000)         \\
\\
Year FE             &          No         &         Yes         &         Yes         &          No         \\
Firm FE             &          No         &          No         &         Yes         &          No         \\
Focal Pub FE        &          No         &          No         &          No         &         Yes         \\
Observations        &     564,360         &     564,360         &     564,360         &     564,360         \\
Avg. Similarity     &       0.679         &       0.679         &       0.679         &       0.679         \\
Adjusted R$^2$      &       0.369         &       0.372         &       0.394         &       0.600         \\
\midrule
\multicolumn{5}{l}{\textbf{Panel B: Subsequent Citing Patents Within the Firm}} \\
                    &\multicolumn{2}{c}{Patent Citing Follow-on or Focal Paper}&\multicolumn{2}{c}{Patent Citing Focal Paper}\\\cmidrule(lr){2-3}\cmidrule(lr){4-5}
                    &\multicolumn{1}{c}{OLS}&\multicolumn{1}{c}{OLS}&\multicolumn{1}{c}{OLS}&\multicolumn{1}{c}{OLS}\\
&(1)&(2)&(3)&(4)\\ \midrule
ln(Follow-on)       &      -0.004\sym{***}&      -0.006\sym{***}&      -0.001\sym{***}&      -0.002\sym{***}\\
                    &     (0.001)         &     (0.001)         &     (0.000)         &     (0.000)         \\
\addlinespace
ln(Focal H-Index)   &      -0.003\sym{***}&      -0.001\sym{*}  &      -0.001         &      -0.000         \\
                    &     (0.001)         &     (0.001)         &     (0.001)         &     (0.001)         \\
\\
Firm FE             &         Yes         &         Yes         &         Yes         &         Yes         \\
Journal FE          &         Yes         &          No         &         Yes         &          No         \\
Year FE             &         Yes         &          No         &         Yes         &          No         \\
Journal-Year FE     &          No         &         Yes         &          No         &         Yes         \\
Observations        &      75,970         &      75,572         &      12,596         &      12,409         \\
Avg. DV             &       0.628         &       0.628         &       0.721         &       0.720         \\
Adjusted R$^2$      &       0.253         &       0.251         &       0.141         &       0.147         \\
\bottomrule
\end{tabularx}
\begin{tablenotes}[flushleft]
\scriptsize
\item \textit{Note.} This table examines the relationship between external follow-on research and textual similarity using a pooled cross-section of publications by U.S.-based publicly-owned firms, published between 1980 and 2015 \citep{AroraBelenzonSheer2020DISCERNDukeInnovationa}. \textit{Panel A} compares the textual similarity of external citing papers to the focal publication versus the citing authors' most recent previous publication. The dependent variable is cosine similarity between text embeddings. A positive coefficient on ``Citing Publication'' indicates that citing papers are more similar to the focal publication than the citing authors' prior work. \textit{Panel B} examines patents from the same firm that cite either the focal publication and/or its follow-on research (up to 3 generations). The dependent variable is the textual similarity between the citing patent abstract and the focal publication abstract. The first patent by filing date is used for each focal publication. Columns 1-2 include patents citing the focal publication or any of its follow-on papers, while columns 3-4 restrict to patents citing only the focal publication. A negative coefficient on ln(Follow-on) indicates that greater external follow-on research is associated with lower patent similarity to the focal publication. All similarity scores are calculated using cosine similarity of embeddings generated by Google's Gemini text-embedding-001 model. Regressions include fixed effects as indicated and control for the log of the highest H-index among the authors of the focal publication. The sample is restricted to publications with complete metadata. Clustered standard-errors in parentheses (Panel A by focal publication, Panel B by firm). Signif. Codes: ***: 0.01, **: 0.05, *: 0.1
\end{tablenotes}
\end{threeparttable}
\end{table}

\section{\label{sec:iv_results}The Effect of External Follow-On Research on Firms' Innovation}

The relationship between external follow-on research and a firm's innovation outcomes is complicated by potential endogeneity due to various unobserved variables. To address this, I propose an instrumental variable based on publishers' quasi-random grouping of manuscripts into journal issues, which creates exogenous variation in scholarly attention to focal publications. This variation serves as the identification mechanism for addressing endogeneity concerns, though the theoretical value of follow-on research itself does not depend on such exogenous variation.

\subsection{\label{sec:iv}Identification Strategy}

\subsubsection{Randomized Journal Allocation}

After submission, academic manuscripts undergo peer review to assess quality and relevance. Upon acceptance, editorial staff prepare and group manuscripts into issues, typically published monthly, bi-monthly, or quarterly. This grouping process, performed by non-academic staff, is unrelated to manuscript content and typically follows a chronological order of acceptance, rendering it quasi-random.\footnote{An exception is special issues and thematic journal issues, which are excluded from my main analysis.}\footnote{These details have been confirmed by executives at leading scientific publishing firms.} Consequently, a manuscript's allocation to a specific journal issue does not indicate its quality relative to adjacent publications.

While this quasi-random grouping created no quality differences across issues, it generated meaningful variation in attention. Journal issues containing articles by prominent authors received disproportionate attention from researchers, creating spillover exposure for all publications in those issues regardless of content or quality. This attention-based mechanism operated through the physical nature of pre-digital academic research. Before the early 2000s shift to online reading, researchers primarily accessed publications by browsing physical journal issues in libraries \citep{Ocasio1997AttentionBasedViewFirm}. Since attention is a limited resource \citep{BikardMarx2020BridgingAcademiaIndustry, ChaiMenon2019BreakthroughRecognitionBias, SimcoeWaguespack2011StatusQualityAttention}, researchers naturally gravitated toward issues featuring prominent authors. This browsing process introduced an element of serendipity to information discovery \citep{FosterFord2003SerendipityInformationSeekinga}, with the physical structure of journals exposing academics to unintended information and potentially inspiring new research directions \citep{MakriBlandford2012ComingInformationSerendipitously}.

The increased exposure translated into measurable differences in citation patterns. \citet{Hudson2007BeKnownCompany} found that issue ``traffic'' and the inclusion of ``major papers'' positively correlate with citations to focal publications in top economics journals. Similarly, \citet{LundMaurya2020RelationshipHighlycitedPapers} observed a positive relationship between citations to ``highly-cited'' papers and other publications in the same information science journal issues. These findings suggest that the prominence of authors in a journal issue drove citations and follow-on research to adjacent publications, irrespective of the inherent quality or content of those articles.

I leverage this quasi-random process to develop a measure of scholarly attention based on the H-index of the top two authors in the same journal issue as the focal publication. The instrument's relevance hinges on the assumption that author prominence in a journal issue drives attention to focal publications, leading to increased follow-on research beneficial for the originating firm's subsequent innovation.\footnote{See Appendix Section \ref{sec:app_iv} for a detailed discussion and validation tests.}

\subsubsection{Instrument Specification}

To proxy for increased attention, I calculate the H-index of all authors of publications in the same journal issue of each focal publication in the data. To ensure that further scientific developments do not affect the H-index, I calculate it based on the year before the focal publication year. Next, I identify the top two authors by H-index for each publication and use the sum as the instrument. Table \ref{tbl:discern_pubspats} column 2 reports first-stage results.\footnote{See Table \ref{tbl:firststage} for alternative specifications.}

The instrument's validity relies on the assumption that other authors' prominence doesn't confound with the focal publication's quality or its authors' prominence. This assumption holds by restricting comparisons within journals and specific time windows, using journal-year fixed effects. Conditional on these effects, the instrument is plausibly unconfounded with outcome variables. Appendix \ref{sec:app_iv} provides validity tests: (1) showing the instrument's lack of correlation with focal authors' H-index, (2) a placebo test using H-index of top authors from random issues, which doesn't predict follow-on research, and (3) demonstrating the instrument's stronger effect before the early 2000s shift to online readership.

\subsection{Baseline Estimation Results}

The econometric specification at the publication level ($i$) is as follows. 

In the first stage, I estimate:

\begin{align}
    \ln(&\text{Follow-On Research})_i =  {} \\
    & \alpha_1 \text{IV}_i + \alpha_2 \ln({\text{Focal H-index}_i}) \notag \\
    &  + \eta_f + \tau_t \times \gamma_j + \epsilon_i \notag 
\end{align}

The second stage is:

\begin{align}
Y_i = {} & \beta_1 \ln(\widehat{\text{Follow-On Research}})_i \\
& + \beta_2 \ln({\text{Focal H-index}_i}) \notag \\
& + \eta_f + \tau_t \times \gamma_j + \epsilon_i \notag
\end{align}

The first stage measures follow-on research as the count of three generations of external citations to the focal publication. The instrument, representing academic attention to the journal issue, is the sum of the two highest H-indexes of other authors in the same issue. First stage results are in Table \ref{tbl:discern_pubspats}, column 2.\footnote{See Table \ref{tbl:firststage} for alternative specifications. A Durbin-Wu-Hausman test rejects exogeneity of follow-on research ($F = 7.57$, $p = 0.006$), confirming the need for instrumental variables. The Kleibergen-Paap cluster-robust F-statistic is 26.8, exceeding conventional thresholds. Anderson-Rubin weak-instrument-robust tests strongly reject the null hypothesis ($\chi^2 = 10.26$, $p = 0.001$), confirming results are not driven by weak instruments.} The second stage uses the outcome of interest as the dependent variable.\footnote{Critically, the dependent variables measure subsequent patents and publications by focal authors regardless of whether they cite the follow-on research pool. This design ensures that results are not mechanically driven by the size of the citation pool. A larger pool of follow-on research does not automatically increase the probability of subsequent firm outputs. Therefore,  this specification eliminates spurious mechanical relationships between pool size and citation probability.} Models include controls for the highest H-index among focal publication authors and firm fixed-effects. Journal-year fixed effects ensure instrument unconfoundedness. Standard-errors are clustered by firm.\footnote{Results are robust to alternative specifications, such as two-way clustering by firm and journal-year.}

Count-dependent variables are typically estimated using PPML or OLS with log-linear specifications. OLS models face challenges with zero counts, often addressed by adding a constant—a practice that introduces arbitrary bias \citep{cohn2022count}. Poisson regressions offer a better solution, but two-staged Poisson regression with fixed effects lacks readily available implementation. Therefore, the main analysis reports OLS and 2SLS regressions of linear probability models, where the dependent variable equals one if the count exceeds zero. For robustness, Appendix \ref{app:additional_results} presents Poisson estimations using PPML and two-staged Poisson with a control function approach, as suggested by \citet{LinWooldridge2019TestingCorrectingEndogeneity}.

Table \ref{tbl:discern_pubs_descstats} provides summary statistics for the publication sample.

\begin{table}[htbp]

\scriptsize
\centering
\caption{\label{tbl:discern_pubs_descstats}
Descriptive Statistics for Publication Sample}
\begin{threeparttable}
%\begin{tabular}{lHrrrrrrrr}
\begin{tabularx}{\textwidth}{l*{8}{Y}}
\toprule
Variable & Mean & SD & Min & p25 & p50 & p75 & Max \\
\midrule
Publication Year & $2,002.0$ & $6.4$ & $1,990$ & $1,997$ & $2,002$ & $2,008$ & $2,012$ \\
Follow-On Research & $12,923.7$ & $39,412.5$ & $1$ & $354$ & $2,081$ & $9,278$ & $2,102,260$ \\
Focal H-Index & $19.3$ & $17.8$ & $0$ & $7$ & $15$ & $27$ & $173$ \\
Top Two Researchers H-Index (IV) & $86.2$ & $48.2$ & $3$ & $50$ & $76$ & $114$ & $420$ \\
Future Pubs & $18.2$ & $41.3$ & $0$ & $0$ & $1$ & $17$ & $933$ \\
Future UIC Pubs & $7.3$ & $21.1$ & $0$ & $0$ & $0$ & $5$ & $858$ \\
Future Conf. Proceedings & $1.0$ & $6.6$ & $0$ & $0$ & $0$ & $0$ & $332$ \\
AMWS Hires & $0.3$ & $1.8$ & $0$ & $0$ & $0$ & $0$ & $58$ \\
AMWS Hires (Award-Winning) & $0.0$ & $0.2$ & $0$ & $0$ & $0$ & $0$ & $6$ \\
Future Patents by Authors & $10.0$ & $30.4$ & $0$ & $0$ & $1$ & $7$ & $1,388$ \\
Future Patents by Authors (gap $\geq$ 3y) & $6.4$ & $23.6$ & $0$ & $0$ & $0$ & $3$ & $1,332$ \\
Future Patents by Authors (gap $\geq$ 5y) & $4.6$ & $19.6$ & $0$ & $0$ & $0$ & $1$ & $1,250$ \\
Future Patents by Authors (UIC) & $0.2$ & $1.9$ & $0$ & $0$ & $0$ & $0$ & $151$ \\
Int. Patents Citing Focal & $0.3$ & $7.1$ & $0$ & $0$ & $0$ & $0$ & $1,888$ \\
Int. Patents Citing Focal or FO & $12.7$ & $89.4$ & $0$ & $0$ & $0$ & $1$ & $3,819$ \\
Ext. Patents Citing Focal & $2.3$ & $18.5$ & $0$ & $0$ & $0$ & $1$ & $3,549$ \\
Patent-Paper Pair & $0.1$ & $0.3$ & $0$ & $0$ & $0$ & $0$ & $1$ \\
University-Industry Collaboration & $0.6$ & $0.5$ & $0$ & $0$ & $1$ & $1$ & $1$ \\
Low Concept Prevalence & $0.6$ & $0.5$ & $0$ & $0$ & $1$ & $1$ & $1$ \\
High Government Funding & $0.4$ & $0.5$ & $0$ & $0$ & $0$ & $1$ & $1$ \\
Future Patents by Authors (Citing FO) & $0.3$ & $4.3$ & $0$ & $0$ & $0$ & $0$ & $467$ \\
Future Patents by Authors (Not Citing FO) & $9.7$ & $29.3$ & $0$ & $0$ & $1$ & $7$ & $1,388$ \\
\bottomrule
%\end{tabular}
\end{tabularx}
\begin{tablenotes}[flushleft]
      \scriptsize
         \item \textit{Note.} This table provides summary statistics for the variables used in the econometric analysis at the publication level. The data is based on the DISCERN database of publications by U.S.-based publicly-owned firms between 1990 and 2012. The sample includes 164516 publications originating from 1522 firms.

\end{tablenotes}
\end{threeparttable}
\end{table}

\subsubsection{Effects on Firms' Investments in Science and Patenting Outcomes}

Table \ref{tbl:discern_pubspats} presents an estimation of the effects of follow-on research on firms' subsequent investments in science, patenting outcomes and employee retention. The analysis focuses on subsequent publications and patents by the focal publications' corporate authors, the firm's patents citing the focal publication and a proxy for job tenure. Results indicate a positive effect of follow-on research on the probability of subsequent publications by focal authors (columns 1 and 3).\footnote{2SLS estimates suggest that, for the average publication, a 1\% increase in follow-on research corresponds to a $\frac{0.123/100}{0.507}=0.24\%$ increase in the probability of at least one additional publication (column 3). A Poisson regression model indicates that an additional publication by focal authors requires a 88\% ($(20.95\times0.054)^{-1}$) increase in follow-on research (Table \ref{tbl:poisson_iv}, column 1). Results are robust to a 2-stage Poisson estimation using a control function approach (Table \ref{tbl:poisson_iv}, column 2). Appendix Figure \ref{fig:baseline_pubs} presents a corresponding binned scatterplot.} The 2SLS estimate implies that a doubling of follow-on research raises the probability of a subsequent publication by roughly 8–9 percentage points, from an average baseline of 50\% to about 59\%. In addition, using publications by the same authors at least 10 years after the focal publication, Columns 8-9 present evidence that follow-on research increases the likelihood of the researcher employee retention by the firm.

Columns 4-7 estimate the effects of follow-on research on firms' patenting outcomes. For patents listing corporate authors of the focal paper as inventors and assigned to the focal firm (columns 4-6), the OLS estimate is positive and significant for patents filed at least three years after the focal publication (column 4). The instrumented model shows a positive but non-significant relation ($p=0.15$, column 5).\footnote{Appendix Figure \ref{fig:baseline_pats} presents a corresponding binned scatterplot to column 4.} For patents filed five or more years after publication, the 2SLS model yields a positive and significant coefficient (column 6),  suggesting that a 1\% increase in follow-on research is associated with a 0.3\% higher likelihood of at least one patent. This implies that a doubling of follow-on research raises the probability of subsequent patenting from about 30\% to 37\%. Column 7 provides additional evidence using an indicator for whether the firm's subsequent patents cite the focal publication as the dependent variable, again showing a positive effect of follow-on research.  These results align with the extended time required for follow-on research to materialize and the long gestation period often required for research to translate into patentable technologies.

Taken together, the results provide consistent evidence that follow-on research strengthens firms’ subsequent innovative performance and engagement in public science, increasing both subsequent publications and patents attributable to the originating firm. In addition, I find evidence that follow-on research increases the likelihood that researcher employees remain with the firm over extended periods.

%\begin{table}[htbp]
\begin{sidewaystable}
\footnotesize \singlespacing
   \centering
   \begin{threeparttable}[b]
      \caption{\label{tbl:discern_pubspats}The Effect of Follow-On Research on Firms' Investments in Science and Patenting Outcomes}
      \bigskip
\begin{tabular}{l*{9}{c}}
\toprule

&\multicolumn{3}{c}{Subsequent Scientific Publications} &\multicolumn{4}{c}{Subsequent Patenting}
&\multicolumn{2}{c}{\shortstack{Employee\\Retention}}\\
\cmidrule(lr){2-4}\cmidrule(lr){5-8}\cmidrule(lr){9-10}

                    &\multicolumn{1}{c}{\shortstack{Pr(Future\\Pub.)}}&\multicolumn{1}{c}{\shortstack{ln(Follow-on\\Research)}}&\multicolumn{1}{c}{\shortstack{Pr(Future\\Pub.)}}&\multicolumn{2}{c}{\shortstack{Pr(Future Pat.,\\ $\geq$ 3y gap)}}&\multicolumn{1}{c}{\shortstack{Pr($\cdot$,\\ $\geq$ 5y gap)}}&\multicolumn{1}{c}{\shortstack{Pr(Pat\\Citation)}}&\multicolumn{2}{c}{\shortstack{Pr(Job Tenure\\ $\geq$ 10y)}}\\\cmidrule(lr){2-2}\cmidrule(lr){3-3}\cmidrule(lr){4-4}\cmidrule(lr){5-6}\cmidrule(lr){7-7}\cmidrule(lr){8-8}\cmidrule(lr){9-10}
                    &\multicolumn{1}{c}{OLS}&\multicolumn{1}{c}{Stage 1}&\multicolumn{1}{c}{Stage 2}&\multicolumn{1}{c}{OLS}&\multicolumn{1}{c}{2SLS}&\multicolumn{1}{c}{2SLS}&\multicolumn{1}{c}{2SLS}&\multicolumn{1}{c}{OLS}&\multicolumn{1}{c}{2SLS}\\
&(1)&(2)&(3)&(4)&(5)&(6)&(7)&(8)&(9)\\ \midrule
ln(Follow-on)       &       0.008\sym{***}&                     &       0.123\sym{**} &       0.007\sym{***}&       0.068         &       0.095\sym{**} &       0.047\sym{*}  &       0.007\sym{***}&       0.113\sym{**} \\
                    &     (0.001)         &                     &     (0.048)         &     (0.001)         &     (0.047)         &     (0.045)         &     (0.027)         &     (0.001)         &     (0.046)         \\
\addlinespace
ln(top researchers&                     &       0.096\sym{***}&                     &                     &                     &                     &                     &                     &                    \\
  H-index)                    &                     &     (0.018)         &                     &                     &                     &                     &                     &                     &                     \\
\addlinespace
ln(Focal H-Index)   &       0.002         &       0.279\sym{***}&      -0.030\sym{**} &       0.012\sym{***}&      -0.005         &      -0.018         &      -0.018\sym{**} &       0.009\sym{***}&      -0.020         \\
                    &     (0.003)         &     (0.010)         &     (0.014)         &     (0.002)         &     (0.013)         &     (0.012)         &     (0.008)         &     (0.003)         &     (0.013)         \\
\\
Firm FE             &         Yes         &         Yes         &         Yes         &         Yes         &         Yes         &         Yes         &         Yes         &         Yes         &         Yes         \\
Journal-Year FE     &         Yes         &         Yes         &         Yes         &         Yes         &         Yes         &         Yes         &         Yes         &         Yes         &         Yes         \\
Observations        &     164,516         &     164,516         &     164,516         &     164,516         &     164,516         &     164,516         &     164,516         &     164,516         &     164,516         \\
Avg. DV             &       0.507         &       7.357         &       0.507         &       0.397         &       0.397         &       0.304         &       0.056         &       0.251         &       0.251         \\
First Stage F-stat  &                     &                     &      26.795         &                     &      26.795         &      26.795         &      26.795         &                     &      26.795         \\
Adjusted R$^2$      &       0.344         &      -0.145         &           .         &       0.348         &           .         &           .         &           .         &       0.238         &           .         \\
\bottomrule
\end{tabular}

      \begin{tablenotes}[flushleft]
      \scriptsize
         \item \textit{Note.} This table presents estimation results for the relationship between external follow-on research and firms' subsequent investments in science and patenting outcomes. The data consists of a pooled cross section of publications by U.S.-based publicly-owned firms, published between 1990 and 2012 \citep{AroraBelenzonSheer2020DISCERNDukeInnovationa}. Follow-on research is the total count of three generations of citations to the focal publication from outside the firm. In columns 1 \& 3, the dependent variable is an indicator for future scientific publications by the corporate authors of the focal publication. 
         The next dependent variables are indicators for a future patent filed at least three (columns 4-5) and five (column 6) years after the publication of the focal paper and an indicator for a future patent by the firm that cites the focal publication (column 7). Last dependent variables are indicators for job tenure, proxied by the existence of a subsequent scientific publication 10 years after the focal publication (columns 8-9).  In 2SLS regressions, the instrumental variable is the sum of H-indexes of the top two authors in the same journal issue as the focal publication. First-stage results are presented in column 2. All regressions include a control for the highest H-index among the authors of the focal publication, as well as firm and journal-year fixed effects. 
         %\item 
         Clustered (Firm) standard-errors in parentheses. Signif. Codes: ***: 0.01, **: 0.05, *: 0.1
      \end{tablenotes}
   \end{threeparttable}
\end{sidewaystable}
%\end{table}

\subsubsection{Effects on Firms' Subsequent Interactions with Academia}

Table \ref{tbl:discern_uic} estimates the effects of follow-on research on firms' subsequent interactions with academics. Results indicate a positive effect on research collaborations with academics (columns 1 and 2).\footnote{A 1\% increase in follow-on research corresponds to a 0.025\% increase in UIC publication probability (column 1), with a 0.28\% local effect in the 2SLS estimate (column 2). Poisson regression suggests an additional UIC publication associates with a 170\% increase in follow-on research. Results hold using a 2-stage Poisson estimation (Table \ref{tbl:poisson_iv} columns 3 \& 4).} In addition, follow-on research positively correlates with subsequent conference proceedings probability (column 3)\footnote{The 2SLS estimate (column 4) is positive but not statistically significant ($p=0.11$).} and joint firm-public institution patent assignments (column 5).

Moreover, follow-on research increases the probability of firms hiring scientists whose research closely relates to the focal publication (columns 7-8), including award-winning scientists (columns 9-10).\footnote{Corresponding count model estimates are also positive, and the results hold using 2-stage Poisson estimation (Table \ref{tbl:poisson_iv} columns 5 \& 6).} The estimated effects, however, are relatively small, possibly due to the partial availability of data regarding the hiring of scientists by the firms in the sample.

Taking together the results in Tables \ref{tbl:discern_pubspats} and \ref{tbl:discern_uic}, I find evidence that follow-on research increases firms' subsequent investments in science, improves their patenting outcomes, and drives direct ties with academics.

%\begin{table}[htbp]
\begin{sidewaystable}
\footnotesize \singlespacing
   \centering
   \begin{threeparttable}[b]
      \caption{\label{tbl:discern_uic}The Effect of Follow-On Research on Firms' Subsequent Interactions with Academia}
      \bigskip

\begin{tabular}{l*{10}{c}}
\toprule
&\multicolumn{6}{c}{Subsequent Interactions with Academia} &\multicolumn{4}{c}{Hiring of Renowned Scientists (AMWS)}  \\
\cmidrule(lr){2-7}\cmidrule(lr){8-11}
                    &\multicolumn{2}{c}{\shortstack{Pr(Future UIC\\ Publication)}}         &\multicolumn{2}{c}{\shortstack{Pr(Conference\\Proceeding)}}         &\multicolumn{2}{c}{\shortstack{Pr(Future UIC\\Patent)}}  &\multicolumn{2}{c}{Pr(Hire)}               &\multicolumn{2}{c}{\shortstack{Pr(Award-winning\\Hire)}} \\\cmidrule(lr){2-3}\cmidrule(lr){4-5}\cmidrule(lr){6-7}\cmidrule(lr){8-9}\cmidrule(lr){10-11}
                    &\multicolumn{1}{c}{OLS}&\multicolumn{1}{c}{2SLS}&\multicolumn{1}{c}{OLS}&\multicolumn{1}{c}{2SLS}&\multicolumn{1}{c}{OLS}&\multicolumn{1}{c}{2SLS}&\multicolumn{1}{c}{OLS}&\multicolumn{1}{c}{2SLS}&\multicolumn{1}{c}{OLS}&\multicolumn{1}{c}{2SLS}\\
&(1)&(2)&(3)&(4)&(5)&(6)&(7)&(8)&(9)&(10)\\ \midrule
ln(Follow-on)       &       0.011\sym{***}&       0.130\sym{***}&       0.002\sym{***}&       0.051         &       0.002\sym{***}&       0.014         &       0.005\sym{***}&       0.059\sym{**} &       0.002\sym{*}  &       0.035\sym{**} \\
                    &     (0.002)         &     (0.049)         &     (0.001)         &     (0.031)         &     (0.001)         &     (0.020)         &     (0.001)         &     (0.024)         &     (0.001)         &     (0.015)         \\
\addlinespace
ln(Focal H-Index)   &       0.035\sym{***}&       0.001         &       0.005\sym{**} &      -0.008         &       0.009\sym{***}&       0.006         &       0.003\sym{**} &      -0.012\sym{*}  &       0.000         &      -0.009\sym{**} \\
                    &     (0.003)         &     (0.015)         &     (0.002)         &     (0.009)         &     (0.002)         &     (0.005)         &     (0.001)         &     (0.007)         &     (0.001)         &     (0.004)         \\
\\
Firm FE             &         Yes         &         Yes         &         Yes         &         Yes         &         Yes         &         Yes         &         Yes         &         Yes         &         Yes         &         Yes         \\
Journal-Year FE     &         Yes         &         Yes         &         Yes         &         Yes         &         Yes         &         Yes         &         Yes         &         Yes         &         Yes         &         Yes         \\
Observations        &     164,516         &     164,516         &     164,516         &     164,516         &     164,516         &     164,516         &     164,516         &     164,516         &     164,516         &     164,516         \\
Avg. DV             &       0.435         &       0.435         &       0.093         &       0.093         &       0.033         &       0.033         &       0.080         &       0.080         &       0.018         &       0.018         \\
First Stage F-stat  &                     &      26.795         &                     &      26.795         &                     &      26.795         &                     &      26.795         &                     &      26.795         \\
Adjusted R$^2$      &       0.320         &           .         &       0.257         &           .         &       0.128         &           .         &       0.223         &           .         &       0.208         &           .         \\
\bottomrule
\end{tabular}

      \begin{tablenotes}[flushleft]
      \scriptsize
         \item \textit{Notes:} This table presents estimation results for the relationship between external follow-on research and firms' subsequent interactions with academia. The data consists of a pooled cross section of publications by U.S.-based publicly-owned firms, published between 1990 and 2012 \citep{AroraBelenzonSheer2020DISCERNDukeInnovationa}. Follow-on research is the total count of three generations of citations to the focal publication from outside the firm.        
         The dependent variables are indicators for future scientific publications co-authored with academics (columns 1-2), future conference proceedings (columns 3-4), and future co-invented patents by the corporate authors of the focal publication (columns 5-6). In columns 7-8, the dependent variable is an indicator for future employment of a renowned scientist (columns 9-10, award-winning scientist) whose work is highly related to the focal publication.
         In 2SLS regressions, the instrumental variable is the sum of H-indexes of the top two authors in the same journal issue as the focal publication. All regressions include a control for the highest H-index among the authors of the focal publication, as well as firm and journal-year fixed effects. 
         %\item 
         Clustered (Firm) standard-errors in parentheses. Signif. Codes: ***: 0.01, **: 0.05, *: 0.1
      \end{tablenotes}
   \end{threeparttable}
\end{sidewaystable}
%\end{table}

\subsubsection{Matched Controls}

 One possible concern with the results above is that the firm's involvement in scientific research is inconsequential. That is, possibly the effect of follow-on research would be similar to completely external advancements in a given scientific research topic. While counterfactuals for firms' investments in research are unobservable, Table \ref{tbl:discern_controls} provides a step in that direction. For each focal (``internal'') publication in the sample, I match a random publication that is not associated with the focal firm and published in the same journal and year. Next, I observe the focal firm's patent citations to both the original sample and the sample of matched publications. The dependent variable is an indicator equal to one if at least one patent by the focal firm cites the internal (or matched) publication. 
 
Results show a larger estimated coefficient for internal publications compared to the matched sample. Columns 4-5 estimate an interaction term of ln(follow-on) with an internal publication indicator. A positive and significant interaction term is found using OLS (column 4) and a weakly significant estimate using 2SLS (column 5, $p=0.09$).\footnote{For the average publication, the probability of a patent citation to the internal publication following a 1\% increase in follow-on research is 2.5 times ($\frac{0.008+0.045+0.02}{0.029}$) the probability for the matched publication.} These findings suggest that follow-on research is more strongly associated with subsequent patents when firms rely on their own science. However, caution is warranted in interpretation due to firms' selection of research topics.

\begin{table}[htbp]
\footnotesize \singlespacing
   \centering
   \begin{threeparttable}[b]
      \caption{\label{tbl:discern_controls}Comparison of Internal Publications and Matched Controls}
      \bigskip

%\begin{tabular}{l*{5}{c}}
\begin{tabularx}{\textwidth}{l@{}*{5}{Y}}
\toprule
                    &\multicolumn{5}{c}{Pr(Subsequent Firms' Patents Citing Focal Publication)}                                                            \\\cmidrule(lr){2-6}
                    &\multicolumn{1}{c}{\shortstack{OLS\\Internal\\Pubs}}&\multicolumn{1}{c}{\shortstack{OLS\\Matched\\Pubs}}&\multicolumn{1}{c}{OLS}&\multicolumn{1}{c}{OLS}&\multicolumn{1}{c}{2SLS}\\
&(1)&(2)&(3)&(4)&(5)\\
\midrule
ln(Follow-on) $\times$ Internal&                     &                     &                     &       0.011\sym{***}&       0.009\sym{*}  \\
                    &                     &                     &                     &     (0.001)         &     (0.005)         \\
\addlinespace
Internal            &                     &                     &       0.051\sym{***}&       0.049\sym{***}&       0.045\sym{***}\\
                    &                     &                     &     (0.004)         &     (0.004)         &     (0.005)         \\
\addlinespace
ln(Follow-on)       &       0.011\sym{***}&       0.001\sym{***}&                     &       0.001         &       0.018         \\
                    &     (0.001)         &     (0.000)         &                     &     (0.001)         &     (0.017)         \\
\addlinespace
ln(Focal H-Index)   &      -0.008\sym{***}&       0.000         &      -0.002\sym{***}&      -0.004\sym{***}&      -0.009\sym{*}  \\
                    &     (0.001)         &     (0.000)         &     (0.001)         &     (0.001)         &     (0.006)         \\
\\
Firm FE             &         Yes         &         Yes         &         Yes         &         Yes         &         Yes         \\
Journal-Year FE     &         Yes         &         Yes         &         Yes         &         Yes         &         Yes         \\
Observations        &     164,516         &     158,816         &     347,154         &     347,154         &     347,154         \\
Avg. DV             &       0.056         &       0.004         &       0.029         &       0.029         &       0.029         \\
First Stage F-stat  &                     &                     &                     &                     &      25.293         \\
Adjusted R$^2$      &      -0.163         &      -0.171         &      -0.087         &      -0.075         &      .         \\
\bottomrule
\end{tabularx}

      \begin{tablenotes}[flushleft]
      \scriptsize
      \item \textit{Note.} This table compares the sample of internal publications and a sample of matched publications originating outside the firm. For each internal publication, a publication from the same journal year is randomly assigned. The dependent variable is an indicator variable that is equal to one if at least one patent within the firm is citing the publication. Follow-on research is the total count of three generations of citations to the focal publication from outside the firm. In interaction models (columns 4-6), ln(follow-on) is recentered around the sample mean.
         Internal is an indicator variable that is equal to one if the publication is published by the originating firm and equal to zero if it is the matched external publication. 
        In 2SLS regressions, the instrumental variable is the sum of H-indexes of the top two authors in the same journal issue as the focal paper. Focal H-index is the highest H-index among the authors of the focal paper at the year of publication. In column 5, ln(follow-on) and the interaction term are instrumented by the the IV and the interaction of the IV with the indicator for internal publication. Observations are automatically dropped from the full sample due to separation and singletons. 
         %\item 
         Clustered (Firm) standard-errors in parentheses. Signif. Codes: ***: 0.01, **: 0.05, *: 0.1
      \end{tablenotes}
   \end{threeparttable}
\end{table}

\subsection{\label{sec:mech}Mechanisms Driving the Value of Follow-On Research}

Follow-on research might offer multiple benefits to firms. One, it can provide valuable inputs for subsequent innovation.
Additionally, follow-on citations that do not represent useful inputs can still benefit originating firms by validating the quality of internally-produced scientific findings, potentially influencing internal resource allocation and innovation direction.

To further explore these effects, I develop proxies for the two ways firms can utilize follow-on research: (1) as an input into invention, proxied by subsequent patents citing follow-on research and (2) as external validation, proxied by subsequent patents not citing follow-on research. These proxies build on the notion that patent citations are required by law when prior art is used in the inventive process. Table \ref{tbl:mech_simple} reports estimation results. Columns 1-2 present evidence that follow-on research is used in subsequent inventive activity. Columns 3-4 present the complementary result: that follow-on research also drives invention when not used as an input. I interpret these results as evidence that follow-on research is useful beyond serving as inputs, potentially providing validation of the quality of the firm's own science and affecting resource allocation within the firm.\footnote{Importantly, while the relations in columns 1-2 might be driven by mechanical exposure effects, these effects work against the results presented in columns 3-4.}
Appendix Section \ref{sec:mech_extended} further explores these results, interacting the extent of follow-on research with the focal authors' prominence. I find suggestive evidence that follow-on research as validation is more important for non-prominent employees, where uncertainty around the quality of their work is presumably higher.

\begin{table}[htbp]
\footnotesize \singlespacing
   \centering
   \begin{threeparttable}[b]
      \caption{\label{tbl:mech_simple}Ways in Which Follow-On Research Is Privately Useful}
      \bigskip

%\begin{tabular}{l*{8}{c}}
\begin{tabularx}{\textwidth}{l@{}*{4}{Y}}
\toprule
                    &\multicolumn{2}{c}{Pr(Patent, Citing FO)}  &\multicolumn{2}{c}{Pr(Patent, Not Citing FO)}\\\cmidrule(lr){2-3}\cmidrule(lr){4-5}
                    &\multicolumn{1}{c}{OLS}&\multicolumn{1}{c}{2SLS}&\multicolumn{1}{c}{OLS}&\multicolumn{1}{c}{2SLS}\\
&(1)&(2)&(3)&(4)\\ \midrule
ln(Follow-on)       &       0.012\sym{***}&       0.054\sym{**} &       0.005\sym{***}&       0.092\sym{**} \\
                    &     (0.002)         &     (0.022)         &     (0.001)         &     (0.045)         \\
\addlinespace
ln(Focal H-Index)   &       0.002\sym{*}  &      -0.009         &       0.006\sym{***}&      -0.018         \\
                    &     (0.001)         &     (0.006)         &     (0.002)         &     (0.013)         \\
\\
Firm FE             &         Yes         &         Yes         &         Yes         &         Yes         \\
Journal-Year FE     &         Yes         &         Yes         &         Yes         &         Yes         \\
Observations        &     164,516         &     164,516         &     164,516         &     164,516         \\
Avg. DV             &       0.039         &       0.039         &       0.300         &       0.300         \\
First Stage F-stat  &                     &      26.795         &                     &      26.795         \\
Adjusted R$^2$      &       0.133         &      .         &       0.366         &      .         \\
\bottomrule
%\end{tabular}
\end{tabularx}
      \begin{tablenotes}[flushleft]
      \scriptsize
      \item \textit{Note.} This table explores how the effect of external follow-on research affects patenting outcomes in different ways. The dependent variables are indicators for subsequent patents by the focal authors, filed at least five years after the focal publication. In columns 1-2, the dependent variable is an indicator for a subsequent patent that cites the follow-on research. In columns 3-4, the dependent variable is an indicator for a subsequent patent that does not cite the follow-on research. 
         %\item 
         Clustered (Firm) standard-errors in parentheses. Signif. Codes: ***: 0.01, **: 0.05, *: 0.1
      \end{tablenotes}
   \end{threeparttable}
\end{table}

\subsection{Conditions That Drive the Private Value of Follow-On Research}

The private value of follow-on research is likely to be moderated by firms' scientific and inventive capabilities, and by characteristics of the external scientific community. I explore such relations in Table \ref{tbl:hetero}. In columns 1-2, I explore the role of firms' possession of complementary intellectual property rights (IPR). According to \citet{Teece1986ProfitingTechnologicalInnovation}, firms that possess complementary assets, such as intellectual property, manufacturing capabilities, or distribution channels, are better positioned to capture value from innovation. In the context of follow-on research, firms with related patents have greater control over the commercialization of related technology, allowing them to more effectively appropriate value from external scientific advances that build on their published work. To proxy for IPR, I search for patent-paper pairs, defined as patents by the focal authors filed at most two years after the focal publication that share textual concepts with the focal publication.\footnote{To ensure that I do not include these patents in the construction of the dependent variable, I limit the outcomes to patents filed three years or more after the focal publication.} I find a positive interaction term in both the OLS and 2SLS estimation, suggesting that complementary assets assist firms in subsequently benefiting from follow-on science. 

Second, I explore the role of firms' decision to collaborate on the focal publication with external academics (UIC). On the one hand, external researchers can extend the firms' absorptive capacity and increase the use of follow-on research. On the other hand, \citet{BikardVakiliTeodoridis2018WhenCollaborationBridges} show that university-industry collaboration can enable a productive division of labor, whereby academic partners focus on scientific exploration while industry partners concentrate on commercialization. However, outsourcing research through temporary collaborations may weaken firms' internal scientific capabilities, reducing their ability to absorb and build upon follow-on research in subsequent years. In Table \ref{tbl:hetero} columns 3 and 4, I find results that are in line with the latter. Compared to strictly in-house research, follow-on related to collaborated focal publications is less likely to lead to a subsequent patent by the focal authors. 

Next, I contrast mature scientific areas with abundant related research to nascent areas where external exploration is scarce and limited. In areas where many academics are operating, the firm is potentially facing less need (or ability) to induce additional external inquiry. I construct a measure of scientific prevalence by counting the appearance of textual concepts related to the focal publications in previous publications outside the firm (up to three years apart). Columns 5 and 6 report the estimation results. Follow-on research is more valuable when it originates in areas less previously explored than in mature and well-developed areas. These results suggest that follow-on research is more likely to benefit the firm when it redirects academics to work on new and less-explored research areas.

Lastly, I compare scientific areas where government funding is relatively abundant to areas with less support from the government. According to \citet{BabinaHeHowellEtAl2023CuttingInnovationEngine}, government funding increases the public-good nature of science and makes it more available for appropriation (in this case, by the focal firm). Therefore, the availability of government funding will make follow-on research more valuable for the focal firm. I construct a measure of government funding availability by observing funding acknowledgments of prior research related to the focal publication (as identified in the previous results above). For each focal publication, I calculate the percentage of prior related works that acknowledge government funding and split the sample based on the median. Columns 7-8 report the results.\footnote{The interaction coefficient in column 8 is marginally significant with a p-value of 0.102.} I find evidence for a positive moderation of government funding availability.\footnote{Appendix Table \ref{tbl:hetero_npl2foc} presents corresponding estimations for the effect on subsequent patents by the firm that cite the focal publication. In general, the results are qualitatively similar but with weaker statistical significance.}

\begin{table}[htbp]
\footnotesize \singlespacing
   \centering
   \begin{threeparttable}[b]
      \caption{\label{tbl:hetero}Heterogeneity in Subsequent Patenting}
      \bigskip

%\begin{tabular}{l*{8}{c}}
\begin{tabularx}{\textwidth}{l@{}*{8}{Y}}
\toprule
%                    &\multicolumn{2}{c}{\shortstack{Complementary\\IP Rights}}&\multicolumn{2}{c}{\shortstack{Knowledge\\Outsourcing}}  &\multicolumn{2}{c}{\shortstack{Scientific Concept\\Prevalence}}&\multicolumn{2}{c}{\shortstack{Govt. Funding\\Availability}}\\\cmidrule(lr){2-3}\cmidrule(lr){4-5}\cmidrule(lr){6-7}\cmidrule(lr){8-9}
%&Pr(Patent)&Pr(Patent)&Pr(Patent)&Pr(Patent)&Pr(Patent)&Pr(Patent)&Pr(Patent)&Pr(Patent)\\
%                    \cmidrule(lr){2-2}\cmidrule(lr){3-3}\cmidrule(lr){4-4}\cmidrule(lr){5-5}\cmidrule(lr){6-6}\cmidrule(lr){7-7}\cmidrule(lr){8-8}
&\multicolumn{4}{c}{Firm Capabilities}&\multicolumn{4}{c}{Scientific Community}\\
\cmidrule(lr){2-5}\cmidrule(l){6-9}
&\multicolumn{8}{c}{Pr(Subsequent Patent by Focal Authors, $\geq$3y gap)}\\
\cmidrule(){2-9}
                    &\multicolumn{1}{c}{OLS}&\multicolumn{1}{c}{2SLS}&\multicolumn{1}{c}{OLS}&\multicolumn{1}{c}{2SLS}&\multicolumn{1}{c}{OLS}&\multicolumn{1}{c}{2SLS}&\multicolumn{1}{c}{OLS}&\multicolumn{1}{c}{2SLS}\\
&(1)&(2)&(3)&(4)&(5)&(6)&(7)&(8)\\ \midrule
\multicolumn{9}{l}{\textbf{Complementary IP Rights}}\\
\addlinespace
ln(Follow-On) $\times$ &       0.006\sym{***}&       0.024\sym{**} &                     &                     &                     &                     &                     &                     \\
   \ \ \ \ PPP                 &     (0.002)         &     (0.010)         &                     &                     &                     &                     &                     &                     \\
\addlinespace
PPP                 &       0.303\sym{***}&       0.270\sym{***}&                     &                     &                     &                     &                     &                     \\
                    &     (0.016)         &     (0.027)         &                     &                     &                     &                     &                     &                     \\
\addlinespace
\multicolumn{9}{l}{\textbf{Knowledge Outsourcing}}\\
\addlinespace
ln(Follow-On) $\times$ &                     &                     &      -0.005\sym{***}&      -0.020\sym{**} &                     &                     &                     &                     \\
\ \ \ \ UIC                    &                     &                     &     (0.001)         &     (0.008)         &                     &                     &                     &                     \\
\addlinespace
UIC                 &                     &                     &      -0.066\sym{***}&      -0.064\sym{***}&                     &                     &                     &                     \\
                    &                     &                     &     (0.006)         &     (0.005)         &                     &                     &                     &                     \\

\addlinespace
\multicolumn{9}{l}{\textbf{Scientific Concept Prevalence}}\\
\addlinespace
ln(Follow-On) $\times$ &                     &                     &                     &                     &       0.001         &       0.023\sym{**} &                     &                     \\
  \ \ \ \ Low Prev.                  &                     &                     &                     &                     &     (0.001)         &     (0.009)         &                     &                     \\
\addlinespace
Low Prevalence      &                     &                     &                     &                     &       0.015\sym{***}&       0.008\sym{*}  &                     &                     \\
                    &                     &                     &                     &                     &     (0.003)         &     (0.005)         &                     &                     \\

\addlinespace
\multicolumn{9}{l}{\textbf{Government  Funding Availability}}\\
\addlinespace
ln(Follow-On) $\times$ &                     &                     &                     &                     &                     &                     &       0.003\sym{**} &       0.025         \\
\ \ \ \ Govt. Funding                    &                     &                     &                     &                     &                     &                     &     (0.001)         &     (0.015)         \\
\addlinespace
Govt Funding        &                     &                     &                     &                     &                     &                     &       0.022\sym{***}&       0.007         \\
                    &                     &                     &                     &                     &                     &                     &     (0.004)         &     (0.009)         \\

\addlinespace
ln(Follow-On)       &       0.004\sym{***}&       0.075         &       0.010\sym{***}&       0.072         &       0.007\sym{***}&       0.057         &       0.006\sym{***}&       0.060         \\
                    &     (0.001)         &     (0.047)         &     (0.001)         &     (0.046)         &     (0.001)         &     (0.045)         &     (0.001)         &     (0.045)         \\
\addlinespace
ln(Focal H-Index)   &       0.010\sym{***}&      -0.010         &       0.023\sym{***}&       0.007         &       0.012\sym{***}&      -0.006         &       0.011\sym{***}&      -0.006         \\
                    &     (0.002)         &     (0.013)         &     (0.002)         &     (0.013)         &     (0.002)         &     (0.013)         &     (0.002)         &     (0.013)         \\
\addlinespace
%Constant            &       0.341\sym{***}&                     &       0.514\sym{***}&                     &       0.492\sym{***}&                     &       0.493\sym{***}&                     \\
%                    &     (0.005)         &                     &     (0.005)         &                     &     (0.006)         &                     &     (0.006)         &                     \\
\\
Firm FE             &         Yes         &         Yes         &         Yes         &         Yes         &         Yes         &         Yes         &         Yes         &         Yes         \\
Journal-Year FE     &         Yes         &         Yes         &         Yes         &         Yes         &         Yes         &         Yes         &         Yes         &         Yes         \\
Observations        &     164,516         &     164,516         &     164,516         &     164,516         &     164,516         &     164,516         &     164,516         &     164,516         \\
Avg. DV             &       0.397         &       0.397         &       0.397         &       0.397         &       0.397         &       0.397         &       0.397         &       0.397         \\
First Stage F-stat  &                     &      13.819         &                     &      13.007         &                     &      13.155         &                     &      13.594         \\
Adjusted R$^2$      &       0.380         &      .         &       0.352         &      .         &       0.349         &      .         &       0.349         &      .         \\
\bottomrule
\end{tabularx}

      \begin{tablenotes}[flushleft]
      \scriptsize
      \item \textit{Note.} This table explores heterogeneity in the relationship between external follow-on research and firms' subsequent patenting outcomes (see Table \ref{tbl:discern_pubspats} for baseline results). In columns 1-2, follow-on research is interacted with an indicator that the focal publication is a part of a patent-paper pair (PPP). In columns 3-4, follow-on research is interacted with an indicator that the focal publication is a university-industry collaboration (UIC). In columns 5-6, follow-on research is interacted with an indicator for below-average prevalence of prior works sharing the same textual concepts. In columns 7-8, follow-on research is interacted with an indicator for above-average government funding acknowledgments in related publications. In all models, log(follow-on) is recentered around the sample mean. In 2SLS models, ln(follow-on) and the interaction term are instrumented by the the IV and the interaction of the IV with the indicator variable.
         %\item 
         Clustered (Firm) standard-errors in parentheses. Signif. Codes: ***: 0.01, **: 0.05, *: 0.1
      \end{tablenotes}
   \end{threeparttable}
\end{table}

\subsection{Comparing OLS and 2SLS Estimations}

The 2SLS coefficient estimates presented in this paper are notably larger than their OLS counterparts. This pattern warrants careful interpretation, as it may initially appear counterintuitive if one expects quality confounders to bias OLS upward.

Several considerations help explain why this difference does not undermine the identification strategy. First, the direction of OLS bias is theoretically indeterminate when multiple omitted variables operate simultaneously \citep{Basu2020BiasOLSEstimators}. While unobserved paper quality would bias OLS upward, countervailing factors could produce downward bias. For instance, if external scientists strategically avoid areas where firms are actively developing, or if a division of labor means that papers attracting citations address problems of limited relevance to the firm's core R\&D agenda. When omitted variables pull in opposing directions, the net bias cannot be signed without strong auxiliary assumptions.

Additionally, the IV estimate identifies a local average treatment effect (LATE) that may legitimately exceed the average treatment effect. As shown in Appendix Section \ref{sec:acr}, the instrument's effect varies with focal author prominence, meaning the 2SLS estimate reflects effects for a specific subset of papers. Moreover, the instrument likely influences only ``marginal'' citations from serendipitous discovery, which may have different effects than ``core'' citations from researchers already aware of the work. Marginal citations may disproportionately originate from researchers in peripheral fields who bring different theoretical frameworks and identify novel applications. This is consistent with theories of recombinant innovation suggesting that advances often arise at disciplinary boundaries \citep{Fleming2001RecombinantUncertaintyTechnological}. Citations from such researchers may generate disproportionate value precisely because they connect the firm's research to unexpected domains.

Appendix Section \ref{sec:iv_sim} demonstrates through simulation that when an instrument affects only a subset of observations with heterogeneous effects, OLS can be biased downward relative to IV as a mechanical consequence of aggregation.

\subsection{Additional Analyses}

Appendix Section \ref{sec:app_res} provides additional results. Namely, I explore heterogeneity analyses of the effect of follow-on research on subsequent patents by the firm citing the focal publication, correlations on a sample of firms' conference proceedings, variation of the baseline results by scientific fields and main industries, and a comparison of the effects of corporate versus academic follow-on research.

\section{\label{sec:firmlevel}Follow-On Research and Firms' Innovative Performance}

In addition to estimating the effects of follow-on research at the publication level, I explore relations between the variables at the firm level.\footnote{Unfortunately, the instrumental variable that I introduce at the publication level is unsuitable for aggregation up to the firm level. Therefore, I present correlational evidence in this section and acknowledge potential bias due to unobserved variables.} Specifically, I study the relationship between the lagged stock of realized follow-on research and firms' annual scientific publications, employment of scientists, and patenting. If follow-on research drives subsequent scientific investments and patenting outcomes at the publication level, these relations should also be observed in a firm-level panel analysis. The econometric specification is as follows:
 
\begin{equation}
\begin{aligned}
    \ln(Y)_{ft} = 
    &\beta_1 \ln(\text{FO}_{f,t-1})+\\
    &\beta_2 \ln(\text{Int. Patents Citing FO}_{f,t-1})+ \\
    &\beta_3 \ln(\text{Ext. Patents Citing FO}_{f,t-1})+ \\
    &\gamma X_{f,{t-1}} + \eta_f + \tau_t + \epsilon_{f,t}
    \end{aligned}
\end{equation}
 
 The dependent variables are annual counts of scientific publications, patents, and stocks of scientists employed by firm $f$ at year $t$. The main variable on the right-hand side is an accumulated stock of external follow-on publications. The follow-on publications are aggregated by their own publication year to indicate realized follow-on research up to year $t$. The full specifications include stocks of realized patents by the focal firm and others that cite the follow-on research. 
 I use the standard 15\% depreciation value to reduce stocks over time. All models include firm and year fixed effects, along with time-varying controls for assets and R\&D investments. Standard-errors are clustered at the firm level.\footnote{Descriptive statistics are provided in Table \ref{tbl:discern_firm_descstats}.}

Table \ref{tbl:firmlevel} presents the estimation results. Using Poisson pseudo-maximum likelihood estimation (PPML), I find a positive and statistically significant relationship between stocks of follow-on research and subsequent scientific publications, employment of AMWS-listed scientists, award-winning scientists, and annual patenting (columns 1, 3, 5, and 7). I then add stocks of firms' and others' patents that cite the follow-on research. The results reveal a stronger relationship between the stock of used follow-on research and subsequent innovative outcomes. The negative coefficients on external patenting suggest that greater external appropriation of the follow-on research is associated with lower subsequent innovation by the focal firm (columns 2, 4, 6, and 8). While these results cannot be interpreted causally, they indicate that follow-on research, particularly the portion cited by firms' patents, correlates with firms' subsequent scientific investments and invention outcomes.
 
 %\begin{landscape}
\begin{table}[htbp]
\footnotesize \singlespacing
   \centering
   \begin{threeparttable}[b]
      \caption{\label{tbl:firmlevel}Follow-On Research and Innovation Outcomes at the Firm-Year Level}
      \bigskip

\begin{tabular}{l*{8}{c}}
\toprule
                    &\multicolumn{2}{c}{ }           &\multicolumn{4}{c}{Employed AMWS Scientists}            &\multicolumn{2}{c}{}                \\
                    %\cmidrule(lr){2-3}\cmidrule(lr){4-7}\cmidrule(lr){8-9}
                    \cmidrule(lr){4-7}
                    &\multicolumn{2}{c}{Annual Publications}           &\multicolumn{2}{c}{All}   &\multicolumn{2}{c}{Award-Winning}          &\multicolumn{2}{c}{Annual Patents}                \\
                    \cmidrule(lr){2-3}\cmidrule(lr){4-5}\cmidrule(lr){6-7}\cmidrule(lr){8-9}
                    &\multicolumn{1}{c}{PPML}&\multicolumn{1}{c}{PPML}&\multicolumn{1}{c}{PPML}&\multicolumn{1}{c}{PPML}&\multicolumn{1}{c}{PPML}&\multicolumn{1}{c}{PPML}&\multicolumn{1}{c}{PPML}&\multicolumn{1}{c}{PPML}\\
&(1)&(2)&(3)&(4)&(5)&(6)&(7)&(8)\\ \midrule
ln(FO Stock)$_{t-1}$&       0.220\sym{***}&       0.230\sym{***}&       0.076\sym{***}&       0.081\sym{***}&       0.078\sym{**} &       0.095\sym{**} &       0.082\sym{***}&       0.039         \\
                    &     (0.044)         &     (0.043)         &     (0.026)         &     (0.027)         &     (0.039)         &     (0.039)         &     (0.032)         &     (0.035)         \\
\addlinespace
ln(Firm's pat.&                     &       0.101\sym{**} &                     &       0.022         &                     &       0.023         &                     &       0.140\sym{***}\\
  stock citing FO)$_{t-1}$       &                     &     (0.051)         &                     &     (0.035)         &                     &     (0.043)         &                     &     (0.030)         \\
\addlinespace
ln(Ext. pat. &                     &      -0.130\sym{***}&                     &      -0.030         &                     &      -0.042         &                     &      -0.068\sym{***}\\
   stock citing FO)$_{t-1}$                 &                     &     (0.037)         &                     &     (0.026)         &                     &     (0.033)         &                     &     (0.026)         \\
\\
Time-Varying Controls&         Yes         &         Yes         &         Yes         &         Yes         &         Yes         &         Yes         &         Yes         &         Yes         \\
Firm FE             &         Yes         &         Yes         &         Yes         &         Yes         &         Yes         &         Yes         &         Yes         &         Yes         \\
Year FE             &         Yes         &         Yes         &         Yes         &         Yes         &         Yes         &         Yes         &         Yes         &         Yes         \\
Observations        &      35,156         &      35,156         &      26,306         &      26,306         &      12,386         &      12,386         &      50,176         &      50,176         \\
Avg. DV             &      11.618         &      11.618         &       9.119         &       9.119         &       2.655         &       2.655         &      24.899         &      24.899         \\
Psuedo R$^2$        &       0.909         &       0.911         &       0.899         &       0.899         &       0.778         &       0.779         &       0.899         &       0.902         \\
\bottomrule
\end{tabular}

      \begin{tablenotes}[flushleft]
        \scriptsize
        \item \textit{Notes:} This table presents estimation results for the relationship between follow-on research, subsequent scientific investments, and patenting outcomes by the originating firms. 
         The data consists of a firm-year panel of U.S.-based publicly-owned firms between 1980 and 2015 \citep{AroraBelenzonSheer2020DISCERNDukeInnovationa}. In columns 1-2, the dependent variable is a count of scientific publications authored by the focal firm in year $t$. 
         In columns 3-4, the dependent variable is a count of AMWS scientists employed by the firm at year $t$. In columns 5-6, the count includes only award-winning scientists among AMWS. In columns 7-8, the dependent variable is a count of patents filed by the focal firm at year $t$. Follow-on is the stock of external papers that cite the firms' scientific publications, aggregated by the year of publication. 
         %Used follow-on is the stock of external papers that is cited by the firms' own patents, while unused follow-on is the remainder.
         Firm's patents citing follow-on are stocks of patents by the focal firm that cite the follow-on research. External patents citing follow-on are stocks of patents unrelated to the focal firm that cite the follow-on research.
         All stocks (other than scientist employment) are depreciated using an annual 15\% depreciation constant. 
         Time-varying control variables include lagged firm assets and R\&D stocks. 
         Indicator variables for zero counts are included. 
         Observations are automatically dropped from the complete sample due to either singletons or separation by fixed effects.
         \item Clustered (Firm) standard-errors in parentheses. Signif. Codes: ***: 0.01, **: 0.05, *: 0.1
      \end{tablenotes}
   \end{threeparttable}
\end{table}

\section{\label{sec:discussion} Discussion and Conclusions}

Firms' incentives to participate in public research have received increased interest in recent years \citep{ AroraBelenzonSheer2021KnowledgeSpilloversCorporate, RotoloCameraniGrassanoEtAl2022WhyFirmsPublish}. Early works have argued that investments in science benefit firms' R\&D processes by improving their combinative capabilities \citep{KogutZander1992KnowledgeFirmCombinative, FlemingSorenson2004ScienceMapTechnological} and absorptive capacity \citep{CohenLevinthal1990AbsorptiveCapacityNew, Rosenberg1990WhyFirmsBasic}. The literature has also suggested that firms engage with the scientific community to benefit from resources external to the firm \citep{CockburnHenderson1998AbsorptiveCapacityCoauthoring}. Nonetheless, the disclosure of findings through scientific publications has been primarily analyzed through a labor market lens, as a concession firms provide to scientist employees who value academic engagement \citep{Stern2004ScientistsPayBe}, and as a cost associated with potential knowledge spillovers to rivals. Exceptions are arguments that suggest that corporate science can influence the direction of academic research outside the firm \citep{Hicks1995PublishedPapersTacit}. Such influence became increasingly important as the innovation ecosystem experienced a division of innovative labor \citep{AroraGambardella1994ChangingTechnologyTechnological} and an increased reliance on public science \citep{fleming2019government}.

This study demonstrates that firms benefit from external follow-on research that builds upon their prior scientific publications. While only 7\% of firms' publications are directly cited in their patents, an additional 33\% are cited by external follow-on research eventually referenced in the firm's patents. Using a novel instrumental variable exploiting plausibly exogenous variation in scholarly attention to corporate publications, I find positive effects of external follow-on research on firms' subsequent scientific publishing, researcher employee retention, patenting outcomes, and engagement with academics through collaborations and hiring. These effects are moderated by firms' resources and capabilities, and the nature of the scientific area. Follow-on research serves as both valuable input for subsequent inventions and quality validation of internal science. 

Prior work has emphasized that publication strengthens scientists' outside options and exposes firms to imitation risks, implying that disclosure is primarily a cost to be managed \citep{Stern2004ScientistsPayBe, GansMurrayStern2017ContractingDisclosureScientific}. My findings suggest a complementary perspective: disclosure generates countervailing benefits through the scientific community's response. In the current setting, follow-on research increased scientist retention on average, suggesting that the shared value created through external knowledge accumulation can offset the enhanced mobility that publication affords. And rather than being unrelated to subsequent innovation, as reputation or IP-based motivations would imply, follow-on research drives publishing and patenting. These patterns point to the external scientific community as a source of returns that should factor into firms' disclosure decisions.

I contribute to the literature on corporate science and firms' disclosure strategies \citep{AroraBelenzonSheer2021KnowledgeSpilloversCorporate, SimethCincera2016CorporateScienceInnovation, GansMurrayStern2017ContractingDisclosureScientific}. By showing that publication can mobilize external research in ways that benefit the originating firm, this paper highlights returns to disclosure beyond scientist satisfaction. More broadly, while prior work has emphasized how firms benefit from access to already-existing external knowledge \citep{CohenLevinthal1990AbsorptiveCapacityNew, FlemingSorenson2004ScienceMapTechnological}, this study highlights firms' ability to influence future academic output in privately beneficial ways. The findings also relate to the literatures on open innovation and knowledge spillovers \citep{Chesbrough2003OpenInnovationNew, Henkel2006SelectiveRevealingOpen}. This paper supports the view that such engagement can drive down R\&D costs, enhance the value of complementary assets, and mitigate uncertainty associated with investments in science.

Several limitations merit acknowledgment. The analysis conditions on the existence of a corporate publication, not exploring the decision to publish or comparing outcomes to non-publication scenarios. As a result, influence on academics' work is not directly observed; rather, the findings show that if such influence exists, firms are likely to benefit. Future work could focus on how corporate publications affect academics and identify their incentives to build upon such work. Additionally, interaction coefficients in the heterogeneity analysis reveal patterns but cannot be interpreted as causal effects. Lastly, identification is at the publication level. While firm-level correlational evidence is provided, a different empirical setup would be required to demonstrate causal effects on firms' financial and overall innovative performance.

Notwithstanding these limitations, the immediate implications of this study are that corporate R\&D managers should consider the potential benefits that could originate from the scientific community when making decisions regarding investments in science and disclosure of findings. Managers should balance disclosure decisions with both imitation risk and academic complementors in mind, potentially concentrating disclosure in areas where both the employee and the firm benefit. These decisions should align with other related firm choices such as geographic proximity to universities \citep{Sohn2021HowLocalIndustry}, the funding of academic research \citep{BabinaHeHowellEtAl2023CuttingInnovationEngine} and direct interactions \citep{BaruffaldiPoege2022StarsHowFirms}. However, it is also important to note that benefits might take years to mature, and that they might not offset completely the risks of knowledge spillovers to rivals. Under the right conditions, influence on the scientific community can result in valuable inputs that reduce the firms' R\&D costs and improve innovation outcomes. 

The evidence also suggests that the presence of strong academic institutions can drive firms to participate in public research. Without sufficient incentives, firms might refrain from investments in science or choose secrecy over openness. Academics willing to engage with corporate researchers are one of the drivers of firms' contributions to the open scientific discussion. Whether firms' influence on the direction of scientific inquiry is socially and scientifically desirable remains a subject for future work.

\clearpage

% BIBLIOGRAPHY
{
  \hypersetup{urlcolor=black}
\clearpage
\singlespacing
\small
\normalem
\printbibliography
}

\clearpage

% APPENDIX
\begin{appendices}
\normalsize

\startcontents[appendices]
{\small\onehalfspacing\printcontents[appendices]{}{1}{}}

\clearpage

\begin{refsection}
\section{Case Studies\label{sec:studycases}}

\small
\onehalfspacing

\subsection{Overview}

This section provides examples of follow-on research and its value for the originating firms. These examples were obtained by closely following scientific and patent citations within the data and combining them with complementary details from online web searches and correspondence with the authors.

In the first example, follow-on research by academics was used as an input into subsequent innovation by the originating firm. The firm engaged with the scientific community by regularly publishing scientific findings. Consequentially, external findings that cited the firms' publications were further developed by the firm and incorporated in its patents. The example describes an incremental advancement to organic light-emitting diode (OLED) technology that was developed at the Hong Kong University of Science and Technology and subsequently implemented in a patent by Universal Display Company (UDC).

In the second example, follow-on research by academics lead to a collaboration between them and the originating firm. The example describes the development of Morpholino Oligomers, a type of molecular structure that can bind to genetic material. These structures were initially developed by AVI BioPharma (subsequently Sarepta Therapeutics), a private biotech company based in Cambridge, Massachusetts. Following this development, a scientific research group at the University of Western Australia found a therapeutic opportunity for Morpholinos in treating Duchenne, a type of muscular dystrophy. The findings were eventually licensed, further developed, and commercialized by Sarepta.

The third example is a case from software development. Here, the focal firm described an application that served as a motivation for academics to develop relevant upstream algorithms. Such external developments then served as a baseline in one of the firm's own patents to test the performance of their technology and compare it to technologies that were available to others. The example describes the development of algorithms for video object segmentation (breaking up video footage into objects) by Adobe Inc. Later, a patent by the firm used several external algorithms to establish the case for better performance of their own technology.

The fourth example is a case from nanotechnology. In this example, the focal firm made a scientific discovery and intended to use it in specific applications. Academics at other institutions produced follow-on findings that were applicable to a completely different set of products. As a result, various assignees filed patents that cited the follow-on findings. However, while the original discovery was patented by the focal firm, it seems not to have used the follow-on findings in subsequent innovation. Possibly, the firm lacked interest in other product lines that were unrelated to its core business, or lacked the complementarities needed to benefit from them. The example describes the development of a technology that allows the tuning of liquid on nanostructured surfaces by a research group at Bell Labs.

In the fifth example, a firm's publications were hardly cited by external researchers, and the follow-on research that did emerge was not mentioned by the originating firm's patents. Possibly, in this case the firm's incentives to engage with the scientific community were not related to the potential usefulness of follow-on research.

\subsection{\label{sec:udc}Example from Electronics: Silver Film Improves Color Saturation in OLEDs}

Universal Display Corporation was founded in 1994 by Sherwin Seligsohn with the goal of developing displays that are based on organic light-emitting diodes (OLED).\footnote{About UDC. \textit{Universal Display Corporation}. Retrieved July 14, 2022, from \href{https://oled.com/about/}{https://oled.com/about/}} Benefiting from long-term research contracts with Princeton University, the company became a leader in OLED technology. Today, almost all OLED products incorporate proprietary technologies owned by UDC. Throughout the years, UDC focused on advancing OLED technologies through investments in scientific research and collaborations with other organizations.

Researchers at UDC frequently publish the firm's scientific findings. The example described here involves four of the firm's papers, published between the years 2000 and 2002: \citet{BurrowsForrestZhouEtAl2000OperatingLifetimePhosphorescent} found that long lifetimes are an intrinsic property of phosphorescent OLEDs; %\citet{LuSturm2002OptimizationExternalCoupling} study three different modes of light emitted from OLEDs. Understanding of these modes can improve the total external light emission from different applications; 
\citet{ChwangKwongBrown2002GradedMixedlayerOrganic} studied the performance of graded mixed-layer OLEDs, a design that the authors suggested could extend the device's lifetime and make them applicable to flat panel displays; \citet{LuWeaverZhouEtAl2002HighefficiencyTopemittingOrganic} studied top-emitting OLEDs, a design that was proved to be 20\% more efficient than equivalent bottom-emitting OLEDs; Lastly, \citet{BurrowsGuForrestEtAl2000SemitransparentCathodesOrganic} studied semitransparent cathodes in OLEDs that can be applied to various use-cases.

Common to all publications above, is that they were cited directly or indirectly by \citet{PengZhuSunEtAl2005EfficientOrganicLightemitting}, which was authored by researchers at the Hong Kong University of Science and Technology. The citations suggest that the researchers built on UDC's prior findings in their work. In addition, \citet{PengZhuSunEtAl2005EfficientOrganicLightemitting} and three of the publications above were published in Applied Physics Letters (the other two were published in the Journal of Applied Physics). This relation further supports the notion that researchers at UDC and at the  Hong Kong University of Science and Technology were part of the same scientific community.

In their paper, \citet{PengZhuSunEtAl2005EfficientOrganicLightemitting} studied several metals as alternatives to the use of indium-tin oxide (ITO), the material that was typically used before as the anode material. They found that silver can serve as an effective alternative to ITO. In their experiments, silver anode resulted in improved current voltage and optical performance. They suggested, theoretically, that the use of semitransparent silver anodes can enhance light extraction efficiency. However, the authors acknowledged that further investigation is necessary.

In 2007, two years after the publication of \citet{PengZhuSunEtAl2005EfficientOrganicLightemitting}, a researcher at UDC filed for a patent that built on their proposal to use semitransparent silver in OLEDs.\footnote{D'Andrade, B. (2013). Saturated color organic light emitting devices (United States Patent No. US8476822B2).} The patent suggested to use silver (or other relevant metals) as a color saturation enhancement layer between the two electrodes of the OLED device:

\blockquote{\textit{It is believed that certain metals, such as aluminum and chromium, are generally thought of as undesirable for placement in organic light emitting devices between the anode and the organic layers. See, Peng et al., Efficient organic light emitting diode using semitransparent silver as anode, Applied Physics Letters 87, 173505, p. 1 (2005), teaching that high work function metals are desirable to lower barriers for hole injection; conversely, low work function metals are not desirable. Surprisingly, it has been found that these metals may be used as thin layers between the anode and the organic layers of an organic light emitting device as color Saturation enhancement layers... Silver is also a preferred material for use as a color Saturation enhancement layer.}} 

To summarize, in this example, an academic research group was influenced by the scientific publications originating from UDC. Then, their work \citep{PengZhuSunEtAl2005EfficientOrganicLightemitting} served as an input into subsequent innovation by the firm.\footnote{The details in this example were verified with the corresponding author of \citet{PengZhuSunEtAl2005EfficientOrganicLightemitting}.}

\subsection{\label{sec:sarepta}Example from Bio-pharmaceuticals: Morpholino Oligomers and the Treatment of Duchenne Disease}

During the 1970s several research groups started working on antisense therapeutics strategies for binding to genetic material \citep{Summerton2016HistoryPropertiesMorpholino}. These strategies, it was suggested, could offer treatment to a wide range of conditions, including viral diseases, cancers and genetic defects. The first patent in this area was filed in 1977 by Summerton and Bartlett.\footnote{Summerton, J. E., \& Bartlett, P. A. (1978). Nucleic acid crosslinking agent and affinity inactivation of nucleic acids therewith (United States Patent No. US4123610A).} In 1980, Summerton founded Antivirals Incorporated (later AVI BioPharma and subsequently Sarepta Therapeutics), the first company that focused on developing and commercializing these treatment methods. In 1985, with advice from Dwight Weller, Summerton developed a molecule structure that radically departed from previous designs. This design, named \textit{morpholino oligomers}, used far cheaper materials and was easier to produce in comparison to previously existing designs. Between 1985 and the mid 1990s, AVI BioPharma and other research groups further developed and enhanced the morpholino molecular structure. In studies on cultured human cells, morpholinos outperformed the competition in both efficacy and specificity. 

Once the promise of morpholino oligomers has been sufficiently established, Summerton and Weller published a review article that covered details on the design, preparation and properties of these structures \citep{SummertonWeller1997MorpholinoAntisenseOligomers}. They also started exploring therapeutic applications, such as increasing hemoglobin production in blood cells of  thalassemic patients \citep{LacerraSierakowskaCarestiaEtAl2000RestorationHemoglobinSynthesis}.\footnote{Thalassemia is an inherited blood disorder that causes hemoglobin deficiencies.}

In the early 2000s, The development of morpholino oligmers was picked up by a research group at the Centre for Neuromuscular and Neurological Disorders in the University of Western Australia, who were working on developing an antisense-based therapy to Duchenne muscular dystrophy (DMD). The group noted:

\blockquote{\textit{
One chemistry that is gaining wide recognition for use in antisense applications is the morpholino oligonucleotide developed by Summerton and Weller. These authors developed the morpholino structural type with the intention that this chemistry could provide several advantages in the clinical application(s) of antisense therapeutics, such as strong nucleic acid binding, resistance to nucleases, minimal nonantisense effects, high aqueous solubility and relatively low synthesis costs \citep{GebskiMannFletcherEtAl2003MorpholinoAntisenseOligonucleotide}.
}}

However, an important challenge that faced the researchers was the delivery of the morpholinos into the cell nucleus. To overcome this challenge, the researchers at the University of Western Australia suggested to anneal the morpholinos with additional DNA/RNA molecules, or `leashes'. Along with filing patents to protect their inventions, they described their findings in scientific publications:\footnote{Fletcher, S., McClorey, G., \& Wilton, S. (2004). \textit{Antisense Oligonucleotides for Inducing Exon Skipping and Methods of Use Thereof} (AU2004903474A0).}

\blockquote{\textit{
The uncharged backbone compromises delivery, for non-ionic AOs cannot easily be delivered into cultured cells using delivery agents such as cationic liposomes. To circumvent this difficulty, we investigated the use of single stranded (anionic) nucleic acid ‘leashes’ which were annealed to the morpholino AO, allowing the AO : leash duplex to be complexed with Lipofectin. \citep{GebskiMannFletcherEtAl2003MorpholinoAntisenseOligonucleotide}.
}}

Following the findings of \citet{GebskiMannFletcherEtAl2003MorpholinoAntisenseOligonucleotide}, Sarepta further developed methods for treating Duchenne using morpholino oligomers and studied additional applications to other related diseases. In 2011, Sarepta filed a patent for the treatment of myotonic dystrophy, another type of muscular dystrophy along with Duchenne:

\blockquote{\textit{
The parent application disclosed and claimed the use of these two CPPs\footnote{Cell penetrating peptides} for targeting antisense oligonucleotides to muscle tissue, in treating certain muscle pathologies. For example, in treating Duchenne muscular dystrophy (DMD) \ldots The present invention applies this strategy additionally to the treatment of myotonic dystrophy MD1 and MD2 in muscle tissue, including skeletal and heart muscle tissue.\footnote{Moulton, H. M., \& Kole, R. (2014). \textit{Compound and method for treating myotonic dystrophy} (United States Patent No. US8741863B2). }
}}

In this patent, the inventors acknowledge the complementarities between the focal invention and the prior scientific findings by the research group at the University of Western Australia:

\blockquote{\textit{
The oligonucleotide-(RXRR(B/X)R)$_2$XB conjugate compounds of the invention may be used in conjunction with homing peptides selective for the target tissue, to further enhance muscle-specific delivery. An example of this approach can be found in the application of muscle-binding peptides (Samoylova and Smith, 1999; Vodyanoy et al., U.S. Appn. Pubn. No. 2003064.0466) coupled to antisense oligomers designed to be therapeutic treatments for Duchenne muscular dystrophy (DMD) (Gebski, Mann et al. 2003; Alter, Lou et al. 2006) (PCT Pubn. No. WO2006000057).
}}

In 2013, the complementarities between the inventions at Sarepta and the Duchenne treatments developed at the University of Western Australia eventually lead to an exclusive licensing agreement for the commercialization of the treatment:\footnote{Sarepta Therapeutics and University of Western Australia Announce Exclusive Worldwide Licensing Agreement for Exon-Skipping Program in Duchenne Muscular Dystrophy. Retrieved July 14, 2022, from \href{https://investorrelations.sarepta.com/news-releases/news-release-details/sarepta-therapeutics-and-university-western-australia-announce}{https://investorrelations.sarepta.com/news-releases/news-release-details/sarepta-therapeutics-and-university-western-australia-announce}
}

\blockquote{\textit{
Sarepta has an exclusive, worldwide licensing agreement with the University of Western Australia (UWA) for intellectual property rights to support the development of exon-skipping drug candidates for the treatment of Duchenne muscular dystrophy (DMD). The agreement grants Sarepta rights to UWA's extensive patent portfolio in DMD and enables the Company to expand its exon-skipping pipeline with new candidates to address the majority of patients with DMD worldwide. \footnote{Sarepta Strategic Partnerships. Retrieved July 14, 2022, from \href{https://www.sarepta.com/science/strategic-partners}{https://www.sarepta.com/science/strategic-partners}}
}}

At last, in 2016, the U.S. Food and Drug Administration (FDA) approved Sarepta's product as the first drug for treating patients with Duchenne:

\blockquote{\textit{
The U.S. Food and Drug Administration today approved Exondys 51 (eteplirsen) injection, the first drug approved to treat patients with Duchenne muscular dystrophy (DMD). Exondys 51 is specifically indicated for patients who have a confirmed mutation of the dystrophin gene amenable to exon 51 skipping, which affects about 13 percent of the population with DMD.}\footnote{FDA grants accelerated approval to first drug for Duchenne muscular dystrophy. FDA; Retrieved July 14, 2022, from \href{https://www.fda.gov/news-events/press-announcements/fda-grants-accelerated-approval-first-drug-duchenne-muscular-dystrophy}{https://www.fda.gov/news-events/press-announcements/fda-grants-accelerated-approval-first-drug-duchenne-muscular-dystrophy}}

}

Today, Duchenne is one of the core disease areas in Sarepta's portfolio and ongoing research is conducted to further develop treatments for additional Duchenne subtypes as well as other related diseases. In 2021, Sarepta's total revenue from their portfolio of treatments and collaborations surpassed \$700 million.\footnote{Sarepta Therapeutics Announces Fourth Quarter and Full-Year 2021 Financial Results and Recent Corporate Developments. Retrieved July 14, 2022, from \href{https://investorrelations.sarepta.com/news-releases/news-release-details/sarepta-therapeutics-announces-fourth-quarter-and-full-year-2021}{https://investorrelations.sarepta.com/news-releases/news-release-details/sarepta-therapeutics-announces-fourth-quarter-and-full-year-2021}
}

\subsection{\label{sec:adobe}Example from Software Development: Image Segmentation Algorithms}

Adobe Inc. is a computer software company that was founded in 1982 in California. It is a leader in specialized software for a wide range of creative content creation. up until the 2010s, the company focused on developing tools for professional artists, designers and video editors. The introduction of cameras on mobile phones, advancements in computer processing power and the rise of online sharing platforms (e.g. Youtube) enabled the general public to take part in content creation, and demand for simple and intuitive editing tools increased. \citet{GoldmanGontermanCurlessEtAl2008VideoObjectAnnotation} is a scientific publication by Adobe developers that suggested that Adobe was interested at the time in developing tools for video annotation and composition that are more user-friendly. However, the development of these tools required a complex preprocessing step that is known as video object segmentation. This process allows the software to automatically identify objects in the video and separate them from other objects and the background:

\blockquote{\textit{
In this paper we propose a framework using (2D) video object motion to enable novel approaches to user interaction\ldots In particular, we propose novel interfaces for three tasks that are under-served by present-day video interfaces: annotation, navigation, and image composition.}

\ldots 

\textit{To achieve these interactions, our system first analyzes the video in a fully automatic preprocessing step that tracks the motion of image points across the video and segments those tracks into coherently moving groups.
}}

\citet{BrendelTodorovic2009VideoObjectSegmentation}, a publication by researchers at Oregon State University, cited \citet{GoldmanGontermanCurlessEtAl2008VideoObjectAnnotation} as a motivation for developing better video object segmentation methods. The authors claimed that the method used by Adobe suffered from several drawbacks and offered an improved solution:

\blockquote{\textit{
This paper presents an approach to unsupervised video object segmentation (VOS). Our goal is to delineate the boundaries of all moving and static objects occurring in an arbitrary video\ldots VOS is a prerequisite step of a wide range of higher level vision algorithms, including activity recognition video summarization and retrieval, and nonphotorealistic video rendering.}

\ldots

\textit{Currently, the two predominant approaches to VOS are tracking interest points, and perceptual grouping of pixels from all frames. There is a number of unsatisfying aspects about both of them. Point-based approaches group the trajectories of keypoints with similar motions. However, tracking points yields only a confidence map of the objects' vicinity – not segmentation.}

\ldots 

\textit{In this paper, we adopt an alternative, hybrid formulation. 
}}

Next, \citet{GrundmannKwatraHanEtAl2010EfficientHierarchicalGraphbased}, a collaboration of researchers at Georgia Institute of Technology and Google Research, identified three main challenges in video object segmentation: temporal coherence, automatic processing and scalability. They adopted a method of image segmentation and generalized it for video processing. Importantly, they claimed that their method outperforms previous methods, including the one developed by  \citet{BrendelTodorovic2009VideoObjectSegmentation}:

\blockquote{\textit{
Tracking-based video segmentation methods generally define segments at frame-level and use motion, color and spatial cues to force temporal coherence. Following the same line of work, \citet{BrendelTodorovic2009VideoObjectSegmentation} used contour cues to allow splitting and merging of segments to boost the tracking performance.}

\ldots

\textit{Our novel video segmentation algorithm addresses all of the above challenges. We build a 3-D graph from the video volume and generalize Felzenszwalb and Huttenlocher’s graph-based image segmentation to obtain an initial oversegmentation of the video volume into relatively small space-time regions.
}}

\citet{JoulinBachPonce2012MulticlassCosegmentation} offered methods for cosegmentation, an additional development related to the methods discussed above, in which availability of multiple images (such as in a video sequence) offers a type of supervision that can improve the segmentation process. They found that their methods performed well with the dataset provided by \citet{GrundmannKwatraHanEtAl2010EfficientHierarchicalGraphbased}:

\blockquote{\textit{
The aim of cosegmentation methods is to simultaneously divide a set of images assumed to contain instances of K different object classes into regions corresponding to these classes. Note that in this context, an ``object'' may refer to what is usually called a ``thing'' (a car, a cow, etc.) but might also be a texture (grass, rocks), or other ``stuff'' (a building, a forest)\ldots The proposed approach has been implemented and tested on several datasets including video sequences.}

\ldots

\textit{Our experiments with iCoseg suggest that our method is particularly well suited to keyframes from the same video shot, since these are likely to feature the same objects under similar illumination. This is confirmed with our experiments with two short video clips taken from the Hollywood-2 and Grundmann datasets.
}}

The last part of this example is \citet*{CohenPriceAhmed2015SemanticObjectProposal}, 
a patent filed in 2013 by researchers at Adobe Inc. The patent claimed priority on a method in which a user provides an input (such as a ``dog'') and the system automatically segments the corresponding object from within the video:

\blockquote{\textit{
Techniques are disclosed herein that enable digital images to be segmented based on a user's semantic input. In other words, given an input image of a person walking a dog adjacent to a tree, a user can simply provide the semantic input ``dog'' and the system will segment the dog from the other elements in the image. If the user provides other semantic input, such as ``person'' or ``tree'', the system will segment the person or the tree, respectively, from the same image.
}}

The inventors used prior developments in object segmentation, including the methods developed by \citet{JoulinBachPonce2012MulticlassCosegmentation}, as benchmarks for their own approach:

\blockquote{\textit{
The Jaccard similarity coefficient $J_s$ corresponding to segmentation using an example embodiment disclosed herein was compared with a corresponding coefficient resulting from segmentation using four different cosegmentation techniques\ldots  The compared cosegmentation algorithms are described in: Joulin et al., ``Multi-Class Cosegmentation''. Proceedings of IEEE ComputerVision and Pattern Recognition (CVPR 2012), pp. 542-549 (2012) (``Joulin-1'');}

\ldots

\textit{The results of the foregoing comparison are listed in Table A. In particular, Table A illustrates that the tested example embodiment provides a segmentation that is significantly more accurate than the compared cosegmentation techniques in most applications. 
}}

To summarize, in 2008 Adobe developed tools to meet the demand for user-friendly and intuitive video editing. These tools required to solve a complex problem of video segmentation, and the methods available then suffered from various issues. Research groups outside of Adobe were aware of the demand for better methods and developed various solutions to these problems. Eventually, Adobe patented additional techniques and used external solutions as a baseline for comparison.

\subsection{Example from Nanotechnology: Dynamic Tuning of Liquids on Nanostructured Surfaces}

In 1996, the world-renowned Bell Laboratories were split from AT\&T and were placed under a new company named Lucent Technologies. Under Lucent, researchers at Bell Labs continued to conduct scientific research in a wide array of areas. In 2004, a research group led by Prof. Krupenkin at Bell Labs discovered a electrical method to dynamically control the behaviour of liquids on nanostructured surfaces. The original publication by \citet{KrupenkinTaylorSchneiderEtAl2004RollingBallComplete} was accompanied by a paired patent \citep{KornblitKrupenkinMandichEtAl2016ApparatusControllingMovement}. In this study, the research group noted the wide range of potential applications of these findings:

\blockquote{\textit{
In this work, we propose a new approach that allows us to achieve effective electrowetting on nanostructured superhydrophobic surfaces\ldots The ability to dynamically change the interaction between the liquid and the nanostructured substrate potentially opens a wide range of exciting new applications. The particular areas of interest include microfluidics, lab-on-a-chip devices, chemical microreactors, thermal management of microelectronics, drag reduction systems, and optical communications, as well as many others.
}}

The work on tunable nanostructured surfaces won Prof. Krupenkin the American Chemical Society Industrial Innovation Award in 2007.\footnote{Krupenkin Group. Retrieved on August 11, 2022 from http://www.krupenkin.com/people/people.aspx} While the invention could be potentially applied to multiple products, it seems that at that time Lucent envisioned its use in developing technologies to enhance power cell batteries. In a press release, Bell Labs announced a partnership with mPhase Technologies to use the discovery for battery development:

\blockquote{\textit{
Bell Labs scientists and engineers recently made a significant breakthrough in microfluidics that enables dynamic control of surfaces when interacting with a liquid - a key enabler for making ``Smart Batteries'' a reality. Fine control of liquids at the micro and macro scale will allow scientists to create batteries that can be activated upon demand.\footnote{MPhase and Bell Labs to Develop Nanotech Power Cell Batteries - New Technology. AZoNano.com, 22 Mar. 2004. Retrieved July 14, 2022, from  https://www.azonano.com/article.aspx?ArticleID=654. 
}
}}

The same research team at Lucent also filed for patents that use the original discovery in new battery technologies
\citep{HodesKolodnerKrupenkinEtAl2010ReserveCellarrayNanostructured}:

\blockquote{\textit{
A battery having an electrode with at least one nanostructured Surface is disclosed wherein the nanostructured Surface is divided into cells and is disposed in a way Such that an electrolyte fluid of the battery is prevented from contacting the portion of electrode associated with each cell. When a Voltage is passed over the nanostructured Surface associated with a particular cell, the electrolyte fluid is caused to penetrate the nanostructured surface of that cell and to contact the electrode, thus activating the portion of the battery associated
with that cell.
}}

Meanwhile, the original paper by \citet{KrupenkinTaylorSchneiderEtAl2004RollingBallComplete} was cited externally 274 times since its publication. These papers were later cited 14,077 times. In the third generation of citations there are over 180 thousand citations. While these publications were cited by over 2,400 unique patents, none of these patents are assigned to Bell Labs or Lucent. This is an indication that follow-on research was not used by Bell Labs in related subsequent innovation. 

It is hard to know the exact reason why Bell Labs did not use the follow-on research originating from their own discovery. However, a close examination of the topics in follow-on publications could reveal a possible explanation -- the subsequent technologies were unrelated to their lines of business. Numerous studies that cite \citet{KrupenkinTaylorSchneiderEtAl2004RollingBallComplete} focused on developing techniques to enhance ``lab-on-a-chip'' applications. For example, The Wheeler Microfluidics Laboratory, a research group at Toronto University, published several follow-on papers to  \citet{KrupenkinTaylorSchneiderEtAl2004RollingBallComplete}. In one such paper, \citet{LukMoWheeler2008PluronicAdditivesSolution} suggested that Pluronics additives can solve stickiness issues in digital microfluidics. The publication was accompanied by several patents assigned to the same research group (e.g. \citet{WheelerJebrail2011DigitalMicrofluidicMethod}). These findings and inventions might not have been of interest for researchers and companies that were focused on battery-enhancing technologies.

This example suggests that firms with a wide range of product lines, adaptable business models and advanced commercialization capabilities could more easily benefit from the breadth of applications that external follow-on research could provide. Firms that are focused on specific products might benefit from licensing their upstream inventions in a market for technology, but might be limited in incorporating external research in their own subsequent inventions.

\subsection{Example from Radio Engineering: Dielectric Loaded Hybrid Mode Horn Antennas}

Lockheed Martin is an American corporation that is a  leader in developing aerospace and defense technologies. Among firms with the largest number of scientific publications, Lockheed has a very low percentage of publications that are cited by external follow-on publications and then used in the firm's own patents (only 15\%, compared with 41\% for Amgen Inc. and 31\% for Hewlett-Packard). A close look at the scientific publications by Lockheed could provide an explanation. 

For example, \citet{LierKishk2005NewClassDielectricloaded} is a collaborative publication by a researcher at Lockheed and a researcher at the University of Mississippi. It describes a new model for a very specific type of antenna. This antenna could be used on planes, satellites and reflector antennas:

\blockquote{\textit{
A new class of hybrid mode horn antennas, which can be designed for a specific gain or sidelobe requirement and low cross-polarization, has been presented\ldots  It could be particularly useful in millimeter wave applications since the design is compliant with small size manufacturing. Finally, the flat top pattern design makes it a candidate earth coverage horn on-board satellites and a candidate feed for reflector antennas with enhanced directivity.
}}

This publication was accompanied by several patents by the same author-inventor \citep{Lier2013LowIndexMetamaterial, Lier2009ArtificialDielectricAntenna, LierKatz2009HornAntennaWaveguide, LierKatz2011HornAntennaWaveguide, Lier2012AntennaArrayMetamaterial}. The patents provided protection to a wide set of related inventions: a low index metamaterial, artificial dielectric antenna elements, horn antennas and antenna arrays. 

While internally useful, the original publication received only 16 external scientific citations (171 across three generations). None of these publications were mentioned in any of Lockheed Martin's subsequent patents and only a handful were cited by Lockheed in subsequent scientific publications. An explanation for this is that possibly this research area is dominated by corporate researchers. Few academics at public research institutions have the relevant knowledge and interest in developing these findings outside the original firm. Therefore, it is likely that Lockheed published the focal publication without expecting to engage with useful external follow-on findings.

\afterpage{\clearpage}
\printbibliography
\end{refsection}

\clearpage

\begin{refsection}
\section{\label{sec:data_appendix}Data Construction} 

\setcounter{table}{0}
\renewcommand{\thetable}{B\arabic{table}}
\setcounter{figure}{0}
\renewcommand{\thefigure}{B\arabic{figure}}
\small
\onehalfspacing

This section provides a comprehensive overview of the data sources, sample construction procedures, variable definitions utilized in the research, and descriptive statistics. Table \ref{tbl:vars} provides a summary of variable definitions used in the analysis.

\subsection{Data Sources}

The primary data source used in this research is the Duke Innovation \& SCientific Enterprises Research Network database (DISCERN, \citet{AroraBelenzonSheer2020DISCERNDukeInnovationa}). DISCERN provides a match between patents and scientific publications to US-based publicly-traded firms between the years 1980 and 2015. For the construction of DISCERN, firm data were obtained from Compustat 2018 (firm-level data), ORBIS (subsidiaries), SDC Platinum (M\&A activity), and WRDS CRSP (name changes). These data were matched to patent data from PatStat and scientific publication data from Web of Science (WoS). DISCERN is unique because it provides a match between patents and scientific publications to firms while considering firm structure and ownership changes. About a third of the firms in the sample changed their names within the sample years. Accounting for these changes improves the matching accuracy and provides a comprehensive baseline for studying firms' scientific and innovative activities.

Since creating DISCERN, several new data sources for scientific publications have become available. These sources provide key variables for the analyses presented in this work. Specifically, Dimensions was launched by Digital Science In January 2018.\footnote{\url{https://www.dimensions.ai/}} It includes linked research information from over 128 million publications to over 99,000 journals, along with records of grants, datasets, patents, policy papers, clinical trials, and more \citep{HerzogHookKonkiel2020DimensionsBringingBarriers, HookPorterHerzog2018DimensionsBuildingContext}. Scientific publication data in Dimensions are sourced from Crossref and PubMed. Dimensions provides multiple data enhancements such as affiliation and researcher disambiguation, concept mapping, and more. Several recent works compared the coverage of Dimensions to previously available sources (such as WoS) and found it to have adequate coverage (e.g., 
 \citet{SinghSinghKarmakarEtAl2021JournalCoverageWeb, Martin-MartinThelwallOrduna-MaleaEtAl2021GoogleScholarMicrosoft}).

 Along with Web of Science and Dimensions, I use Microsoft Academic Graph as a third source of scientific publications data. Microsoft launched the Academic Graph (MAG) in 2016 and provided open access to the complete dataset \citep{WangShenHuangEtAl2020MicrosoftAcademicGraph}.\footnote{In May 2021, Microsoft terminated the Academic Services project and future development of MAG. OurResearch, a non-profit organization founded in 2012, superseded MAG by releasing the OpenAlex project \citep{PriemPiwowarOrr2022OpenAlexFullyopenIndex}.} Importantly, MAG enables the construction of the citations count that is eventually linked to patent citations to science, obtained from \citet{MarxFuegi2022RelianceSciencePatenting}.

 Several complementary data sources are also included in the analysis. The American Men and Women of Science (AMWS) is a biographical directory of renowned North American scientists in the physical, biological, and related sciences. Entrants are scientists who have made significant contributions in their fields. \citet{BelenzonCioaca2021GuaranteedPublicDemand} have acquired 17 electronic versions of the AMWS directory, covering editions published from 2005 through 2021. These editions include information about 240,800 living and deceased scientists. With the research assistance of Hansen Zhang, these data were matched to the DISCERN firms and Dimensions publications using textual similarity matching. I incorporate the matched data to study firms' hiring of scientists.

\subsection{\label{sec:sample_const} Sample Construction}

The DISCERN data set includes 796,068 unique WoS records matched to 3,134 firms and published between 1980-2015.\footnote{In some cases, publications are matched to multiple firms, so the total number of observations is 822,529.} First, I keep the 582,107 records of articles and proceedings papers and drop other types (such as books, editorial materials, letters, reviews, and meeting abstracts). Next, I rely on a fuzzy textual match (using TF-IDF) to link WoS records with their equivalent MAG identifiers. Using DOIs obtained from MAG, I join the sample with Dimensions records. The matched sample includes 463,027 records with both MAG and Dimensions identifiers.

I applied additional filters for the sample I use in the publication level analysis. In the earlier years of the sample, I encountered substantial issues of missing data fields and inconsistent quality of disambiguated author identifiers. To improve the quality and avoid truncation on both sides of the sample, I restrict the data to the years between 1990 and 2012. Next, I remove cases where journal volume and issue data are unavailable. I also remove outlier journals, such as journals with more than 24 issues per year (or less than 3) and issues with more than 100 articles (or less than 5). In addition, since the ``first-in-first-out'' allocation rule of accepted manuscripts into issues does not apply to conference proceedings and special journal issues, I limit the sample to standard journal publications. First, I drop conference proceedings, special issue publications, and supplementary materials based on indicators obtained from WoS. Second, to complement these indicators, I use the average H-index of all authors in a journal issue to locate some unmarked outlier issues and remove them. In the next step, since the econometric specification requires a log count of citations, I remove publications with zero citations. Lastly, I remove singleton cases due to the inclusion of fixed effects. My final sample at the publication level includes 164,516 observations (156,486 unique publications, since some publications are coauthored by researchers from multiple firms) matched to 1,522 firms.

\subsection{Variable Construction}

Below I detail the construction of the variables used in the analyses. 

\begin{figure}[h]
\caption{\label{fig:followon}Illustration of Follow-On Research}
\captionsetup{width=0.8\textwidth}
\centering
\includegraphics[width=0.8\textwidth]{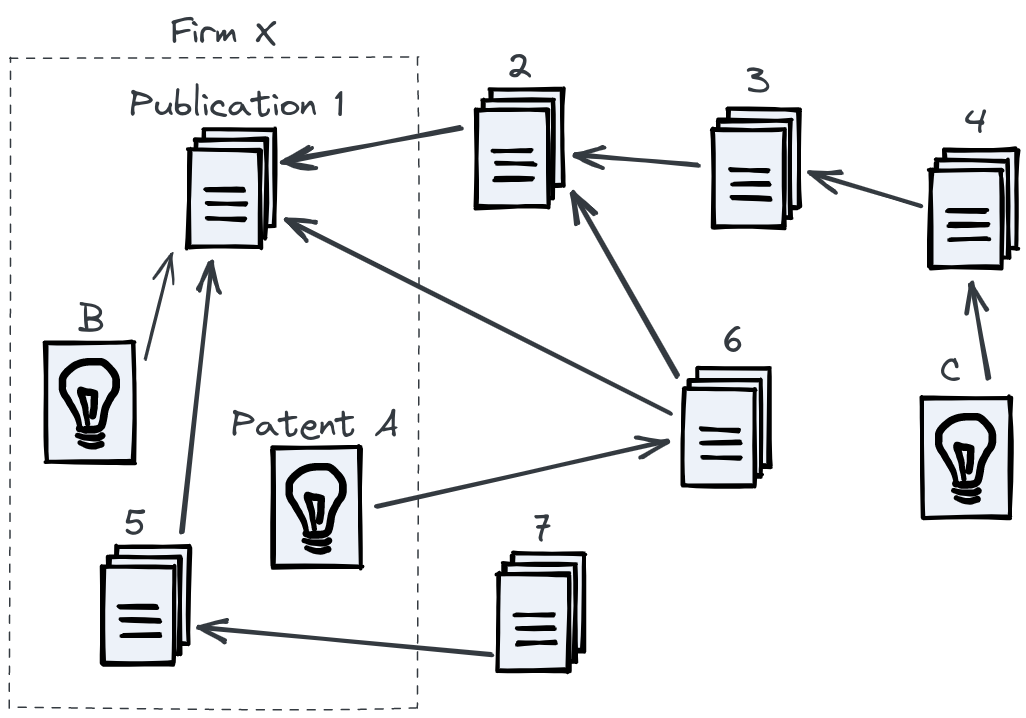}
\end{figure}

\begin{description}[style=unboxed,leftmargin=0cm]
\item[Follow-On Research] I use citation data from MAG to count three generations of scientific citations to the focal publication. First, I restrict citing records to journal articles and conference proceedings published up to 2015. Next, I drop all citations within the firm by filtering the citations using my complete sample of corporate publications. I identify external direct citations and then rerun the match to identify second and third-generation citations. For each citing document, I keep the shortest citation route. I use the total count of citations to measure follow-on research originating outside the firm.
Figure \ref{fig:followon} illustrates the construction of the measure. Publication 1 is the focal paper published by firm X. Publications 2, 3, and 4 are counted as first, second, and third-generation follow-on research, respectively. I count publication 6  as a first-generation follow-on. Publication 5 is internal to the firm and, therefore, not counted as a follow-on. Publication 7 is not counted as a follow-on to publication 1 (however, it is counted as a direct follow-on to publication 5). Overall, publication 1 has four follow-on publications.
\item[Patent citations to science (NPL citations)] I use data from \citet{MarxFuegi2022RelianceSciencePatenting} to identify patents that cite scientific publications. I keep records related to USPTO patents with a confidence score equal to or greater than 3, filed up to 2015. I match NPL citations to my sample using the provided MAG identifiers. When considering patents citing follow-on research, I keep the shortest route between the patent and the focal publication. In Figure \ref{fig:followon}, patent A is an illustration of a patent by the same firm that cites follow-on research. Patent B directly uses science within the firm. Patent C is a patent by others that cite external research that is considered a follow-on to publication 1.
\item[Author H-index] I calculate temporal H-index measures both for the construction of the instrumental variable (other authors in the same journal issue) and as control variables (top author of focal publication). I use Dimensions journal, issue, and volume data to identify other publications in the same journal issue as the focal publication. To create the H-index measure, I use the Dimensions disambiguated researcher identifiers to first identify prior published works of all authors in the journal issue (published up to the focal publication year). Next, for each prior work, I identify all scientific citations received up to the same focal publication year. I count the citations for each prior work and then apply the H-index algorithm described in Section \ref{sec:hindex}. I use the top H-index among the authors of the focal publication as a control variable. For the instrument, I sum the top two H-indexes among all authors in the journal issue after excluding the authors of the focal publication.
\item[Future publications and patents by focal authors] I use Dimensions disambiguated author identifiers to count subsequent scientific publications published by authors of the focal publication and related to the same firm. However, this method did not work for subsequent patents due to an incomplete match of inventors and scientific authors in the Dimensions data. Therefore, to identify patents by the focal authors, I conducted a textual match using the author's last name and first initial for inventors of USPTO patents assigned to the same firm. 
\item[AMWS Hiring] I incorporate a match between AMWS, DISCERN and Dimensions publications. The data includes the employment years of AMWS scientists by the firms in my sample. Overall, 20,552 employed individuals are identified (26,385 records, as some individuals move between firms). However, my analysis requires identifying the employment of scientists whose work is related to the focal publication. To identify such relations, I restrict the sample to 6,673 individuals for whom I have scientific publication records. For these publications, I use the Dimensions concepts variable to identify granular research topics. The concepts are extracted by Dimensions from titles and abstracts of publications and their relevance is assigned using the pointwise mutual information algorithm against the publications field of research domain (FOR). I use a cutoff of 0.75 to identify highly relevant concepts.\footnote{For example, a random list of concepts includes: target molecules, axial magnetic field, periodic solutions, amorphous silicon thin films, user attention, genetic interactions, polymer adsorption, coal-fired power plants, material removal rate, coverage algorithm, radical cation, laser pulse shape, nitrogen content, cell wall integrity, surface-enhanced Raman scattering, photophysical properties.} Next, I match the scientist's employment data using the concepts from the focal publication and the concepts related to the scientist's works. If there is an overlap in concepts and the employment term begins after the focal publication year, I consider the individual's hiring by the firm as related to the focal publication.
\item[University-industry collaborations] I identify scientific publications and patents as university-industry collaborations (UIC) using Dimensions organization identifiers and the GRID data set. Records affiliated with the firm and with an organization of type ``Education'' are considered UIC. In an alternative specification, I identify UICs using the raw affiliation text and searching for `univ$\mid$colleg$\mid$hosp'. The results of the analyses are similar for both specifications.
\item[Patent-paper pairs] I identify patent-paper pairs (PPP) using several steps. First, similarly to identifying future patents, I conduct a textual match of authors' last names and first initials, where both the focal publication and the patent are assigned to the same firm. Second, I restrict the patent filing date to be within two years of the focal publication date. Third, I require textual concept overlap using Dimensions concepts (as described above) between the patent and the focal publication. The resulting pairs are authored within the same firm by at least one shared author-inventor and share unique textual concepts.
\item[Scientific concept prevalence] First, I count Dimensions scientific concept appearances in non-corporate scientific publications by year. Second, I sum the concept counts in the previous three years for each concept related to the focal publication. Third, I aggregate the counts to the focal publication level. The resulting measure indicates if the concepts in the focal publication were prevalent in prior works by the scientific community.
\item[Government funding] I identify US government-funded publications using the funding acknowledgment field in Dimensions data. Similarly to the method described above for constructing the scientific concept prevalence, I count prior concept appearance only for government-funded publications. The ratio between government-funded concept appearances and the total concept appearance count is my measure of the availability of government funding for related scientific work.
\end{description}

\clearpage

\begingroup
\scriptsize
\singlespacing
\begin{longtable}{p{\dimexpr.175\textwidth-2\tabcolsep}
                  p{\dimexpr.1\textwidth-2\tabcolsep}
                  p{\dimexpr.25\textwidth-2\tabcolsep}
                  p{\dimexpr.475\textwidth-2\tabcolsep}}
\caption{\label{tbl:vars}Summary of Variables}\\
\toprule
Variable Name &
  Type &
  Measure of &
  Definition \\
\midrule
\endfirsthead

\multicolumn{4}{@{}l}{Table \thetable, continued}\\
\addlinespace
\toprule
Variable Name &  Type &  Measure of &  Definition \\ \midrule 
\endhead

\midrule
\multicolumn{4}{r@{}}{\em\scriptsize (continued on next page)}\\
\endfoot

\bottomrule
\endlastfoot

%% Body of table
Pr(Publication)
&Indicator
&Firm's subsequent scientific investments
&Equals one if the corporate authors of focal publication publish a subsequent scientific paper, and zero otherwise.\\
Pr(Univ. Collab)
&Indicator
&Firm's subsequent direct ties with academics
&Equals one if the corporate authors of focal publication publish a subsequent scientific paper with external academics, and zero otherwise.\\
Pr(Conference Proc.)
&Indicator
&Firm's subsequent participation in academics conferences
&Equals one if the corporate authors of focal publication publish a subsequent conference proceeding, and zero otherwise.\\
Pr(AMWS Hire)
&Indicator
&Firm's hiring of a renowned scientist whose work is related to the focal publication
&Equals one if the firm hires a related AMWS scientist in the years after the focal publication, and zero otherwise.\\
Pr(AMWS Award-winning Hire)
&Indicator
&Firm's hiring of an award-winning renowned scientist whose work is related to the focal publication
&Equals one if the firm hires an award-winning related AMWS scientist in the years after the focal publication, and zero otherwise.\\
Pr(Patent, $\geq3y$ gap)
&Indicator
&Firm's subsequent patenting outcomes
&Equals one if any of the corporate authors of the focal publication file a subsequent patent at least three years after the focal publication year, and zero otherwise.\\
Pr(Patent, $\geq5y$ gap)
&Indicator
&Firm's subsequent patenting outcomes
&Equals one if any of the corporate authors of the focal publication file a subsequent patent at least five years after the focal publication year, and zero otherwise.\\
Pr(Job Tenure $\geq10y$)
&Indicator
&Job tenure length following publication
&Equals one if any of the authors of the focal publication publish a subsequent paper at the same firm at least 10 years after the focal publication year, and zero otherwise.\\
Pr(Patent, Univ. Collab.)
&Indicator
&Firm's subsequent patenting with external academics
&Equals one if any of the corporate authors of the focal publication file a subsequent patent that is co-assigned to a public research institution, and zero otherwise.\\
Pr(Patent Citation to Focal)
&Indicator
&Firm's subsequent patenting outcomes
&Equals one if a patent assigned to the firm and filed after the focal publication year cites the focal publication, and zero otherwise.\\
Pr(Patent Citation to Focal or FO)
&Indicator
&Firm's subsequent patenting outcomes
&Equals one if a patent assigned to the firm and filed after the focal publication year cites the focal publication or any of the follow-on publications, and zero otherwise.\\
Pr(Patent, $\geq5y$ gap, Citing FO)
&Indicator
&Firm's subsequent patenting, follow-on research is an input
&Equals one if any of the corporate authors of the focal publication file a subsequent patent that cites any of the follow-on research, and zero otherwise.\\
Pr(Patent, $\geq5y$ gap, Not Citing FO)
&Indicator
&Firm's subsequent patenting, follow-on research provides quality validation
&Equals one if any of the corporate authors of the focal publication file a subsequent patent that does not cite the follow-on research, and zero otherwise.\\
ln(Follow-On)
&Continuous
&External research that follows the focal publication
&A logged count of external scientific publications that cite the focal publication, up to three generations away.\\
ln(Focal H-index)
&Continuous
&Prominence of the leading author of the focal publication
&A logged H-index measure calculated at the focal publication year based on prior publications and citations.\\
ln(Top Two Researchers H-index)
&Continuous
&Prominence of the top two authors in the same journal issue of the focal publication
&A logged sum of the H-index measures of the top two authors in a journal issue, after excluding the authors of the focal publication. Calculated at the focal publication year based on prior publications and citations.\\
Patent-Paper Pair (PPP)
&Indicator
&Possession of complementary IP rights
&Equals one in the presence of a patent assigned to the same firm, invented by at least one of the authors of the focal publication, filed within two years of the focal publication year, and shares a textual concept.\\
University-Industry Collaboration (UIC)
&Indicator
&Knowledge Outsourcing
&Equals one when at least one of the authors of the focal publication is affiliated with an educational institution, based on Dimensions GRID organization data.\\
Low Scientific Concept Prevalence
&Indicator
&Nascent research domain
&Equals one for below-median count of related scientific concept appearance in external publications in the three years prior to the focal publication year.\\
High Government Funding Availability
&Indicator
&Availability of government funding for the scientific community in related works
&Equals one for above-median ratio between government-funded concept count and the total concept count in the three  years prior to the focal publication year. \\
\end{longtable}
\endgroup

\printbibliography
\end{refsection}

\clearpage

\begin{refsection}

\section{\label{app:additional_results}Additional Results} 

\setcounter{table}{0}
\renewcommand{\thetable}{C\arabic{table}}
\setcounter{figure}{0}
\renewcommand{\thefigure}{C\arabic{figure}}

\small
\onehalfspacing

\subsection{Supplementary Descriptive Analysis}

\subsubsection{\label{app:desc_pats}Patents using internal and follow-on research}

Following the analysis in Section \ref{sec:res_desc}, a similar result emerges by classifying corporate patents based on their use of science. Table \ref{tbl:discern_desc_pats} presents descriptive statistics at the NPL citation and patent levels. According to columns 1 and 2, out of 6.6 million citations from corporate patents to science, only 2\% are to publications by the same firm. An additional 18\% are to follow-on research, and the remaining citations are to external research unrelated to the firm. Columns 3-6 aggregate NPL citations to the patent level. Out of 492,871 science-based patents,\footnote{I define a science-based patent as a patent with at least one citation to a scientific publication.} only 12\% directly cite a scientific publication by the firm. An additional 22.5\% do not cite internal science but cite external follow-on research. After accounting for truncation, I find that more than 40\% of corporate science-based patents are directly or indirectly related to firms' contributions to public research. Lastly, among patents that directly cite internal publications (columns 7-8), about 35\% also cite external follow-on research.

\subsubsection{Time Trends in the Use of Follow-On Research}

\begin{figure}[h]
\caption{\label{fig:pubyear_cumnpl}Time Trends in the Use of Follow-On Research}
\captionsetup{width=0.8\textwidth}
\centering
\includegraphics[width=0.6\textwidth]{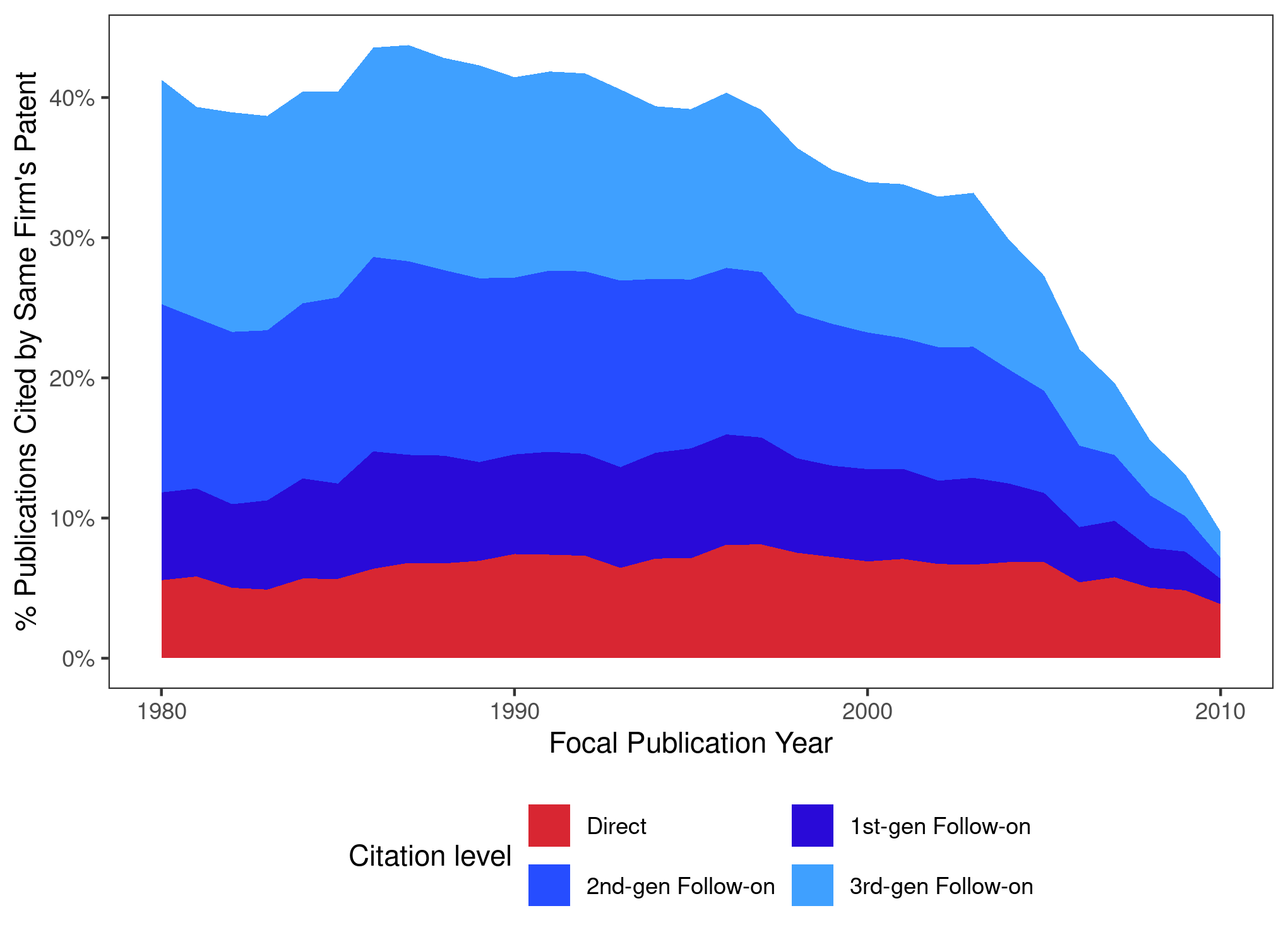}
\caption*{\footnotesize \textit{Note:} This figure shows time trends in the distribution of corporate publications by publication year and eventual direct and indirect patent citations. Publications that are directly cited are publications for which there is patent by the same firm that cites them. Publications that are indirectly cited are publications for which there is external follow-on research that is eventually cited by a patent by the same firm. A patent is counted once for each focal publication based on the shortest citation route. The decline in later years is due to truncation of the data in 2015.}
\end{figure}

Figure \ref{fig:pubyear_cumnpl} complements Figure \ref{fig:cumnpl} and Table \ref{tbl:discern_desc_pubs}. It presents time trends in the distribution of corporate publications by publication year and eventual direct and indirect patent citations. Note that the decline in later years is due to truncation of the data in 2015. The figure suggests that, at least up to the early 2000s, there is no clear change in the rate of direct and indirect patent citations to firms' publications throughout the sample years. A breakdown of these trends by field (available upon request) provides similar results, with slight variation across fields. 

\subsubsection{\label{app:desc_samename}Same-name inventors in NPL citations}

\begin{figure}[h]
\caption{\label{fig:pct_samename_npl}Percentage of NPL Citations to Follow-on Research with Same-Name Authors}
\captionsetup{width=0.8\textwidth}
\centering
\includegraphics[width=0.6\textwidth]{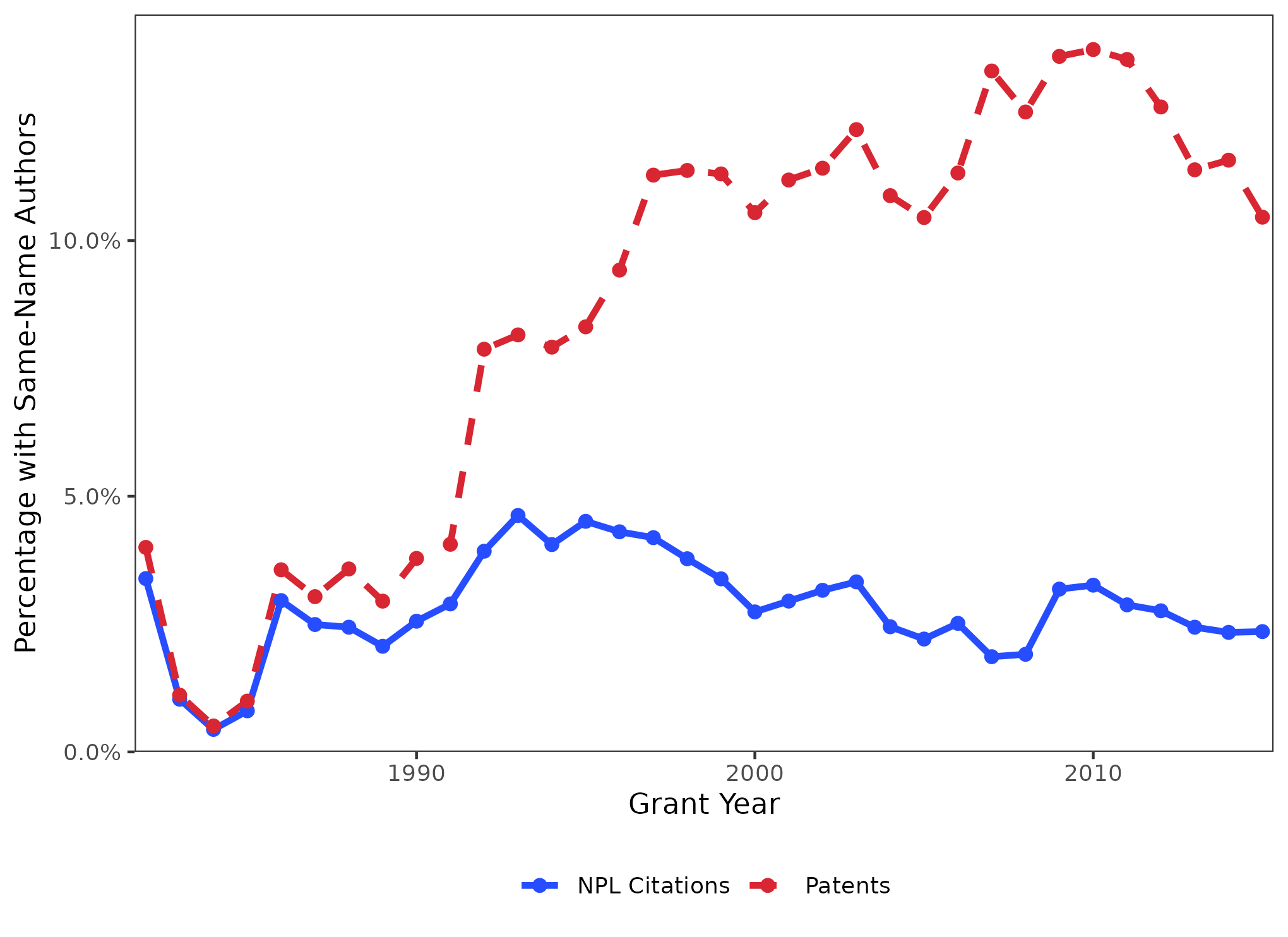}
\caption*{\footnotesize \textit{Note:} This figure shows the percentage of NPL citations and patents with same-name authors by patent grant year. Same-name matches are identified when the citing author's last name matches the patent inventor's last name and the first two characters of their first names match. The solid line shows the percentage of distinct NPL citations (patent–MAG ID pairs) with at least one same-name author. The dashed line shows the percentage of patents with at least one NPL citation featuring a same-name author.}
\end{figure}

To assess whether firms’ use of follow-on research reflects independent learning or requires collaboration with the cited external scientists, I examine the overlap between patent inventors and the authors of the scientific publications cited in their NPL references. Specifically, I identify cases in which a patent inventor’s last name and the first two characters of their first name match those of an author on a cited publication. Among patent citations to follow-on research, 2.65\% (37k out of 1.4m) include at least one inventor whose name matches that of an author of the cited external research, and 11.5\% (19k out of 168k) of patents contain at least one such citation. These figures provide an indication of how follow-on scientific knowledge is incorporated into corporate invention, whether through internalized learning and scientific consumption or by establishing direct ties with external academics. Figure \ref{fig:pct_samename_npl} illustrates these shares over time, showing modest but persistent inventor–author overlap.

\subsection{\label{sec:app_res}Supplementary Regression Analysis}

\subsubsection{Binned Scatterplots of Baseline Results}

\begin{figure}[h]
\caption{\label{fig:baseline_pubs}Follow-On Research and Subsequent Scientific Publication}
\captionsetup{width=0.8\textwidth}
\centering
\includegraphics[width=0.6\textwidth]{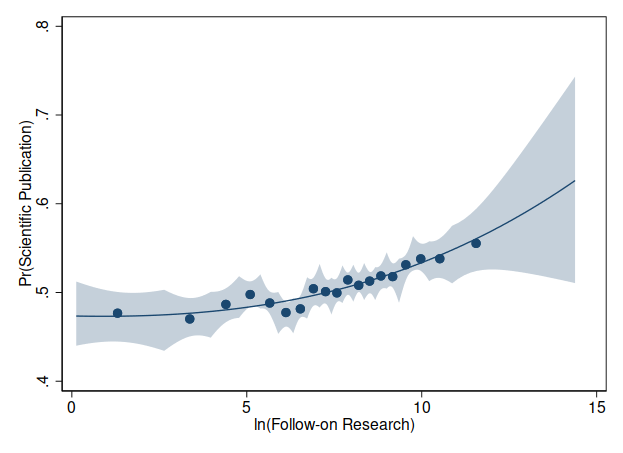}
\caption*{\footnotesize \textit{Note:} This figure presents a binned scatterplot of the relation between logged follow-on citations and the probability of subsequent scientific publishing by the focal authors. The values are fitted by controlling for the logged H-index of the focal author, firm fixed effects, and journal-year fixed effects.}
\end{figure}

\begin{figure}[h]
\caption{\label{fig:baseline_pats}Follow-On Research and Subsequent Patenting}
\captionsetup{width=0.8\textwidth}
\centering
\includegraphics[width=0.6\textwidth]{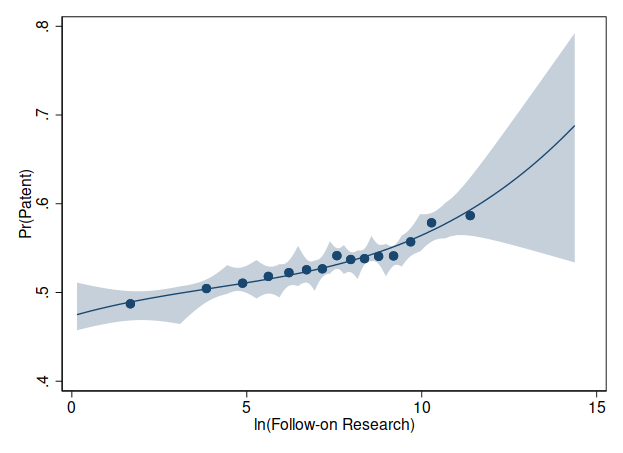}
\caption*{\footnotesize \textit{Note:} This figure presents a binned scatterplot of the relation between logged follow-on citations and the probability of subsequent patenting by the focal authors. The values are fitted by controlling for the logged H-index of the focal author, firm fixed effects, and journal-year fixed effects.}
\end{figure}

Figures \ref{fig:baseline_pubs} and \ref{fig:baseline_pats} present binned scatterplots that visualize the relationship between follow-on research and subsequent scientific publishing and patenting by focal authors. These figures complement the regression results in Table \ref{tbl:discern_pubspats} by illustrating the functional form of the estimated relationships. The scatterplots display residualized values after controlling for the logged H-index of the focal author, firm fixed effects, and journal-year fixed effects. 

\subsubsection{Poisson Estimation of Count Models}

Models with counts as the dependent variable are best estimated using Poisson regressions \citep{Wooldridge2010EconometricAnalysisCross}. Currently, standard methods are available for the estimation of two-stage Poisson (e.g., ivpoisson in Stata) and the estimation of Poisson with fixed effects (e.g., ppmlhdfe in Stata, fepois from the fixest package in R). However, there is currently no widely accepted implementation of a two-stage Poisson regression that allows the inclusion of a large number of fixed effects. Therefore, for my main analyses, I present a binary outcome (equal to one for a positive count) and a linear probability model estimated with standard OLS and 2SLS. 

I complement these results with  single-staged PPML and two-staged Poisson estimations using the control function approach, which is discussed by \citet{LinWooldridge2019TestingCorrectingEndogeneity}. The control function approach addresses endogeneity issues by including the residuals from the first-stage regression as an additional explanatory variable in the second stage. I follow \citet{BelletDeNeveWard2023DoesEmployeeHappiness} in implementing the two-step estimation procedure and adjust the standard errors through bootstrapping. Table \ref{tbl:poisson_iv} present analyses corresponding to the baseline results using Poisson fixed effect estimations of count models and the two-stage Poisson estimations. I find positive effects across the different specifications. However, for patents, the two-stage estimation yields positive coefficients that do not reach statistical significance. Overall, these findings provide additional support for the main results presented in the paper.

An additional robustness check pertains to cases where the dependent variable is constant across all observations within a fixed effect group. In Poisson estimation, observations are automatically dropped to avoid singletons and separation. One source of separation is a constant dependent variable within a fixed effect group. Table \ref{tbl:discern_patspubs_filtered} presents OLS and 2SLS results (specifications corresponding to the baseline results) for the observations that remained in the Poisson estimation samples. Overall, these results are similar to the results presented in the paper.

\subsubsection{Heterogeneity in Subsequent NPL Citations}

Table \ref{tbl:hetero} explores heterogeneity in the estimations of the effect of follow-on research on the likelihood of subsequent patent by the focal authors. In the baseline results (Table \ref{tbl:discern_pubspats}), I also present an analysis of the effect of follow-on research on the likelihood that subsequent patents by the firm (not restricted to the focal authors) will cite the focal publication. Table \ref{tbl:hetero_npl2foc} presents a corresponding heterogeneity analysis. Overall, the results are qualitatively similar but weaker in statistical significance.

\subsubsection{\label{sec:proceedings}Firms' Conference Proceedings Sample}

Firms' conference proceedings are dropped from the baseline sample because they do not follow the same assignment procedures into journal issues (see Section \ref{sec:sample_const} for details). However, proceedings are an important part of firms' scientific output, especially in some specific fields (e.g., computer science). Table \ref{tbl:proceedings} presents an analysis of firms' conference proceedings that were dropped from the main sample. The positive estimated correlations are in line with the baseline results in the paper.

\subsubsection{Variation by Field and Industry}

Table \ref{tbl:industry} presents estimation results for subsamples based on firms' main industries. Industry classifications are based on 2-digit SIC codes. Note that the subsamples not equal in size. Since the subsamples are much smaller than the full sample, in none of them the instrument has enough power to predict follow-on research. I therefore present OLS regressions. The results indicate that in most industries, there is a positive and statistically significant relation between follow-on research and subsequent scientific publication and patenting by the focal authors. Focusing on the chemical industry (including medicine), I find a significantly stronger relation compared to other industries (columns 6 and 13). 

Table \ref{tbl:field} presents estimation results for subsamples based on publications' assigned research field. Fields are determined by Dimensions.ai based on Australian and New Zealand Standard Research Classification (ANZSRC) 2020. Similar to the case of industries, subsamples are too small for 2SLS estimation. The results indicate a positive relation to subsequent publishing and patenting across different fields. However, differences across fields are not statistically significant (e.g., columns 6 and 13).

\subsubsection{\label{sec:fo_univ_corp}Corporate vs Academic Follow-On Research}

An interesting extension to the baseline results would be to explore how different sources of follow-on affect the focal firm. For example, it seems likely that follow-on research that originates for universities will have a different effect than follow-on research that originates from other firms. I attempt to explore these differences in Table \ref{tbl:corpfo}. I split the count of follow-on research by source, based on Dimensions.ai classification of institution type. Academic follow-on research is research that originates from academic and other non-private institutions. Corporate follow-on research originates from other firms. 

In both OLS and 2SLS regressions, I find very similar coefficients (with the exception of columns 3 and 5). Note, however, that in this case it is hard to defend the exclusion restriction of the instrument. For example, in column 6, while the instrument increases citations to corporate follow-on research, it also affects non-corporate follow-on research. There is not enough variation in the data to separately include both variables in the regression. It therefore seems that a different design is required in order to explore this source of heterogeneity.

\subsubsection{\label{sec:mech_extended}Variation by the Prominence of Focal authors}

It is possible that the extent to which follow-on research is useful as an input and as validation varies by the extent the of uncertainty around the firm's own science. When the quality of the science is more uncertain, it is likely that validation from the external environment will be more useful. To explore these relations, I build on the analysis presented in Section \ref{sec:mech}. I argue that when focal publication authors are less prominent, firms face greater uncertainty regarding the quality of their work. Conversely, for prominent focal authors, follow-on research is more likely to be beneficial as an input, possibly through their greater capacity to absorb external knowledge \citep{CohenLevinthal1990AbsorptiveCapacityNew}. 

Table \ref{tbl:mech_interactions} reports estimation results, interacting the effect of follow-on research with an indicator for above-median focal author prominence (measured by the top H-index among focal authors). The dependent variables are indicators for subsequent patents by focal authors, classified based on citations to follow-on research. While estimates are somewhat noisy, they provide evidence supporting the discussed mechanisms. Follow-on research appears more useful as an input when focal authors have an above-median H-index (columns 1-2), whereas it is more useful as validation when focal authors have a below-median H-index (columns 3-4). These results suggest that follow-on research can be valuable in multiple ways, with the mechanisms driving private value depending on firm and author attributes.

\subsubsection{\label{app:firm_desc}Descriptive Statistics for Firm-Level Analysis}

Table \ref{tbl:discern_firm_descstats} provides descriptive statistics for the firm-level panel analysis presented in Section \ref{sec:firmlevel}. The sample consists of 35,156 firm-year observations of 2,338 unique DISCERN firms observed over 1981--2015. The dependent variables include annual counts of scientific publications, patents, and the number of scientists employed by each firm. The main independent variables are depreciated stocks of follow-on research, as well as stocks of internal and external patents that cite the follow-on publications. All stock variables are computed using a 15\% annual depreciation rate.

 \begin{landscape}
\begin{table}[htbp]

\footnotesize

\centering

\caption{\label{tbl:discern_desc_pats}
{Firms' Patent Citations to Own Scientific Publications}
}

\begin{threeparttable}

\begin{tabular}{lHrrrrrrrr}
\toprule
\multicolumn{1}{l}{} &  & \multicolumn{2}{c}{NPL Citations} & \multicolumn{6}{c}{Patents}\\
\cmidrule(lr){3-4}\cmidrule(lr){5-10}
\multicolumn{1}{l}{} &  & \multicolumn{2}{c}{}&\multicolumn{2}{c}{All}&\multicolumn{2}{c}{Granted 2000-2015} & \multicolumn{2}{c}{Directly Citing}\\
\cmidrule(lr){5-6} \cmidrule(lr){7-8}\cmidrule(lr){9-10}
\multicolumn{1}{l}{} &  &
\multicolumn{1}{c}{Count}&
\multicolumn{1}{c}{Percent}&
\multicolumn{1}{c}{Count}&
\multicolumn{1}{c}{Percent}&
\multicolumn{1}{c}{Count}&
\multicolumn{1}{c}{Percent}&
\multicolumn{1}{c}{Count}&
\multicolumn{1}{c}{Percent}\\
&&\multicolumn{1}{c}{(1)}&\multicolumn{1}{c}{(2)}&\multicolumn{1}{c}{(3)}&\multicolumn{1}{c}{(4)}&\multicolumn{1}{c}{(5)}&\multicolumn{1}{c}{(6)}&\multicolumn{1}{c}{(7)}&\multicolumn{1}{c}{(8)}\\
\midrule
   Direct citation to a firm's publication && $132,558$ & $2.02\%$ & $58,301$ & $11.83\%$ & $46,029$ & $13.04\%$ & $58,301$ & $100.00\%$\\

\multicolumn{4}{l}{Indirect citation to a firm's publication (citing follow-on research)} \\
 \hspace{2em}  1st Generation && $283,606$ & $4.31\%$ & $38,150$ & $7.74\%$ & $33,536$ & $9.50\%$ & $16,021$ & $27.48\%$ \\
 \hspace{2em}  2nd Generation && $570,557$ & $8.68\%$ & $41,662$ & $8.45\%$ & $37,628$ & $10.66\%$ & $13,018$ & $22.33\%$ \\
 \hspace{2em}  3rd Generation && $397,358$ & $6.05\%$ & $31,164$ & $6.32\%$ & $27,742$ & $7.86\%$ & $11,738$ & $20.13\%$ \\
 %\cmidrule(lr){3-8}
 \hspace{2em}  Any && $1,251,521$ & $19.04\%$ & $110,976$ & $22.52\%$ & $98,906$ & $28.02\%$ & $20,374$ & $34.95\%$ \\
 \\

 \midrule
 Citing a firm's pub. (direct or indirect) && $1,384,079$ & $21.06\%$ & $169,277$ &  $34.35\%$ & $144,935$ & $41.06\%$ & $58,301$ & $100\%$ \\
  Not Citing a firm's pub. & &  $5,188,566$ & $78.94\%$ & $323,594$ & $65.65\%$ & $208,069$ & $58.94\%$ & $0$ & $0\%$\\
Total &  & $6,572,645$ & $100.00\%$ & $492,871$ & $100.00\%$ & $353,004$ & $100.00\%$ & $58,301$ & $100\%$\\
\bottomrule
\end{tabular}
\begin{tablenotes}[flushleft] \scriptsize
\item This table presents summary statistics of citations from firms' patents to scientific publications (NPL). The dataset includes 492,871 patents with a total of 6,572,645 NPL citations. In columns 1-6, a patent is categorized by the shortest route to an internal scientific publication. 58,301 (12\%) patents directly cite a scientific publication published by the same firm. An additional 110,976 (22.5\%) patents cite external scientific publications that have an internal publication as an upstream reference,  up to the 3rd generation of references. 323,594 patents (65.7\%) do not have a citation to an upstream internal scientific publication. To account for truncation, columns 5 and 6 present a subset of patents published after the year 2000. Among these patents, about 41\% cite a firm's publication either directly or indirectly. Columns 7-8 only include patents that directly cite internal scientific publications. Among these patents, all citation routes are considered and patents are classified based on whether they include a citation to different levels of follow-on research. The results indicate that about 35\% of directly-citing patents also cite external follow-on research.
\end{tablenotes}
\end{threeparttable}

\end{table}

 \end{landscape}

 %\begin{landscape}

\begin{landscape}
\begin{table}[htbp]
\footnotesize \singlespacing
   \centering
   \begin{threeparttable}[b]
      \caption{\label{tbl:poisson_iv}  Two-Stage Poisson Estimation Using a Control Function Approach}
      
\bigskip
\begin{tabular}{l*{8}{c}}
\toprule
&\multicolumn{2}{c}{\shortstack{Subsequent\\Publications}}&\multicolumn{2}{c}{\shortstack{Subsequent\\UIC Publications}}&\multicolumn{2}{c}{\shortstack{Scientist\\Hiring}}&\multicolumn{2}{c}{\shortstack{Subsequent\\Patents}}\\
\cmidrule(lr){2-3}\cmidrule(lr){4-5}\cmidrule(lr){6-7}\cmidrule(lr){8-9}
&\multicolumn{1}{c}{PPML}&\multicolumn{1}{c}{2S-PPML}&\multicolumn{1}{c}{PPML}&\multicolumn{1}{c}{2S-PPML}&\multicolumn{1}{c}{PPML}&\multicolumn{1}{c}{2S-PPML}&\multicolumn{1}{c}{PPML}&\multicolumn{1}{c}{2S-PPML}\\
%                    &\multicolumn{1}{c}{PPML}&\multicolumn{1}{c}{2S-PPML}&\multicolumn{1}{c}{PPML}&\multicolumn{1}{c}{2S-PPML}&\multicolumn{1}{c}{PPML}&\multicolumn{1}{c}{2S-PPML}&\multicolumn{1}{c}{PPML}&\multicolumn{1}{c}{2S-PPML}\\
&(1)&(2)&(3)&(4)&(5)&(6)&(7)&(8)\\ \midrule
ln(Follow-on)       &       0.054\sym{***}&       0.545\sym{**} &       0.068\sym{***}&       0.775\sym{**} &       0.076\sym{***}&       0.729         &       0.032\sym{***}&       0.567         \\
                    &     (0.006)         &     (0.261)         &     (0.007)         &     (0.350)         &     (0.004)         &     (0.507)         &     (0.012)         &     (0.480)         \\
\addlinespace
ln(Focal H-Index)   &       0.221\sym{***}&       0.084         &       0.319\sym{***}&       0.122         &       0.029\sym{***}&      -0.152         &       0.060\sym{**} &      -0.089         \\
                    &     (0.020)         &     (0.074)         &     (0.021)         &     (0.098)         &     (0.010)         &     (0.142)         &     (0.025)         &     (0.136)         \\
\\
Firm FE             &         Yes         &  Yes        &         Yes         & Yes         &         Yes        &  Yes           &         Yes         &   Yes       \\
Journal-Year FE     &         Yes         &  Yes        &         Yes         & Yes         &        Yes         &  Yes           &         Yes         &   Yes       \\
Observations        &     142,659         &     142,659         &     137,339         &     137,339         &      61,576         &      61,576         &     148,745         &     148,745         \\
Avg. DV             &      20.951         &       20.951        &       8.698         &        8.698             &       0.775         &     0.775                &      11.094         &       11.094              \\
First Stage F-stat  &                     &      30.061         &                     &      30.061         &                     &      30.061         &                     &      30.061         \\
Pseudo R$^2$        &       0.489         &                     &       0.490         &                     &       0.551         &                     &       0.570         &                     \\
\bottomrule
\end{tabular}
\begin{tablenotes}[flushleft]
\scriptsize
\item \textit{Notes:} This table reports estimation results of a two-staged Poisson estimation. Standard-errors are adjusted using a clustered bootstrap with 200 replications.
\item Clustered (Firm) standard-errors in parentheses. Signif. Codes: ***: 0.01, **: 0.05, *: 0.1
\end{tablenotes}
\end{threeparttable}
\end{table}

\end{landscape}

\begin{landscape}

%\begin{table}[htbp]
\begin{sidewaystable}
\footnotesize \singlespacing
   \centering
   \begin{threeparttable}[b]
      \caption{\label{tbl:discern_patspubs_filtered}The Effect of Follow-On Research on Firms' Investments in Science and Patenting Outcomes, Filtered Sample}
      \bigskip
\begin{tabular}{l*{9}{c}}
\toprule

&\multicolumn{3}{c}{Subsequent Scientific Publications} &\multicolumn{4}{c}{Subsequent Patenting}
&\multicolumn{2}{c}{\shortstack{Employee\\Retention}}\\
\cmidrule(lr){2-4}\cmidrule(lr){5-8}\cmidrule(lr){9-10}

                    &\multicolumn{1}{c}{\shortstack{Pr(Future\\Pub.)}}&\multicolumn{1}{c}{\shortstack{ln(Follow-on\\Research)}}&\multicolumn{1}{c}{\shortstack{Pr(Future\\Pub.)}}&\multicolumn{2}{c}{\shortstack{Pr(Future Pat.,\\ $\geq$ 3y gap)}}&\multicolumn{1}{c}{\shortstack{Pr($\cdot$,\\ $\geq$ 5y gap)}}&\multicolumn{1}{c}{\shortstack{Pr(Pat\\Citation)}}&\multicolumn{2}{c}{\shortstack{Pr(Job Tenure\\ $\geq$ 10y)}}\\\cmidrule(lr){2-2}\cmidrule(lr){3-3}\cmidrule(lr){4-4}\cmidrule(lr){5-6}\cmidrule(lr){7-7}\cmidrule(lr){8-8}\cmidrule(lr){9-10}
                    &\multicolumn{1}{c}{OLS}&\multicolumn{1}{c}{Stage 1}&\multicolumn{1}{c}{Stage 2}&\multicolumn{1}{c}{OLS}&\multicolumn{1}{c}{2SLS}&\multicolumn{1}{c}{2SLS}&\multicolumn{1}{c}{2SLS}&\multicolumn{1}{c}{OLS}&\multicolumn{1}{c}{2SLS}\\
&(1)&(2)&(3)&(4)&(5)&(6)&(7)&(8)&(9)\\ \midrule
ln(Follow-on)       &       0.009\sym{***}&                     &       0.134\sym{***}&       0.009\sym{***}&       0.059         &       0.093\sym{*}  &       0.112\sym{**} &       0.008\sym{***}&       0.132\sym{***}\\
                    &     (0.002)         &                     &     (0.051)         &     (0.001)         &     (0.052)         &     (0.055)         &     (0.052)         &     (0.001)         &     (0.050)         \\
\addlinespace
ln(top researchers H-index)&                     &       0.101\sym{***}&                     &                     &                     &                     &                     &                     &                     \\
                    &                     &     (0.019)         &                     &                     &                     &                     &                     &                     &                     \\
\addlinespace
ln(Focal H-Index)   &       0.001         &       0.279\sym{***}&      -0.033\sym{**} &       0.013\sym{***}&      -0.001         &      -0.014         &      -0.042\sym{***}&       0.010\sym{***}&      -0.024\sym{*}  \\
                    &     (0.004)         &     (0.011)         &     (0.015)         &     (0.002)         &     (0.015)         &     (0.015)         &     (0.015)         &     (0.003)         &     (0.014)         \\
\\
Firm FE             &         Yes         &         Yes         &         Yes         &         Yes         &         Yes         &         Yes         &         Yes         &         Yes         &         Yes         \\
Journal-Year FE     &         Yes         &         Yes         &         Yes         &         Yes         &         Yes         &         Yes         &         Yes         &         Yes         &         Yes         \\
Observations        &     142,659         &     142,659         &     142,659         &     133,064         &     133,064         &     113,663         &      65,222         &     142,659         &     142,659         \\
Avg. DV             &       0.584         &       7.407         &       0.584         &       0.489         &       0.489         &       0.437         &       0.140         &       0.289         &       0.289         \\
First Stage F-stat  &                     &                     &      29.064         &                     &      24.604         &      23.141         &      18.421         &                     &      29.064         \\
Adjusted R$^2$      &       0.258         &      -0.125         &           .         &       0.274         &           .         &           .         &           .         &       0.223         &           .         \\
\bottomrule
\end{tabular}

      \begin{tablenotes}[flushleft]
      \scriptsize
         \item \textit{Note.} This table is a filtered version of Table \ref{tbl:discern_pubspats}. The samples include observations that are kept after PPML estimation. Observations are automatically dropped to avoid singletons and separation.
         Clustered (Firm) standard-errors in parentheses. Signif. Codes: ***: 0.01, **: 0.05, *: 0.1
      \end{tablenotes}
   \end{threeparttable}
\end{sidewaystable}
%\end{table}
\end{landscape}

\begin{table}[htbp]
\footnotesize \singlespacing
   \centering
   \begin{threeparttable}[b]
      \caption{\label{tbl:hetero_npl2foc}Heterogeneity in Subsequent Patenting, Patent to Paper Citations}
      \bigskip

%\begin{tabular}{l*{8}{c}}
\begin{tabularx}{\textwidth}{l@{}*{8}{Y}}
\toprule
%                    &\multicolumn{2}{c}{\shortstack{Complementary\\IP Rights}}&\multicolumn{2}{c}{\shortstack{Knowledge\\Outsourcing}}  &\multicolumn{2}{c}{\shortstack{Scientific Concept\\Prevalence}}&\multicolumn{2}{c}{\shortstack{Govt. Funding\\Availability}}\\\cmidrule(lr){2-3}\cmidrule(lr){4-5}\cmidrule(lr){6-7}\cmidrule(lr){8-9}
%&Pr(Patent)&Pr(Patent)&Pr(Patent)&Pr(Patent)&Pr(Patent)&Pr(Patent)&Pr(Patent)&Pr(Patent)\\
%                    \cmidrule(lr){2-2}\cmidrule(lr){3-3}\cmidrule(lr){4-4}\cmidrule(lr){5-5}\cmidrule(lr){6-6}\cmidrule(lr){7-7}\cmidrule(lr){8-8}
&\multicolumn{4}{c}{Firm Capabilities}&\multicolumn{4}{c}{Scientific Community}\\
\cmidrule(lr){2-5}\cmidrule(l){6-9}
%&\multicolumn{8}{c}{Pr(Subsequent Patent by Focal Authors, $\geq$3y gap)}\\
&\multicolumn{8}{c}{Pr(Subsequent Firm Patent Citing Focal Publication)}\\
\cmidrule(){2-9}
                    &\multicolumn{1}{c}{OLS}&\multicolumn{1}{c}{2SLS}&\multicolumn{1}{c}{OLS}&\multicolumn{1}{c}{2SLS}&\multicolumn{1}{c}{OLS}&\multicolumn{1}{c}{2SLS}&\multicolumn{1}{c}{OLS}&\multicolumn{1}{c}{2SLS}\\
&(1)&(2)&(3)&(4)&(5)&(6)&(7)&(8)\\ \midrule
\multicolumn{9}{l}{\textbf{Complementary IP Rights}}\\
\addlinespace
ln(Follow-On) $\times$ &       0.023\sym{***}&       0.031\sym{***}&                     &                     &                     &                     &                     &                     \\
  \ \ \ \ PPP                  &     (0.002)         &     (0.006)         &                     &                     &                     &                     &                     &                     \\
\addlinespace
PPP                 &       0.105\sym{***}&       0.088\sym{***}&                     &                     &                     &                     &                     &                     \\
                    &     (0.005)         &     (0.011)         &                     &                     &                     &                     &                     &                     \\

\addlinespace
\multicolumn{9}{l}{\textbf{Knowledge Outsourcing}}\\
\addlinespace
ln(Follow-On) $\times$ &                     &                     &      -0.007\sym{***}&      -0.005         &                     &                     &                     &                     \\
  \ \ \ \ UIC                  &                     &                     &     (0.001)         &     (0.004)         &                     &                     &                     &                     \\
\addlinespace
UIC                 &                     &                     &      -0.042\sym{***}&      -0.043\sym{***}&                     &                     &                     &                     \\
                    &                     &                     &     (0.004)         &     (0.004)         &                     &                     &                     &                     \\

\addlinespace
\multicolumn{9}{l}{\textbf{Scientific Concept Prevalence}}\\
\addlinespace
ln(Follow-On) $\times$ &                     &                     &                     &                     &       0.000         &       0.008\sym{*}  &                     &                     \\
   \ \ \ \ Low Prev.                 &                     &                     &                     &                     &     (0.001)         &     (0.004)         &                     &                     \\
\addlinespace
Low Prevalence      &                     &                     &                     &                     &       0.004\sym{***}&       0.001         &                     &                     \\
                    &                     &                     &                     &                     &     (0.001)         &     (0.003)         &                     &                     \\

\addlinespace
\multicolumn{9}{l}{\textbf{Government  Funding Availability}}\\
\addlinespace
ln(Follow-On) $\times$ &                     &                     &                     &                     &                     &                     &       0.001         &       0.009         \\
\ \ \ \ Govt. Funding                    &                     &                     &                     &                     &                     &                     &     (0.001)         &     (0.006)         \\
\addlinespace
Govt Funding        &                     &                     &                     &                     &                     &                     &       0.004\sym{*}  &      -0.003         \\
                    &                     &                     &                     &                     &                     &                     &     (0.002)         &     (0.005)         \\

\addlinespace
ln(Follow-On)       &       0.008\sym{***}&       0.046\sym{*}  &       0.015\sym{***}&       0.048\sym{*}  &       0.011\sym{***}&       0.043\sym{*}  &       0.011\sym{***}&       0.045\sym{*}  \\
                    &     (0.001)         &     (0.027)         &     (0.001)         &     (0.027)         &     (0.001)         &     (0.026)         &     (0.001)         &     (0.027)         \\
\addlinespace
ln(Focal H-Index)   &      -0.009\sym{***}&      -0.020\sym{**} &      -0.001         &      -0.010         &      -0.008\sym{***}&      -0.018\sym{**} &      -0.008\sym{***}&      -0.018\sym{**} \\
                    &     (0.001)         &     (0.008)         &     (0.001)         &     (0.008)         &     (0.001)         &     (0.008)         &     (0.001)         &     (0.008)         \\

\addlinespace
%Constant            &       0.341\sym{***}&                     &       0.514\sym{***}&                     &       0.492\sym{***}&                     &       0.493\sym{***}&                     \\
%                    &     (0.005)         &                     &     (0.005)         &                     &     (0.006)         &                     &     (0.006)         &                     \\
\\
Firm FE             &         Yes         &         Yes         &         Yes         &         Yes         &         Yes         &         Yes         &         Yes         &         Yes         \\
Journal-Year FE     &         Yes         &         Yes         &         Yes         &         Yes         &         Yes         &         Yes         &         Yes         &         Yes         \\
Observations        &     164,516         &     164,516         &     164,516         &     164,516         &     164,516         &     164,516         &     164,516         &     164,516         \\
Avg. DV             &       0.056         &       0.056         &       0.056         &       0.056         &       0.056         &       0.056         &       0.056         &       0.056         \\
First Stage F-stat  &                     &      13.819         &                     &      13.007         &                     &      13.155         &                     &      13.594         \\
Adjusted R$^2$      &       0.097         &      .         &       0.078         &      .         &       0.070         &      .         &       0.070         &      .         \\
\bottomrule
\end{tabularx}

      \begin{tablenotes}[flushleft]
      \scriptsize
          \item \textit{Notes:} This table accompanies table \ref{tbl:hetero}. In this table, the dependent variable is an indicator for a subsequent patent by the focal firm that cites the focal publication.
         \item Clustered (Firm) standard-errors in parentheses. Signif. Codes: ***: 0.01, **: 0.05, *: 0.1
      \end{tablenotes}
   \end{threeparttable}
\end{table}

\begin{table}[htbp]
\footnotesize \singlespacing
   \centering
   \begin{threeparttable}[b]
      \caption{\label{tbl:proceedings}Corporate Conference Proceedings}
      \bigskip

\begin{tabular}{l*{4}{c}}
\toprule
                    &\multicolumn{1}{c}{\shortstack{Subsequent\\Publications}}&\multicolumn{1}{c}{\shortstack{Pr(Subsequent\\Publication)}}&\multicolumn{1}{c}{\shortstack{Subsequent\\Patents}}&\multicolumn{1}{c}{\shortstack{Pr(Subsequent\\Patent)}}\\\cmidrule(lr){2-2}\cmidrule(lr){3-3}\cmidrule(lr){4-4}\cmidrule(lr){5-5}
                    &\multicolumn{1}{c}{PPML}&\multicolumn{1}{c}{OLS}&\multicolumn{1}{c}{PPML}&\multicolumn{1}{c}{OLS}\\
&(1)&(2)&(3)&(4)\\ \midrule
ln(Follow-On)       &       0.054\sym{***}&       0.003         &       0.037\sym{***}&       0.005\sym{***}\\
                    &     (0.010)         &     (0.003)         &     (0.006)         &     (0.002)         \\
\addlinespace
ln(Focal H-Index)   &       0.297\sym{***}&      -0.004         &       0.083\sym{***}&       0.004         \\
                    &     (0.041)         &     (0.009)         &     (0.020)         &     (0.005)         \\
\\
Firm FE             &         Yes         &         Yes         &         Yes         &         Yes         \\
Conference FE       &         Yes         &         Yes         &         Yes         &         Yes         \\
Observations        &      22,253         &      22,253         &      24,604         &      24,604         \\
Avg. DV             &      12.946         &       0.452         &      33.585         &       0.789         \\
Adjusted R$^2$      &                     &       0.279         &                     &       0.252         \\
Pseudo R$^2$        &       0.603         &                     &       0.620         &                     \\

\bottomrule
\end{tabular}

      \begin{tablenotes}[flushleft]
      \scriptsize
         \item \textit{Notes:} This table presents the baseline results for a sample of corporate conference proceedings that are dropped from the main analysis. The data consists of a pooled cross-section of proceedings by U.S.-based publicly-owned firms, published between 1990 and 2012 \citep{AroraBelenzonSheer2020DISCERNDukeInnovationa}. Follow-on research is the total count of three generations of citations to the focal publication from outside the firm. The dependent variables are counts and indicators for subsequent scientific publications (columns 1-2) and patents (columns 3-4) by the corporate authors of the focal papers. All regressions include a control for the highest H-index among the authors of the focal publication, as well as firm and conference fixed effects. 
         \item Clustered (Firm) standard-errors in parentheses. Signif. Codes: ***: 0.01, **: 0.05, *: 0.1
      \end{tablenotes}
   \end{threeparttable}
\end{table}

\begin{landscape}
\begin{table}[htbp]
\scriptsize \singlespacing
   \centering
   \begin{threeparttable}[b]
      \caption{\label{tbl:industry}Variation by Main Industry}
      \bigskip

\begin{tabular}{l*{14}{c}}
\toprule
                    &\multicolumn{7}{c}{Pr(Subsequent Scientific Publication by Focal Authors)}                                                                                                           &\multicolumn{7}{c}{Pr(Subsequent Patent by Focal Authors, $\geq$ 3y gap)}                                                                                                                 \\\cmidrule(lr){2-8}\cmidrule(lr){9-15}
                     & Chem & Elec & Instr & Serv & Other & All & All & Chem & Elec & Instr & Serv & Other & All & All \\
                    &\multicolumn{1}{c}{OLS}&\multicolumn{1}{c}{OLS}&\multicolumn{1}{c}{OLS}&\multicolumn{1}{c}{OLS}&\multicolumn{1}{c}{OLS}&\multicolumn{1}{c}{OLS}&\multicolumn{1}{c}{2SLS}&\multicolumn{1}{c}{OLS}&\multicolumn{1}{c}{OLS}&\multicolumn{1}{c}{OLS}&\multicolumn{1}{c}{OLS}&\multicolumn{1}{c}{OLS}&\multicolumn{1}{c}{OLS}&\multicolumn{1}{c}{2SLS}\\
&(1)&(2)&(3)&(4)&(5)&(6)&(7)&(8)&(9)&(10)&(11)&(12)&(13)&(14)\\ \midrule
ln(Follow-on)       &       0.012\sym{***}&       0.004         &       0.002         &       0.011\sym{***}&       0.004\sym{**} &       0.004\sym{**} &       0.106\sym{**} &       0.008\sym{***}&       0.007\sym{***}&       0.002         &       0.012\sym{***}&       0.007\sym{***}&       0.006\sym{***}&       0.082         \\
                    &     (0.002)         &     (0.003)         &     (0.005)         &     (0.003)         &     (0.002)         &     (0.002)         &     (0.049)         &     (0.001)         &     (0.002)         &     (0.002)         &     (0.002)         &     (0.002)         &     (0.001)         &     (0.052)         \\
\addlinespace
ln(Follow-on) $\times$ &                     &                     &                     &                     &                     &       0.009\sym{***}&       0.054         &                     &                     &                     &                     &                     &       0.004\sym{*}  &      -0.048         \\
  \ \ \ \ Chemicals                  &                     &                     &                     &                     &                     &     (0.002)         &     (0.038)         &                     &                     &                     &                     &                     &     (0.002)         &     (0.040)         \\
%\addlinespace
%Chemicals           &                     &                     &                     &                     &                     &                     &                     &                     &                     &                     &                     &                     &                     &                     \\
%                    &                     &                     &                     &                     &                     &                     &                     &                     &                     &                     &                     &                     &                     &                     \\
\addlinespace
ln(Focal H-Index)   &      -0.010\sym{***}&       0.007         &       0.016         &       0.009         &       0.024\sym{***}&       0.001         &      -0.033\sym{**} &       0.012\sym{***}&       0.005         &       0.035\sym{***}&       0.001         &       0.015\sym{*}  &       0.012\sym{***}&      -0.003         \\
                    &     (0.004)         &     (0.007)         &     (0.018)         &     (0.007)         &     (0.007)         &     (0.003)         &     (0.014)         &     (0.003)         &     (0.006)         &     (0.005)         &     (0.006)         &     (0.009)         &     (0.002)         &     (0.013)         \\
%\addlinespace
%Constant            &       0.570\sym{***}&       0.404\sym{***}&       0.260\sym{***}&       0.345\sym{***}&       0.327\sym{***}&       0.519\sym{***}&                     &       0.257\sym{***}&       0.499\sym{***}&       0.487\sym{***}&       0.281\sym{***}&       0.349\sym{***}&       0.369\sym{***}&                     \\
%                    &     (0.016)         &     (0.027)         &     (0.068)         &     (0.025)         &     (0.022)         &     (0.008)         &                     &     (0.013)         &     (0.021)         &     (0.020)         &     (0.019)         &     (0.016)         &     (0.006)         &                     \\
\\
Firm FE             &         Yes         &         Yes         &         Yes         &         Yes         &         Yes         &         Yes         &         Yes         &         Yes         &         Yes         &         Yes         &         Yes         &         Yes         &         Yes         &         Yes         \\
Journal-Year FE     &         Yes         &         Yes         &         Yes         &         Yes         &         Yes         &         Yes         &         Yes         &         Yes         &         Yes         &         Yes         &         Yes         &         Yes         &         Yes         &         Yes         \\
Observations        &      88,170         &      15,848         &      13,471         &      14,671         &      15,292         &     164,516         &     164,516         &      88,170         &      15,848         &      13,471         &      14,671         &      15,292         &     164,516         &     164,516         \\
Avg. DV             &       0.623         &       0.420         &       0.299         &       0.408         &       0.371         &       0.507         &       0.507         &       0.353         &       0.548         &       0.583         &       0.357         &       0.426         &       0.397         &       0.397         \\
First Stage F-stat  &                     &                     &                     &                     &                     &                     &      13.721         &                     &                     &                     &                     &                     &                     &      13.721         \\
Adjusted R$^2$      &       0.312         &       0.295         &       0.380         &       0.348         &       0.340         &       0.345         &      -0.428         &       0.328         &       0.342         &       0.474         &       0.318         &       0.405         &       0.349         &      -0.238         \\

\bottomrule
\end{tabular}
     \begin{tablenotes}[flushleft]
      \scriptsize
         \item \textit{Notes:} This table presents variation in the baseline estimation results by firms' main industry classifications. 
         %The analysis corresponds to Tables \ref{tbl:discern_pubs} and \ref{tbl:discern_pats}. 
         Industry classification is based on 2-digit SIC: Columns 1 and 8 include ``Chemicals And Allied Products'' (SIC 28);  Columns 2 and 9 include ``Industrial And Commercial Machinery And Computer Equipment'' (SIC 35) and ``Electronic And Other Electrical Equipment And Components, Except Computer Equipment'' (SIC 36); Columns 3 and 10 include ``Transportation Equipment'' (SIC 37) and ``Measuring, Analyzing, And Controlling Instruments; Photographic, Medical And Optical Goods; Watches And Clocks'' (SIC 38); Column 4 and 11 include ``Business Services'' (SIC 73) and ``Engineering, Accounting, Research, Management, And Related Services`` (SIC 87); Columns 5 and 12 include all other firms. Columns 6, 7, 13 and 14 include the complete sample and Chemicals is an indicator for SIC 28.
         \item Clustered (Firm) standard-errors in parentheses. Signif. Codes: ***: 0.01, **: 0.05, *: 0.1
      \end{tablenotes}
   \end{threeparttable}
\end{table}
\end{landscape}

\begin{landscape}
\begin{table}[htbp]
\scriptsize \singlespacing
   \centering
   \begin{threeparttable}[b]
      \caption{\label{tbl:field}Variation by Research Field}
      \bigskip

\begin{tabular}{l*{14}{c}}
\toprule
                    &\multicolumn{7}{c}{Pr(Subsequent Scientific Publication by Focal Authors)}                                                                                                           &\multicolumn{7}{c}{Pr(Subsequent Patent by Focal Authors, $\geq$ 3y gap)}                                                                                                                 \\\cmidrule(lr){2-8}\cmidrule(lr){9-15}
                     & Med & Chem & \shortstack{Eng\\\& ICT} & \shortstack{Math\\\& Phys} & Other & All & All & Med & Chem & \shortstack{Eng\\\& ICT} & \shortstack{Math\\\& Phys} & Other & All & All \\
                    &\multicolumn{1}{c}{OLS}&\multicolumn{1}{c}{OLS}&\multicolumn{1}{c}{OLS}&\multicolumn{1}{c}{OLS}&\multicolumn{1}{c}{OLS}&\multicolumn{1}{c}{OLS}&\multicolumn{1}{c}{2SLS}&\multicolumn{1}{c}{OLS}&\multicolumn{1}{c}{OLS}&\multicolumn{1}{c}{OLS}&\multicolumn{1}{c}{OLS}&\multicolumn{1}{c}{OLS}&\multicolumn{1}{c}{OLS}&\multicolumn{1}{c}{2SLS}\\
&(1)&(2)&(3)&(4)&(5)&(6)&(7)&(8)&(9)&(10)&(11)&(12)&(13)&(14)\\ \midrule
ln(Follow-on)       &       0.014\sym{***}&       0.003         &       0.005\sym{***}&       0.002         &       0.011\sym{***}&       0.006\sym{***}&       0.093\sym{**} &       0.009\sym{***}&       0.006\sym{***}&       0.007\sym{***}&       0.004\sym{*}  &       0.008\sym{**} &       0.007\sym{***}&      -0.014         \\
                    &     (0.002)         &     (0.002)         &     (0.002)         &     (0.003)         &     (0.004)         &     (0.002)         &     (0.045)         &     (0.002)         &     (0.002)         &     (0.002)         &     (0.002)         &     (0.003)         &     (0.001)         &     (0.060)         \\
\addlinespace
ln(Follow-on) $\times$ &                     &                     &                     &                     &                     &       0.003         &       0.073         &                     &                     &                     &                     &                     &       0.001         &       0.203         \\
  \ \ \ \ Chem \& Med                  &                     &                     &                     &                     &                     &     (0.002)         &     (0.122)         &                     &                     &                     &                     &                     &     (0.002)         &     (0.151)         \\
%\addlinespace
%Chemicals           &                     &                     &                     &                     &                     &                     &                     &                     &                     &                     &                     &                     &                     &                     \\
%                    &                     &                     &                     &                     &                     &                     &                     &                     &                     &                     &                     &                     &                     &                     \\
\addlinespace
ln(Focal H-Index)   &      -0.011\sym{***}&      -0.011\sym{*}  &       0.016\sym{***}&       0.014\sym{*}  &       0.019\sym{**} &       0.002         &      -0.034\sym{*}  &       0.010\sym{***}&       0.012\sym{**} &       0.012\sym{***}&       0.028\sym{***}&       0.019\sym{**} &       0.012\sym{***}&      -0.015         \\
                    &     (0.004)         &     (0.006)         &     (0.006)         &     (0.007)         &     (0.009)         &     (0.003)         &     (0.018)         &     (0.003)         &     (0.005)         &     (0.004)         &     (0.008)         &     (0.008)         &     (0.002)         &     (0.018)         \\
%\addlinespace
%Constant            &       0.520\sym{***}&       0.617\sym{***}&       0.332\sym{***}&       0.332\sym{***}&       0.278\sym{***}&       0.521\sym{***}&                     &       0.208\sym{***}&       0.445\sym{***}&       0.447\sym{***}&       0.385\sym{***}&       0.186\sym{***}&       0.370\sym{***}&                     \\
                    &     (0.017)         &     (0.017)         &     (0.020)         &     (0.028)         &     (0.033)         &     (0.008)         &                     &     (0.015)         &     (0.016)         &     (0.011)         &     (0.020)         &     (0.030)         &     (0.006)         &                     \\
\\
Firm FE             &         Yes         &         Yes         &         Yes         &         Yes         &         Yes         &         Yes         &         Yes         &         Yes         &         Yes         &         Yes         &         Yes         &         Yes         &         Yes         &         Yes         \\
Journal-Year FE     &         Yes         &         Yes         &         Yes         &         Yes         &         Yes         &         Yes         &         Yes         &         Yes         &         Yes         &         Yes         &         Yes         &         Yes         &         Yes         &         Yes         \\
Observations        &      78,033         &      25,173         &      47,923         &      11,282         &       7,829         &     164,516         &     164,516         &      78,033         &      25,173         &      47,923         &      11,282         &       7,829         &     164,516         &     164,516         \\
Avg. DV             &       0.594         &       0.584         &       0.372         &       0.358         &       0.369         &       0.507         &       0.507         &       0.310         &       0.512         &       0.508         &       0.470         &       0.276         &       0.397         &       0.397         \\
First Stage F-stat  &                     &                     &                     &                     &                     &                     &       3.427         &                     &                     &                     &                     &                     &                     &       3.427         \\
Adjusted R$^2$      &       0.308         &       0.409         &       0.290         &       0.405         &       0.339         &       0.345         &      -0.440         &       0.301         &       0.369         &       0.351         &       0.334         &       0.301         &       0.348         &      -0.516         \\
\bottomrule
\end{tabular}
     \begin{tablenotes}[flushleft]
      \scriptsize
         \item \textit{Notes:} This table presents variation in the baseline estimation results by publications' research field. %The analysis corresponds to Tables \ref{tbl:discern_pubs} and \ref{tbl:discern_pats}. 
         Research fields are determined by Dimensions.ai based on Australian and New Zealand Standard Research Classification (ANZSRC) 2020. Columns 1 and 8 include ``Biological Sciences'' (FOR 31), ``Biomedical and Clinical Sciences'' (FOR 32) and ``Health Sciences'' (FOR 42); Columns 2 and 9 include ``Chemical Sciences'' (FOR 34); Columns 3 and 10 include ``Engineering'' (FOR 40) and ``Information and Computing Sciences'' (FOR 46); Columns 4 and 11 include ``Mathematical Sciences'' (FOR 49) and ``Physical Sciences'' (FOR 51); Columns 5 and 12 include all other publications. Columns 6, 7, 13 and 14 include the complete sample and ``Chem \& Med'' is an indicator for FOR codes 32, 34, 31 and 34. 
         \item Clustered (Firm) standard-errors in parentheses. Signif. Codes: ***: 0.01, **: 0.05, *: 0.1
      \end{tablenotes}
   \end{threeparttable}
\end{table}
\end{landscape}

\begin{landscape}
\begin{table}[htbp]
\scriptsize \singlespacing
   \centering
   \begin{threeparttable}[b]
      \caption{\label{tbl:corpfo}Corporate vs Academic Follow-On Research}
      \bigskip

\begin{tabular}{l*{12}{c}}
\toprule
                    &\multicolumn{6}{c}{Pr(Subsequent Scientific Publication by Focal Authors)}                                                                                                           &\multicolumn{6}{c}{Pr(Subsequent Patent by Focal Authors, $\geq$ 10y gap)}                                                                                                                          \\\cmidrule(lr){2-7}\cmidrule(lr){8-13}
                    &\multicolumn{1}{c}{OLS}&\multicolumn{1}{c}{2SLS}&\multicolumn{1}{c}{OLS}&\multicolumn{1}{c}{2SLS}&\multicolumn{1}{c}{OLS}&\multicolumn{1}{c}{2SLS}&\multicolumn{1}{c}{OLS}&\multicolumn{1}{c}{2SLS}&\multicolumn{1}{c}{OLS}&\multicolumn{1}{c}{2SLS}&\multicolumn{1}{c}{OLS}&\multicolumn{1}{c}{2SLS}\\
&(1)&(2)&(3)&(4)&(5)&(6)&(7)&(8)&(9)&(10)&(11)&(12)\\ \midrule
ln(Follow-On)       &       0.008\sym{***}&       0.123\sym{**} &                     &                     &                     &                     &       0.003\sym{***}&       0.096\sym{**} &                     &                     &                     &                     \\
                    &     (0.001)         &     (0.048)         &                     &                     &                     &                     &     (0.001)         &     (0.040)         &                     &                     &                     &                     \\
\addlinespace
ln(Non-Corporate Follow-On)&                     &                     &       0.008\sym{***}&       0.127\sym{***}&                     &                     &                     &                     &       0.003\sym{***}&       0.101\sym{**} &                     &                     \\
                    &                     &                     &     (0.001)         &     (0.048)         &                     &                     &                     &                     &     (0.001)         &     (0.041)         &                     &                     \\
\addlinespace
ln(Corporate Follow-On)&                     &                     &                     &                     &       0.015\sym{***}&       0.127\sym{**} &                     &                     &                     &                     &       0.003\sym{***}&       0.114\sym{**} \\
                    &                     &                     &                     &                     &     (0.002)         &     (0.056)         &                     &                     &                     &                     &     (0.001)         &     (0.047)         \\
\addlinespace
ln(Focal H-Index)   &       0.002         &      -0.030\sym{**} &       0.002         &      -0.032\sym{**} &      -0.001         &      -0.024\sym{**} &       0.002         &      -0.024\sym{**} &       0.002         &      -0.025\sym{**} &       0.003         &      -0.020\sym{**} \\
                    &     (0.003)         &     (0.014)         &     (0.003)         &     (0.014)         &     (0.003)         &     (0.012)         &     (0.002)         &     (0.011)         &     (0.002)         &     (0.011)         &     (0.002)         &     (0.009)         \\
%\addlinespace
%Constant            &       0.463\sym{***}&                     &       0.466\sym{***}&                     &       0.468\sym{***}&                     &       0.119\sym{***}&                     &       0.120\sym{***}&                     &       0.128\sym{***}&                     \\
%                    &     (0.013)         &                     &     (0.013)         &                     &     (0.011)         &                     &     (0.006)         &                     &     (0.006)         &                     &     (0.005)         &                     \\
\\
Firm FE             &         Yes         &         Yes         &         Yes         &         Yes         &         Yes         &         Yes         &         Yes         &         Yes         &         Yes         &         Yes         &         Yes         &         Yes         \\
Journal-Year FE     &         Yes         &         Yes         &         Yes         &         Yes         &         Yes         &         Yes         &         Yes         &         Yes         &         Yes         &         Yes         &         Yes         &         Yes         \\
Observations        &     164,516         &     164,516         &     164,154         &     164,154         &     154,563         &     154,563         &     164,516         &     164,516         &     164,154         &     164,154         &     154,563         &     154,563         \\
Avg. DV             &       0.507         &       0.507         &       0.507         &       0.507         &       0.515         &       0.515         &       0.143         &       0.143         &       0.143         &       0.143         &       0.147         &       0.147         \\
First Stage F-stat  &                     &      26.795         &                     &      25.457         &                     &      33.458         &                     &      26.795         &                     &      25.457         &                     &      33.458         \\
Adjusted R$^2$      &       0.344         &      -0.382         &       0.345         &      -0.402         &       0.347         &      -0.320         &       0.351         &      -0.495         &       0.351         &      -0.526         &       0.352         &      -0.512         \\
\bottomrule
\end{tabular}
     \begin{tablenotes}[flushleft]
      \scriptsize
         \item \textit{Notes:} This table presents an analysis of follow-on research split by the type of source organization. %The analysis corresponds to the baseline results presented in Tables \ref{tbl:discern_pubs} and \ref{tbl:discern_pats}. 
         Columns 1, 2, 7 and 8 replicate the baseline results. In columns 3, 4,  9 and 10, the independent variable of interest is a count of three generations of follow-on research that originates from academic and other non-private institutions. In columns 5, 6, 11 and 12, the independent variable of interest is a count of three generations of follow-on research that originates from other firms. Institution type is determined based on a classification of authorship affiliations provided by the Dimensions dataset.
         \item Clustered (Firm) standard-errors in parentheses. Signif. Codes: ***: 0.01, **: 0.05, *: 0.1
      \end{tablenotes}
   \end{threeparttable}
\end{table}
\end{landscape}

\begin{table}[htbp]
\footnotesize \singlespacing
   \centering
   \begin{threeparttable}[b]
      \caption{\label{tbl:mech_interactions}Ways in Which Follow-On Research Is Privately Useful}
      \bigskip

%\begin{tabular}{l*{8}{c}}
\begin{tabularx}{\textwidth}{l@{}*{4}{Y}}
\toprule
                    &\multicolumn{2}{c}{Pr(Patent, Citing FO)}  &\multicolumn{2}{c}{Pr(Patent, Not Citing FO)}\\\cmidrule(lr){2-3}\cmidrule(lr){4-5}
                    &\multicolumn{1}{c}{OLS}&\multicolumn{1}{c}{2SLS}&\multicolumn{1}{c}{OLS}&\multicolumn{1}{c}{2SLS}\\
&(1)&(2)&(3)&(4)\\ \midrule
ln(Follow-on) $\times$ H&       0.002\sym{**} &      -0.001         &       0.001         &      -0.015\sym{***}\\
                    &     (0.001)         &     (0.003)         &     (0.001)         &     (0.005)         \\
\addlinespace
High Focal H-Index  &      -0.002         &      -0.004         &      -0.005         &      -0.007         \\
                    &     (0.002)         &     (0.004)         &     (0.004)         &     (0.005)         \\
\addlinespace
ln(Follow-on)       &       0.011\sym{***}&       0.055\sym{**} &       0.005\sym{***}&       0.095\sym{**} \\
                    &     (0.002)         &     (0.022)         &     (0.001)         &     (0.045)         \\
\addlinespace
ln(Focal H-Index)   &       0.003\sym{**} &      -0.008         &       0.009\sym{***}&      -0.014         \\
                    &     (0.001)         &     (0.005)         &     (0.002)         &     (0.012)         \\
\\
Firm FE             &         Yes         &         Yes         &         Yes         &         Yes         \\
Journal-Year FE     &         Yes         &         Yes         &         Yes         &         Yes         \\
Observations        &     164,516         &     164,516         &     164,516         &     164,516         \\
Avg. DV             &       0.039         &       0.039         &       0.300         &       0.300         \\
First Stage F-stat  &                     &      13.126         &                     &      13.126         \\
Adjusted R$^2$      &       0.133         &      -0.319         &       0.366         &      -0.332         \\
\bottomrule
%\end{tabular}
\end{tabularx}
      \begin{tablenotes}[flushleft]
      \scriptsize
      \item \textit{Note.} This table explores how the effect of external follow-on research affects patenting outcomes in different ways. The dependent variables are indicators for subsequent patents by the focal authors, filed at least five years after the focal publication. In columns 1-2, the dependent variable is an indicator for a subsequent patent that cites the follow-on research. In columns 3-4, the dependent variable is an indicator for a subsequent patent that does not cite the follow-on research. In all models, follow-on research is interacted with an indicator for above-average prominence of the top focal author (within each journal). Prominence is measured using H-index values at the time of publication. In all models, log(follow-on) is recentered around the sample mean. In 2SLS models, ln(follow-on) and the interaction term are instrumented by the the IV and the interaction of the IV with the indicator variable.
         %\item 
         Clustered (Firm) standard-errors in parentheses. Signif. Codes: ***: 0.01, **: 0.05, *: 0.1
      \end{tablenotes}
   \end{threeparttable}
\end{table}

\begin{table}[htbp]

\scriptsize
\centering
\caption{\label{tbl:discern_firm_descstats}
Descriptive Statistics for Firm-Level Analysis}
\begin{threeparttable}
%\begin{tabular}{lHrrrrrrrr}
\begin{tabularx}{\textwidth}{l*{8}{Y}}
\toprule
Variable & Mean & SD & Min & p25 & p50 & p75 & Max \\
\midrule
Year & $2,000.0$ & $9.1$ & $1,981$ & $1,993$ & $2,001$ & $2,007$ & $2,015$ \\
Annual Publications & $11.6$ & $63.3$ & $0$ & $0$ & $0$ & $3$ & $1,590$ \\
AMWS Scientist Stock & $6.7$ & $38.2$ & $0$ & $0$ & $0$ & $2$ & $1,014$ \\
Award-Winning Scientist Stock & $0.9$ & $6.6$ & $0$ & $0$ & $0$ & $0$ & $170$ \\
Patents & $34.7$ & $164.5$ & $0$ & $0$ & $2$ & $14$ & $8,842$ \\
Follow-On Stock & $34,653.4$ & $198,857.1$ & $0$ & $1$ & $72$ & $2,509$ & $5,299,716$ \\
$\ln(\text{FO Stock})_{t-1}$ & $4.5$ & $4.1$ & $-1$ & $0$ & $4$ & $8$ & $15$ \\
No Follow-On Stock & $0.3$ & $0.4$ & $0$ & $0$ & $0$ & $1$ & $1$ \\
Firm's Patent Stock Citing FO & $14.7$ & $153.9$ & $0$ & $1$ & $1$ & $1$ & $11,638$ \\
$\ln(\text{Firm's pat. stock citing FO})_{t-1}$ & $0.4$ & $1.2$ & $-5$ & $0$ & $0$ & $0$ & $9$ \\
External Patent Stock Citing FO & $1,658.4$ & $8,988.2$ & $0$ & $1$ & $1$ & $59$ & $253,669$ \\
$\ln(\text{Ext. pat. stock citing FO})_{t-1}$ & $2.2$ & $3.1$ & $-4$ & $0$ & $0$ & $4$ & $12$ \\
$\text{Assets}_{t-1}$ & $2,338.8$ & $11,653.6$ & $0$ & $13$ & $110$ & $884$ & $333,774$ \\
$\text{R\&D}_{t-1}$ & $560.2$ & $2,751.2$ & $0$ & $11$ & $48$ & $190$ & $54,593$ \\
\bottomrule
%\end{tabular}
\end{tabularx}
\begin{tablenotes}[flushleft]
      \scriptsize
         \item \textit{Note.} This table provides summary statistics for the variables used in the econometric analysis at the firm-year level. The data is based on the firm-level panel constructed from the DISCERN database. The sample includes 35156 firm-year observations from 2338 firms.

\end{tablenotes}
\end{threeparttable}
\end{table}

\clearpage

\printbibliography
\end{refsection}

\clearpage

\begin{refsection}

\section{\label{sec:app_iv}Discussion of Instrumental Variable} 

\setcounter{table}{0}
\renewcommand{\thetable}{D\arabic{table}}
\setcounter{figure}{0}
\renewcommand{\thefigure}{D\arabic{figure}}

\small
\onehalfspacing

In this section, I discuss the main assumptions underlying my instrumental variable approach for identifying the effects of external follow-on research on firms' innovation outcomes.

\subsection{Instrument Relevance}

The instrument's relevance relies on the notion of peer effects across publications grouped together in physical journal issues. The elements of this process are discussed in detail in section \ref{sec:iv}. In short, until academic readership moved online in the early 2000s, academics accessed most scientific publications by walking to their institution’s library and checking out physical journal issues. As a result, publications that were grouped in the same journal issue with a publication by a prominent researcher were likely to be circulated more often than others. Therefore, the level of attention to journal publications could have varied irrespectively of the content and quality of a given paper. I argue that serendipitous increases in academic attention sometimes translate into meaningful follow-on research. This research can then be observed through the number of citations the focal publication received. 

\begin{figure}[h]
\caption{\label{fig:firststage}First Stage Relation}
\captionsetup{width=0.8\textwidth}
\centering
\includegraphics[width=0.6\textwidth]{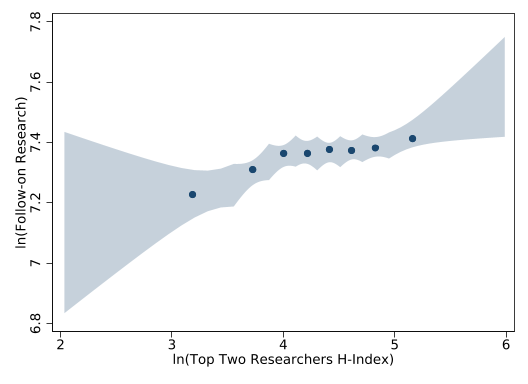}
\caption*{\footnotesize \textit{Note:} This figure presents a binned scatterplot of the relation between the logged sum of the top two researcher H-indexes (the instrument) and logged follow-on citations to the focal publication (endogenous variable). The values in the plot are fitted values after controlling for the logged H-index of the focal author, firm fixed effects and journal-year fixed effects.}
\end{figure}

Table \ref{tbl:firststage} reports the first-stage coefficient estimates for the instrument's relevance. First, I use the top one H-index among all other authors in the same journal issue. Next, I consider the sum of the H-indexes of the top two authors. Since the predictive power is stronger under this specification, I chose it as the instrument across all analyses in the paper. Figure \ref{fig:firststage} presents a corresponding binned scatterplot. In addition to the chosen specification, I report in Table \ref{tbl:firststage}  coefficient estimates of alternative specifications for the instrument. First, I use the count of the authors' publications up to the year before the focal publication year. Second, I use the citation-weighted measure of the same publications. In all cases, I find strong evidence for the relevance of the instrument for the count of follow-on citations.

The Kleibergen-Paap cluster-robust first-stage F-statistic is 26.8, substantially exceeding the conventional threshold of 10. A Durbin-Wu-Hausman test confirms that follow-on research is endogenous (F = 7.57, p = 0.006), justifying the use of instrumental variables. To ensure results are robust to potential weak instrument concerns, I implement Anderson-Rubin tests, which remain valid under arbitrary instrument weakness and clustering. The AR test strongly rejects the null hypothesis that follow-on research has no effect on subsequent publications ($\chi^2 = 10.26$, $p = 0.001$).

\begin{figure}[h]
\caption{\label{fig:ivtrend}First Stage Time Trends}
\captionsetup{width=0.8\textwidth}
\centering
\includegraphics[width=0.6\textwidth]{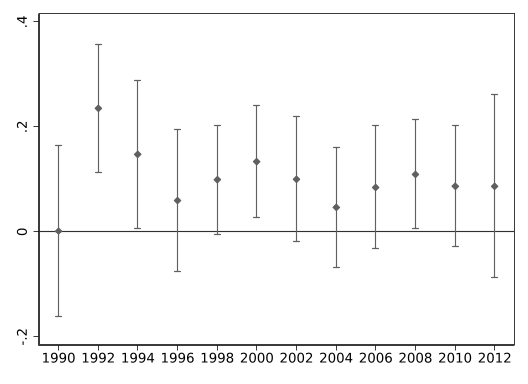}
\caption*{\footnotesize \textit{Note:} This figure presents coefficient estimates for time trends of the first stage. The reported coefficients are of interactions between the instrument and 2-year indicators. The regression includes firm and journal-year fixed effects. Standard errors are clustered by firm.}
\end{figure}

To further support the mechanism that drives the instrument relevance, I explore the time trend of coefficient estimates. Figure \ref{fig:ivtrend} presents the coefficient estimates of the interaction between the instrumental variable and 2-year indicators. As expected, the correlation between the instrument and follow-on citations was stronger during the 1990s, before academic readership moved online. Starting in the 2000s, I observe lower point estimates and larger standard errors. These trends suggest that in later years the journal issue peer effects got weaker. Potentially, these trends are due to increase in online readership.

\subsection{Conditionally Unconfounded Instrument}

Unconfoundedness of the instrument requires that there are no unmeasured common causes between the instrument and the endogenous variable (follow-on citations), and between the instrument and the second stage outcomes of interest. I will discuss these assumptions and provide supporting evidence.

More prominent authors will tend to publish in more prestigious journals. The measurement of prominence through H-indexes varies across years. In addition, over time, some journals became more prestigious and influential compared to others. To account for these differences across journals and time, I condition all models on a strict set of fixed effects. Namely, I compare publications in different journal issues of the same journal and in the same year by including a set of journal-year fixed effects. Within a journal-year and given the first-in-first-out assignment process of manuscripts into journal issues, I claim that it is unlikely that confounders will drive the allocation of accepted manuscripts into specific journal issues. Exceptions to this assignment process are special issues and conference proceedings. Using indicators obtained from Web of Science and Dimensions data, I drop these cases from the sample.

\begin{figure}[h]
\caption{\label{fig:hindexcorr}Author H-index Correlation}
\captionsetup{width=0.8\textwidth}
\centering
\includegraphics[width=0.6\textwidth]{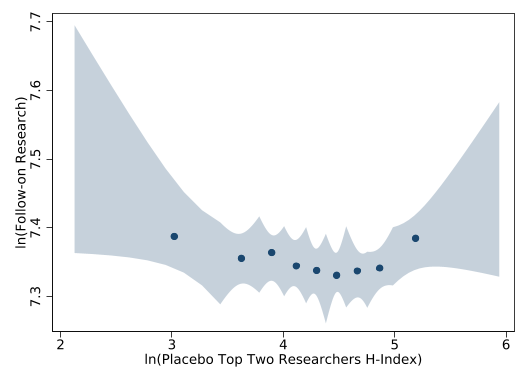}
\caption*{\footnotesize \textit{Note:} This figure presents a binned scatterplot of the relation between the logged sum of top two researcher H-indexes from a random journal issue in the same journal-year (the placebo) and the logged follow-on citations to the focal publication (endogenous variable). The values in the plot are fitted values after controlling for firm and journal-year fixed effects.  A corresponding linear regression reports a statistically insignificant slope estimate of 0.0143 (s.e. = 0.0158).}
\end{figure}

To support the unconfoundedness assumption, Figure \ref{fig:hindexcorr} presents a binned scatterplot of the relation between the instrument and the H-index of the top author of the focal paper. A corresponding linear regression reports a statistically insignificant slope estimate of 0.0093 (s.e. = 0.0099). In both the scatterplot and the regression estimates, there is no evidence that within a journal and year, more prominent authors jointly publish in specific journal issues. Nonetheless, in all model specifications I include the top focal H-index as an additional control.

To further support the validity of the instrument, I perform a placebo test. In this test, I replace my instrument with a corresponding measure of top H-indexes from a randomly-picked journal issue within the same journal and year. In this test, the prominence of authors should not be relevant for follow-on citations. Figure \ref{fig:hindexcorr} presents the test results. A corresponding linear regression reports a statistically insignificant slope estimate of 0.0143 (s.e. = 0.0158). According to the plot, there is no indication that prominence of authors from other journal issues drive attention (and therefore citations) to the focal publication. These results provide additional support for the mechanism that drives the relevance of the instrument.

Taken together, the results discussed above provide support for the assumption that the instrument is conditionally unconfounded.

\subsection{Exclusion Restriction}

The exclusion restriction posits that the prominence of other authors in the same journal issue as the focal paper affects outcomes only through their effects on external follow-on citations. A threat to this assumption could occur if, for example, information frictions within the firm limit internal awareness of the firm's own publications. In that case, the prominence of other authors in the same journal issue can drive more attention by the firm's own scientists, in the same way that it drives external attention to the focal publication. This possibility is highly unlikely, specifically when considering outcomes that directly relate to the focal authors (such as counts of their future publications and patents).

Another possibility is that the prominence of other authors will drive the firms' innovative outcomes through channels that are unaccounted for by the citation counts of follow-on research. For example, this can happen if some follow-on research does not cite the focal publication, but is found to be useful by the firm. While this is a possibility, it does not interfere with the general sense that follow-on academic activity can be beneficial for the firm.

\begin{table}[htbp]
\footnotesize
\centering
\caption{\label{tbl:firststage}
First Stage Regressions}
\begin{threeparttable}
\begin{tabular}{lcccc}
\toprule
                    &\multicolumn{4}{c}{ln(3-Gen Follow-on)}                                                \\\cmidrule(lr){2-5}
                    &\multicolumn{1}{c}{OLS}&\multicolumn{1}{c}{OLS}&\multicolumn{1}{c}{OLS}&\multicolumn{1}{c}{OLS}\\
&(1)&(2)&(3)&(4)\\ \midrule
ln(Top One Researcher H-index)&       0.074\sym{***}&                     &                     &                     \\
                    &     (0.016)         &                     &                     &                     \\
\addlinespace
ln(Top Two Researchers H-index)&                     &       0.096\sym{***}&                     &                     \\
                    &                     &     (0.018)         &                     &                     \\
\addlinespace
ln(Top Two Researchers Pub. Count)&                     &                     &       0.027\sym{***}&                     \\
                    &                     &                     &     (0.010)         &                     \\
\addlinespace
ln(Top Two Researchers Cit. Count)&                     &                     &                     &       0.046\sym{***}\\
                    &                     &                     &                     &     (0.008)         \\
\addlinespace
ln(Focal H-Index)   &       0.279\sym{***}&       0.279\sym{***}&       0.279\sym{***}&       0.279\sym{***}\\
                    &     (0.010)         &     (0.010)         &     (0.010)         &     (0.010)         \\
\\
Firm FE             &         Yes         &         Yes         &         Yes         &         Yes         \\
Journal-Year FE     &         Yes         &         Yes         &         Yes         &         Yes         \\
Observations        &     164,516         &     164,516         &     164,516         &     164,516         \\
Avg. DV             &       7.357         &       7.357         &       7.357         &       7.357         \\
Adjusted R$^2$      &       0.589         &       0.589         &       0.589         &       0.589         \\
\bottomrule
\end{tabular}
\begin{tablenotes}[flushleft]
      \scriptsize
\item This table presents estimation results for the first-stage relationship between the prominence of top researchers in the same journal issue as the focal publication and external follow-on research. In columns 1 and 2, prominence is measured using the authors' H-index in the year prior to publication. In column 3, prominence is measured as the previous publication count. In column 4, prominence is measured as the previous citation-weighted publication count. 
\item Clustered (Firm) standard-errors in parentheses. Signif. Codes: ***: 0.01, **: 0.05, *: 0.1
\end{tablenotes}
\end{threeparttable}
\end{table}

\subsection{\label{sec:acr}Heterogeneity and the Average Causal Response}

The literature on instrumental variables have long acknowledged the possibility of heterogeneity across the studied population \citep{AngristPischke2009MostlyHarmlessEconometrics}. Under heterogeneity in observed and unobserved characteristics, instruments can only be used to estimate a local average treatment effect (LATE) instead of the average treatment effect (ATE). LATE refers to the average effect for a specific subset of the population, defined by their response to the instrumental variable. It may differ from the ATE, which is an estimate of the effect on the entire population. The difference between the LATE and the ATE depends on the degree of response heterogeneity and the strength of the instrumental variable.

The theoretical interpretation of the estimand is further complicated when the endogenous variable (the ``treatment'') and instrumental variable are continuous. \citet{AngristPischke2009MostlyHarmlessEconometrics} offer a generalization of the LATE framework to accommodate variable treatment intensity. The Average causal response (ACR) is a weighted average of the unit causal response, which in turn is the average difference in potential outcomes for compliers at different levels of treatment. When the treatment is fully continuous, IV estimation will recover the average derivative across the range of treatment values. When the instrument itself is continuous, the estimation will produce a weighted average of derivatives across the range of values of the instrument \citep{CornelissenDustmannRauteEtAl2016LATEMTEAlternative}. 

Many of the coefficient estimates of the 2SLS regressions presented in this paper are larger than their OLS counterparts. However, given the continuous nature of the instrument and endogenous variables, it is possible that these differences are due to the weighted nature of the ACR. Therefore, a direct comparison between the 2SLS and OLS estimates might be misleading and an analysis of the direction of bias is not trivial. 

\begin{figure}[h]
\caption{\label{fig:iv_by_H}First Stage, by Focal H-index}
\captionsetup{width=0.8\textwidth}
\centering
\includegraphics[width=0.6\textwidth]{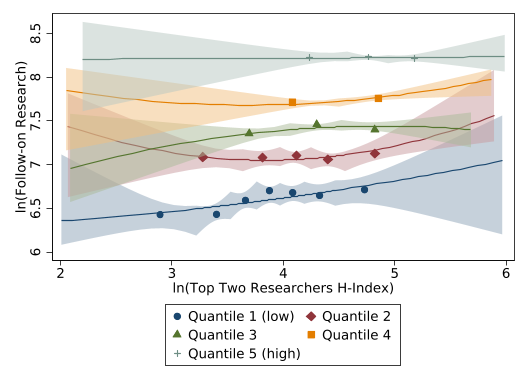}
\caption*{\footnotesize \textit{Note:} This figure presents a binned scatterplot of the relation between the logged sum of the top two researcher H-indexes (the instrument) and logged follow-on citations to the focal publication (endogenous variable), across levels of the focal authors' H-index. The values in the plot are fitted values after controlling for the logged H-index of the focal author, firm fixed effects and journal-year fixed effects.}
\end{figure}

\begin{figure}[h]
\caption{\label{fig:pubs_by_H}Follow-On Research and Subsequent Publications, by Focal H-index}
\captionsetup{width=0.8\textwidth}
\centering
\includegraphics[width=0.6\textwidth]{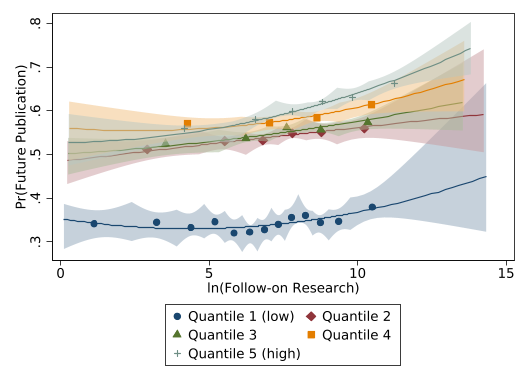}
\caption*{\footnotesize \textit{Note:} This figure presents a binned scatterplot of the relation between the logged follow-on citations and the probability of subsequent scientific publishing by the focal authors, across levels of the focal authors' H-index. The values in the plot are fitted values after controlling for the logged H-index of the focal author, firm fixed effects and journal-year fixed effects.}
\end{figure}

\begin{figure}[h]
\caption{\label{fig:pats_by_H}Follow-On Research and Subsequent Patenting, by Focal H-index}
\captionsetup{width=0.8\textwidth}
\centering
\includegraphics[width=0.6\textwidth]{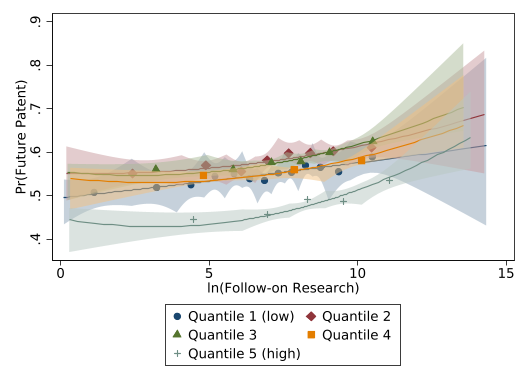}
\caption*{\footnotesize \textit{Note:} This figure presents a binned scatterplot of the relation between the logged follow-on citations and the probability of subsequent patenting by the focal authors, across levels of the focal authors' H-index. The values in the plot are fitted values after controlling for the logged H-index of the focal author, firm fixed effects and journal-year fixed effects.}
\end{figure}

Further analysis of the first stage effects provides evidence for treatment heterogeneity across covariates. Figure \ref{fig:iv_by_H} presents the first stage relation, across five quantiles of the focal authors' H-index. As expected, the effect is stronger for publications by less prominent authors. In addition, for these authors, the level of follow-on seems more strongly correlated with the probability of subsequent scientific publishing (Figure \ref{fig:pubs_by_H}). While these relations seem not to hold for the case of patenting (Figure \ref{fig:pats_by_H}), the focal H-index is only one dimension of potential treatment heterogeneity.

\clearpage
\subsection{\label{sec:iv_sim}Omitted Variable Bias in OLS}

There are various sources of omitted variable bias in OLS estimates. A typical concern is the existence of an unobserved confounder (e.g., scientific quality). However, an additional source of bias, that is often overlooked, is a difference between the functional form of the data generating process and the observed variables. In such case, the direction of bias will depend on parameters of the model and can result in OLS estimates that are smaller than IV estimates, even when potential confounders would predict an upward bias of OLS. Consider the following data generating process:

\begin{align*}
X_1 &= Z_1 + \mu_1 \\
X_2 &= Z_2 + \mu_2 \\
X &= \delta X_1 + (1-\delta) X_2 \\
Y &=  \alpha \delta X_1 + 0.5 (1-\delta)X_2 + \mu_y \\
\end{align*}

The data includes two independent variables, $X_1$ and $X_2$. These variables are functions of instruments $Z_1$ and $Z_2$, respectively. $Y$ is a linear combination of $X_1$, $X_2$, with $\delta$ defining the relative weights. The coefficient on $(1-\delta)X_2$ is set to 0.5 and the coefficient on $\delta X_1$ is $\alpha$. Importantly, the researchers only observe $X$, $Y$ and $Z_2$.  Therefore,  they estimate a 2-stage model as follows:

\begin{align*}
X &= \eta_0 + \eta_1 Z_2 + \xi \\
Y &= \beta_0 + \beta_1 X + \epsilon
\end{align*}

To clearly see the source of bias, consider the following rearrangement:

\begin{align*}
Y &= (\alpha \delta X_1 + 0.5 (1-\delta)X_2 + \mu_y) + (0.5\delta X_1 - 0.5\delta X_1) \\
  &= (\alpha - 0.5)\delta X_1 + 0.5 \delta X_1 + 0.5 (1-\delta)X_2 + \mu_y \\
  &= 0.5X + U
\end{align*}

Where $U \equiv \mu_y + (\alpha-0.5)\delta X_1$. The instrument $Z_2$ is uncorrelated with $U$, and therefore $\widehat \beta_1^{IV}$ would be unbiased regardless of the value of $\alpha$. However, $X$ itself is correlated with $U$ through the joint dependence on $X_1$, and the direction of bias of $\widehat \beta_1^{OLS}$ will depend on the sign of $(\alpha-0.5)$.

\begin{landscape}
\begin{figure}[h]
\caption{\label{fig:iv_sim}OLS and IV Coefficient Estimates, Simulation Results}
\captionsetup{width=0.8\textwidth}
\centering
\includegraphics[width=\linewidth]{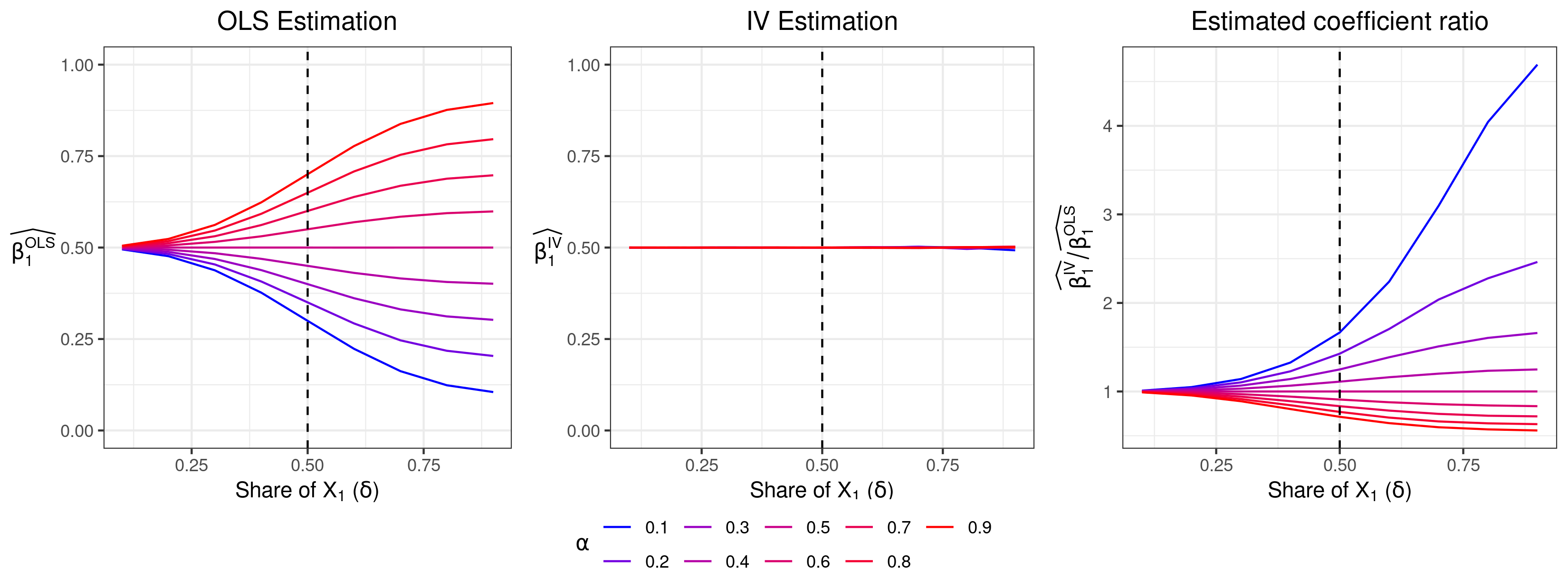}
\caption*{\footnotesize \textit{Note:} This figure presents simulated results comparing OLS and IV coefficient estimates. The parameters used are as follows: $Z_1 \sim \mathcal{N}(0,25)$, $Z_2 \sim \mathcal{N}(0,25)$, $X_1 = 10 Z_1 + \mu_1$, $X_2 = 10 Z_2 + \mu_2$. All noise variables ($\mu_1, \mu_2, \mu_y$) are distributed $\mathcal{N}(0, 100)$. Sample size is $50000$. }
\end{figure}
\end{landscape}

I provide an analysis of simulated data to explore this possibility. Results are presented in Figure \ref{fig:iv_sim}. Clearly, when $\alpha = 0.5$ (i.e., the true coefficients on $X_1$ and $X_2$ are equal) then the OLS estimates are similar to the IV estimates. However, when $\alpha<0.5$, then the OLS coefficient estimates are lower than the corresponding IV estimates, and the magnitude of difference increases with $\delta$ (the share of $X_1$ in $Y$). Alternatively, when $\alpha>0.5$, the OLS coefficient estimates will be larger. 

In the context of this paper, $X$ is a number of citations for each publication and $Z_2$ is the observed instrument, however the instrument might influence only ``marginal'' citations, which are a part of $X$ but unknown in size. Possibly, ``marginal'' and ``core'' citations have different effects on $Y$ (note that the simulation uses normal distributions for simplicity). The implications of this analysis on the interpretation of results presented in the paper are that the functional form of the data generating process of citations can result in OLS coefficient estimates that are smaller than the IV estimates. If ``marginal'' citations have a stronger effect on the outcome, compared to ``core'' citations that are unaffected by the IV, then OLS coefficients will be smaller than the IV estimates. Note that this result can happen regardless of the direction of potential confounders and without treatment heterogeneity.

\printbibliography
\end{refsection}

\stopcontents[appendices]
\end{appendices}

\end{document}